\documentclass[11pt,preprintnumbers,nofootinbib,prd]{revtex4}
\usepackage{graphicx}
\usepackage{amsmath}
\usepackage{mathrsfs}
\usepackage{epsfig}

\newcommand{\be}{\begin{equation}}
\newcommand{\ee}{\end{equation}}
\newcommand{\ba}{\begin{eqnarray}}
\newcommand{\ea}{\end{eqnarray}}
\newcommand{\notE}{E\kern-0.6em\hbox{/}\kern0.05em}
\newcommand{\notEt}{E_{T}\kern-1.21em\hbox{/}\kern0.45em}
\newcommand{\notsusy}{SUSY\kern-1.21em\hbox{/}\kern0.45em}

\begin{document}
\begin{flushright}
MCTP-06-07 \\
NSF-KITP-06-111
\end{flushright}
\title{LHC String Phenomenology}
\author{Gordon L. Kane}
\author{Piyush Kumar}
\author{Jing Shao}
\affiliation{Michigan Center for Theoretical Physics, University
of Michigan, Ann Arbor, MI 48109, USA}
\date{\today}
\begin{abstract}
We argue that it is possible to address the deeper LHC Inverse
Problem, to gain insight into the underlying theory from LHC
signatures of new physics. We propose a technique which may allow
us to distinguish among, and favor or disfavor, various classes of
underlying theoretical constructions using (assumed) new physics
signals at the LHC. We think that this can be done with limited
data $(5-10\,fb^{-1})$, and improved with more data. This is
because of two reasons -- a) it is possible in many cases to
reliably go from (semi)realistic microscopic string constructions
to the space of experimental observables, say, LHC signatures. b)
The patterns of signatures at the LHC are sensitive to the
structure of the underlying theoretical constructions. We
illustrate our approach by analyzing two promising classes of
string compactifications along with six other string-motivated
constructions. Even though these constructions are not complete,
they illustrate the point we want to emphasize. We think that
using this technique effectively over time can eventually help us
to meaningfully connect experimental data to microscopic theory.

\end{abstract}
\maketitle
\newpage
\vspace{-1.2cm} \tableofcontents

\newpage

\section{Introduction}
The start of the LHC will usher in a new era of particle physics.
Hopefully, physics beyond the standard model will be discovered.
Many possibilities for new physics have been proposed and their
phenomenological implications have been studied in detail. When
the LHC starts accumulating data, one would like to answer the
following little studied question -- Assuming a signal for physics
beyond the standard model, how can one determine the nature of new
physics from LHC data? This question, the so-called ``\emph{LHC
Inverse Problem}", has received relatively little attention until
very recently. The LHC Inverse Problem\footnote{While we focus on
the LHC here, if new physics is discovered at the Tevatron, the
approach we advocate will be still valid.} is actually multiple
questions -- a) Is the new physics supersymmetry, or large extra
dimensions or something else , b) What is the spectrum of
particles and the effective theory at collider scales, and c) What
is the structure of the underlying deeper, perhaps short distance,
theory.

Recently, attention has been drawn to parts a) and b) of the LHC
Inverse Problem with encouraging results \cite{Battaglia:2005zf}.
However, part c) of the Inverse Problem has not even been
addressed in a systematic way. This paper is intended to be a step
towards that goal. In order to even have a shot at addressing the
deeper Inverse Problem in a meaningful way, one has to answer the
following two questions in the affirmative:
\begin{itemize}
\item A) Is it possible to reliably go from a ``reasonable''
microscopic construction, such as a specific class of string
constructions, to the ``real world'', say, the space of LHC
signatures? \item B) If yes, then are experimentally measured
observables sensitive to the properties of the underlying
microscopic construction, or equivalently, is it possible to
distinguish different microscopic constructions on the basis of
experimental observables?
\end{itemize}
We would like to propose and explore an approach which allows us
to answer both questions in the affirmative for many
semi-realistic string constructions which can be described within
the supergravity approximation. The basic idea that this can be
done was first proposed in \cite{Binetruy:2003cy}. Our study in
this paper shows that the idea is very promising and it is
possible to realize it in a concrete way.

By studying the pattern of signatures (signatures that are real
experimental observables) for many classes of realistic
microscopic constructions, one may be able to rule out some
classes of underlying theory constructions giving rise to the
observed physics beyond the standard model, and be pointed towards
others. Our results suggest that a lot of this can be done with
limited data and systematically improved with more data and better
techniques.

A traditional way to approach new physics data is to construct an
effective lagrangian that describes the data, in the context of a
general framework. For example, in the case of supersymmetry, one
would write the full supersymmetric soft-breaking lagrangian at
the electroweak scale, determine as many of its parameters as
possible from the data, then try to deduce or guess the shorter
distance theory from the effective theory. This procedure has many
obstacles. One is that even though the underlying theory may have
very few parameters, the effective theory will have many, as is
evident from the case of supersymmetry. A second obstacle is the
ambiguities and degeneracies in determining the effective theory
parameters, which are now well documented. Our approach can be
viewed as an attempt to bypass many of these difficulties by using
the {\it patterns} of LHC signatures in conjunction with a {\it
systematic} analysis of string theory predictions. Of course, the
two approaches in practice would be pursued in parallel, and would
strengthen each other.

For concreteness, in this paper we focus on traditional low-scale
supersymmetry as new physics beyond the standard model and the
underlying theoretical framework of string theory with different
constructions giving rise to low-scale supersymmetry. While there
exist other possibilities for new physics beyond the standard
model such as technicolor \cite{Weinberg:1975gm}, large extra
dimensions \cite{Arkani-Hamed:1998rs}, warped extra dimensions
\cite{Randall:1998uk}, higgsless models \cite{Nomura:2003du},
composite higgs models \cite{Kaplan:1983sm}, little higgs models
\cite{Arkani-Hamed:2002qx}, split supersymmetry
\cite{Arkani-Hamed:2004fb}, etc., low-scale supersymmetry remains
the most appealing - both theoretically and phenomenologically. In
addition, even though some of these other possibilities may be
embedded in the framework of string theory, low-scale
supersymmetry is perhaps the most natural and certainly the most
popular possibility arising from string constructions. Having said
the above, we would like to emphasize that the proposed technique
is completely general and can be used for any new physics arising
from any theoretical framework whatsoever.

The paper is organized as follows. Section \ref{string-pheno}
argues that it is meaningful to do string phenomenology at the
present time, a point of view questioned by some people. This is
followed by examples of semi-realistic string vacua as well as
examples of string-motivated constructions. We then present a
summary of results for the pattern table of these benchmark
constructions in section \ref{results}, so that the reader can see
where we are heading. The details of the procedure are spelled out
in section \ref{procedure}. Section \ref{distinguishibility}
describes the distinguishibility of signatures, with detailed
discussions of the connections of the signatures to the
superpartner spectrum in section \ref{spectrum}, discussions of
the connections of the spectrum to the soft parameters in section
\ref{fromsoft} and discussions of the connections of the soft
parameters to the theoretical structure in section
\ref{fromtheory}. Section \ref{qualitative} discusses how one can
extract general lessons from the analysis of specific classes of
constructions and use them to distinguish theories qualitatively
in terms of a combination of phenomenologically relevant features.
Section \ref{limitations} discusses possible limitations, followed
by conclusions and suggestions for the future in section
\ref{conclude}. In the Appendix, a description of the various
string-motivated constructions used in our study is provided.

\section{Why String Phenomenology?}\label{string-pheno}

Before proceeding to answering question A) in detail, it is worthwhile to explain that it is meaningful to do string phenomenology.
Naively speaking, one could complain that string phenomenology is a useless exercise for the following reasons : a)
There is still no non-perturbative or background independent definition of string theory. b) We have a very poor
understanding of the full M theory moduli space of vacua.

However, the situation is not so hopeless as it may seem. For
instance, even though we may not have a good understanding of the
full M theory moduli space, in recent years there has been a lot
of progress in understanding aspects of moduli stabilization and
supersymmetry breaking in various corners of the M theory moduli
space, at least in the supergravity regime, possibly with some
stringy ($\alpha'$) and quantum corrections. In addition, model
building in heterotic and type II string theories is a healthy
area of research with many semi-realistic examples, and new
approaches to model building are emerging. Therefore, it is
reasonable to expect that it is possible to go from many classes
of string vacua (at least in the supergravity regime, with
reasonable assumptions) to the real world. In the following
sections, we explicitly carry out this procedure for various
examples. The idea, therefore, is to carry out this exercise for
as many classes of vacua for which it is possible to compute
signatures for low energy observables in a reliable way and then
use the correlations in experimental observables to distinguish
among them as well as learn about the microscopic theory.

In our opinion, with recent evidence for a landscape of string
vacua and the absence of a deep underlying principle which selects
a special class of vacua or points to more general classes of
vacua, such a pragmatic approach is a sensible one if one is still
interested in connecting string theory to the real world. Of
course, a better understanding of the structure of the full theory
would sharpen our approach further and make it even more useful.

\section{The Hierarchy Problem as a motivation for realistic String Vacua}\label{Scales}

Before moving on to discuss examples of realistic string vacua, it
is important to understand how one gets small mass scales in
string theory. A priori, a string theoretical construction
naturally contains only the Planck scale and the string scale as
inputs. All other mass scales have to come out of various
combinations of these scales. From experiment, we know that there
exists a scale governed by the mass of the $Z$ and $W$ bosons,
known as the Electroweak scale. In addition, a weakly interacting
massive particle (WIMP) dark matter favored by observations works
well if it is also at the Electroweak scale. The origin and
stability of the Electroweak scale are two of the most challenging
problems in physics, known as the \emph{Hierarchy Problem}. In
order to solve the Hierarchy Problem or at least accommodate the
Hierarchy, one has to obtain a small scale ($\sim$ TeV) in string
theory from an intrinsic high scale like the Planck scale. In the
context of a low energy supersymmetry framework arising from a
string compactification, this means that the soft supersymmetry
breaking parameters have to be around the TeV scale, implying that
they are in the observable range of the LHC.

Many string theory vacua do not have any mechanism to generate and
stabilize the Hierarchy, but many do. The mechanism to obtain a
small scale (generate the Hierarchy) varies among different
classes of string vacua. If one wants a high string scale ($\geq
M_{unif}$), one mechanism to generate hierarchies is by strong
gauge dynamics in the hidden sector. Keeping the string scale
high, a second mechanism is to utilize the discretuum of flux
vacua and obtain a small scale by tuning the flux superpotential
to be very small in Planck units. A third way of obtaining a small
scale is to relax the requirement of a high string scale, making
it sufficiently small\footnote{The precise value will depend on
explicit constructions.}. The two examples of string vacua
reviewed in the next section use the second and third mechanisms
mentioned above. We see therefore that although it is possible to
accommodate a small scale in many classes of string vacua, it is
very hard to {\it explain} the precise value of the small scale by
fundamental principles, at present. The precise values are
governed by experimental and phenomenological considerations. So,
even though the situation is not perfectly satisfactory from a
theoretical point of view, it still allows us to look at
experimental predictions of these special classes of vacua.

Once one obtains a small overall mass scale in string vacua which
have a mechanism to obtain a small scale, the precise spectrum
pattern of superpartners at the small scale is determined by a
possible little hierarchy among the various soft parameters as
well as many phenomenological and experimental constraints. The
most important experimental constraints are getting correct
electroweak symmetry breaking, upper bound on the relic density,
lower bounds on superpartner masses, constraints from rare decays,
etc. Even if we restrict to studying only those classes of vacua
which can give rise to roughly the same overall small scale
($\sim$ TeV), the differences in the underlying structure of the
constructions still cause them to have different patterns of
signatures at the LHC, thereby allowing to distinguish among them.
How this actually works will be seen clearly in the following
sections.

\section{Examples}\label{examples}

As mentioned in section \ref{string-pheno}, in recent years there
has been a lot of progress in understanding dynamical issues of
moduli stabilization and supersymmetry breaking in string/$M$
theory compactifications within the validity of the supergravity
approximation. In order to have the possibility of generating
small masses compared to the Planck scale, to make them stable
against radiative corrections and to be interesting
phenomenologically, all such compactifications preserve
$\mathcal{N}=1$ supersymmetry in four
dimensions\footnote{Compactifications preserving higher
supersymmetry in four dimensions are uninteresting
phenomenologically as they do not give rise to chiral fermions.}.
When combined with gravity, this gives rise to $\mathcal{N}=1$
supergravity in four dimensions. The vacuum structure of
$\mathcal{N}=1$ supergravity is completely specified by three
functions - a holomorphic gauge kinetic function ($f$), a
holomorphic superpotential ($W$) and a real analytic function
called the K\"{a}hler potential. These functions determine the
effective scalar potential and depend on moduli in general. It has
been shown in recent years that in particular classes of string
compactifications, this scalar potential can be reliably minimized
leading to stabilization of most (all) moduli and the breaking of
supersymmetry in a regime in which the supergravity approximation
is valid. It is this class of string compactifications which we
particularly want to turn our attention to, as these classes of
constructions are most amenable to generating a hierarchy between
the Planck and Electroweak scales thereby allowing us to connect
these constructions to real experimental observables. In
addition,superpartner masses and gauge couplings, which determine
production and decay rate,depend on moduli, so if the moduli are
not stabilized the values chosen may not be reliable. To
illustrate our approach, we analyze two particular classes of type
IIB string theory vacua in detail -- KKLT compactifications
\cite{kklt03} and Large Volume compactifications
\cite{Balasubramanian:2005zx}, where the issues of moduli
stabilization and supersymmetry breaking have been well
understood. Other classes of string/$M$ theory vacua which also
have the above desirable features should also be studied in the
future.

\vspace{0.5cm}

\hspace{4cm} \textit{Type IIB KKLT compactifications} (IIB-K)

\vspace{0.5cm}

This class of constructions is a part of the IIB landscape with
all moduli stabilized\cite{kklt03}. Closed string fluxes are used
to stabilize the dilaton and complex structure moduli at a high
scale and non-perturbative corrections to the superpotential are
used to stabilize the lighter K\"{a}hler moduli. One obtains a
supersymmetric anti-deSitter vacuum and D terms
\cite{Burgess:2003ic} or anti D-branes are used to break
supersymmetry and to lift the vacuum to a deSitter one.
Supersymmetry breaking is then mediated to the visible sector by
gravity. The flux superpotential ($W_0$) has to be tuned very
small to get a gravitino mass of $\mathcal{O}$(1-10 TeV). By
parameterizing the lift from a supersymmetric anti-deSitter vacuum
to a non-supersymmetric deSitter vacuum, one can calculate the
soft terms \cite{Choi:2004sx}. The soft terms depend on the
following microscopic input parameters -- \{$W_0,\alpha,n_i$\} or
equivalently \{$m_{3/2},\alpha,n_i$\}, where $\alpha$ is the ratio
$\frac{F^T/(T+\bar{T})}{m_{3/2}}$ and $n_i$ are the modular
weights of the matter fields \cite{Choi:2004sx}. In addition,
$\tan(\beta)$ and sign($\mu$) are fixed by electroweak symmetry
breaking. A feature of this class of constructions is that the
tree level soft terms are comparable to the anomaly mediated
contributions, which are always present and have been calculated
in \cite{GaNeWu99}.

\vspace{0.5cm}

\hspace{3cm} \textit{Type IIB Large Volume Compactifications}
(IIB-L)

\vspace{0.5cm}

This class of constructions also form part of the IIB landscape
with all moduli stabilized. In this case, the internal manifold
admits a large volume limit with the overall volume modulus very
large and all the remaining moduli small
\cite{Balasubramanian:2005zx}. Fluxes again stabilize the complex
structure and dilaton moduli at a high scale, but the flux
superpotential $W_0$ in this case can be $\mathcal{O}(1)$. One
also incorporates perturbative contributions to the K\"{a}hler
potential in addition to non-perturbative contributions to the
superpotential to stabilize the K\"{a}hler moduli. This class of
vacua is more general and includes the KKLT vacua as a special
case, in which $W_0$ is tuned very small \cite{Quevedo06}.
However, when $W_0$ is $\mathcal{O}(1)$, the conclusions are
qualitatively different. We will analyze such a situation, since
then there will be no theoretical overlap between the two classes
of vacua. Now one gets a non-supersymmetric anti-de Sitter vacuum
in contrast to the KKLT case, which can be lifted to a de Sitter
one by similar mechanisms as in the previous case. Since the
volume is very large, the string scale turns out to be quite low.
Assuming a natural value of $W_0$ to be $\mathcal{O}(1)$
\footnote{we actually varied it roughly from 0.1 to 10.}, to get a
gravitino mass of $\mathcal{O}$(1-10 TeV), one needs the string
scale of $\sim 10^{11}$ GeV. Since the string scale is much
smaller than the unification scale, one cannot have standard gauge
unification in these compactifications with $W_0$ as $O(1)$.
Supersymmetry breaking is again mediated to the visible sector by
gravity and soft terms can be calculated \cite{Conlon:2006us}.
Anomaly mediated contributions turn out to be important for some
soft parameters and have to be accounted for. The soft terms
depend on the following microscopic input parameters -
\{$\mathcal{V},n_i$\} or equivalently \{$m_{3/2},n_i$\}, where
$\mathcal{V}$ denotes the volume of the internal manifold and
$n_i$ denote the modular weight of the matter fields.
$\tan(\beta)$ and sign($\mu$) are fixed by electroweak symmetry
breaking.

\vspace{0.3cm}

These two classes of compactifications are good for the following
two reasons :

\begin{itemize}
\item These compactifications stabilize \emph{all} the moduli,
making them massive at acceptable scales. This is good for two
reasons -- a) Light scalars (moduli) are in conflict with
astrophysical observations, and b) Since particle physics masses
and couplings explicitly depend on the moduli, one cannot compute
these couplings unless the moduli are stabilized. Therefore,
unless one stabilizes the moduli, one does not obtain a vacuum.

\item They have a mechanism for generating a small gravitino mass
($O(1-10)$ TeV). This is essential to deal with the {\it hierarchy
problem}. The mechanisms available for generating a small
gravitino mass may not be completely satisfactory though. For
example, the KKLT vacua require an enormous amount of tuning,
while the Large Volume vacua (with $W_0 = O(1)$) do not have
standard gauge unification at $2\times10^{16}$ GeV. Recently, a
class of M theory vacua have been proposed which stabilize all the
moduli, naturally explain the hierarchy and are also consistent
with standard gauge unification \cite{Acharya:2006ia}.
\end{itemize}

There do not exist MSSM-like matter embeddings in the KKLT and
Large Volume classes of vacua at present. However, since many
examples of MSSM-like matter embeddings have been constructed in
simpler type II orientifold constructions, one hopes that it will
be possible to also construct explicit MSSM-like matter embeddings
in these vacua as well in the future. Therefore, we take the
following approach in our analysis -- we {\it assume} the
existence of an MSSM matter embedding on stacks of D7
branes\footnote{for concreteness.} in these vacua and analyze the
consequences for low energy observables. Having said that, it is
important to understand that the assumption of an MSSM-like matter
embedding has been made only for conceptual and computational
simplicity -- a) Any model of low energy supersymmetry must at
least have the MSSM matter spectrum for consistency, so assuming
the MSSM seems to be a reasonable starting point. b) In addition,
most of the software tools and packages available are optimized
for the MSSM. In principle, the approach advocated is completely
general and can be applied to any theoretical construction. The
main point we want to emphasize is that \emph{it is possible to
answer questions (A) and (B) (in the Introduction) affirmatively
for many classes of realistic string constructions.} Choosing a
different class of vacua or the above vacua with a different
matter embedding will change the results, but not the properties
that it is possible to go from classes of semi-realistic string
vacua to experimental observables and that classes of string vacua
can be distinguished on the basis of their experimental
observables.

In order to illustrate better the fact that our approach works for
any given theoretical construction, we also include some other
classes of constructions in our analysis. A brief description of
these constructions can be found in the Appendix. These
constructions are {\it inspired} from microscopic string
constructions and include some of their model building and some of
their moduli stabilization features, although not in a completely
convincing and comprehensive manner. Also, in these constructions
the supersymmetry breaking mechanism is not specified explicitly,
it is only parameterized. These constructions serve as nice toy
constructions making it easy to connect these constructions to low
energy phenomenology quickly and efficiently. Therefore, even
though from a strictly technical point of view they only have
educational significance, they are still very helpful in bringing
home the point we want to emphasize.

All the string constructions studied in this work have a thing in
common -- the soft supersymmetry breaking terms at the string
scale are determined in terms of a few parameters. This is in
stark contrast to completely phenomenological models such as
mSUGRA or minimal gauge mediation, where the soft supersymmetry
breaking terms are chosen \emph{by hand} instead of being
determined from a few underlying parameters.

Although for all studied string constructions the soft terms are
determined in terms of a few underlying parameters, the KKLT and
Large Volume constructions differ from the others in the sense
that for these constructions, the parameters which determine the
soft terms are intimately connected to the underlying microscopic
theoretical structure compared to the other constructions. From a
practical point of view though, once the soft terms are
determined, then one can treat all the constructions at par as far
as the analysis of low energy observables is concerned. This also
applies if one is only interested in understanding the origin of
the specific pattern of signatures of a given construction from
its spectrum and soft parameters, as is done in sections
\ref{spectrum} and \ref{fromsoft}. However, in order to understand
the origin of the soft parameters from the structure of the
underlying theoretical construction, it makes more sense to
analyze the KKLT and Large Volume constructions as they are
microscopically better defined\footnote{in the sense that they
provide an explicit mechanism of supersymmetry breaking and moduli
stabilization.} and because they provide a better representation
of phenomenological characteristics of classes of string vacua.
This will be done in section \ref{fromtheory}.

The string-motivated constructions considered in the analysis are
the following:
\begin{itemize}\footnotesize{
\item HM-A -- Heterotic M theory constructions with one modulus.
\item HM-B -- Heterotic M theory constructions with five-branes.
\item HM-C -- Heterotic M theory constructions with more than one
moduli. \item PH-A -- Weakly coupled heterotic string
constructions with non-perturbative corrections to the K\"{a}hler
potential. \item PH-B --  Weakly coupled heterotic string
constructions with a tree level K\"{a}hler potential and multiple
gaugino condensates. \item II-A -- Type IIA constructions on
toroidal orientifolds with Intersecting D branes.}
\end{itemize}

\section{The ``String" Benchmark Pattern Table - Results} \label{results}

Before we explain, we briefly summarize the results so that the
reader can see the goals. The results for the pattern table are
summarized in Table \ref{resultstable}. The rows and columns
constitute eight ``string'' constructions\footnote{One should be
aware of the qualifications made in the previous section.}
analyzed in our study.

\begin{table}[h!]
{\begin{center}
\begin{tabular}{|l|c|p{1.5cm}|p{1.5cm}|p{1.5cm}|p{1.5cm}|p{1.5cm}|p{1.5cm}|p{1.5cm}|p{1.5cm}|}
\hline
  && HM-A & HM-B & HM-C & PH-A & PH-B & II-A & IIB-K & IIB-L
\\ \hline\hline HM-A & & -- & PY & PY & Yes & Yes & Yes & Yes & Yes
\\ HM-B && & -- & PY & Yes & Yes & Yes & Yes & Yes
\\ HM-C && &  & -- & PY & Yes & Yes & PY & PY
\\ PH-A && &  &  & -- & Yes & Yes & Yes & PY
\\ PH-B && &  &  &  & -- & Yes & Yes & Yes
\\ II-A && &  &  &  &  & -- & Yes & Yes
\\
\hline\hline IIB-K && &  &  &  &  &  &  -- & Yes
\\ IIB-L && &  &  &  &  &  &  &  --
\\
\hline
\end{tabular}
\end{center}}
{\caption{\label{resultstable}\footnotesize{\bf The String Pattern
Table Results} \newline A ``Yes" for a given pair of constructions
indicates that the two constructions are distinguishable in a
robust way, while a ``No" indicates that the two models are not
distinguishable with available data (5 $fb^{-1}$ in this case). A
"Probably Yes (No)" means that the two models are (aren't)
distinguishable in large regions of their parameter spaces.}}
\end{table}

For each construction, we go from the ten or eleven dimensional
string/M theory to its four dimensional effective theory and then
to its LHC signatures. For the string constructions we study, the task of
deducing the effective four-dimensional lagrangian has already been accomplished.
Therefore, we use results for the description of the effective four-dimensional
theories from literature. However, barring the KKLT constructions \footnote{The collider
phenoemnology of Large Volume constructions had not been studied at the time of writing, it was
studied soon afterwards.}, none of the other constructions have been studied to the extent that
predictions for LHC observables can be made. In this work, we have studied the phenomenological
consequences of each of these constructions in detail, and computed their LHC signatures.
An LHC signature by definition is one that is really observable at a hadron collider,
e.g. number of events for $n$ leptons, $m$ jets and $\notEt$, and various ratios of
numbers of events, but not (for example) masses of superpartners
or $\tan \beta$. Signatures are typically of two kinds - counting
signatures, as mentioned in the examples above and distribution
signatures, e.g. the effective mass distribution, invariant mass
distribution of various objects, etc.

The results shown in Table \ref{resultstable} are deduced by
calculating signatures for the various constructions and looking
for signatures that are particularly useful in distinguishing
different constructions, shown in Table \ref{patterntable}. The
details of the procedure involved and the kind of signatures used
will be explained unambiguously in later sections. The results are
shown here so that the interested reader can see the goals. The
rows depict the string constructions used in our study while the
columns consist of useful signatures, which will be defined
precisely later. The \emph{pattern table} (Table
\ref{patterntable}) has been constructed for 5 $fb^{-1}$ of data
at the LHC, which is roughly two years' worth of initial LHC
running. Since everyone is eager to make progress, we focus on
getting early results. More data will allow doing even better.
From Table \ref{resultstable}, it can be seen that most of the
pairs can be distinguished from each other, encouraging optimism
about the power and usefulness of this analysis.

\begin{table}[h!]
{\begin{center}
\begin{tabular}{|l|c|p{1.7cm}|p{1.7cm}|p{1.7cm}|p{1.7cm}|p{1.7cm}|p{1.7cm}|p{1.7cm}|} \cline{1-9}
Signature && A & B & C & D & E & F & G
\\ \hline \hline Condition && $> 1200$ & $> 25$ & $> 1.6 $ & $> 0.54 $ & $> 0.05 $ & $ > 160 $GeV & $> 0.58 $
\\\hline \hline HM-A && OC& OC& OC& OC & OC & Both & OC
\\ \cline{1-9} HM-B && Both & Both & Both & Both & Both& Both &
Both
\\ \cline{1-9} HM-C && Both & Both & OC & Both & Both & Both & Both
\\ \cline{1-9} PH-A && ONC & N.O. & OC & Both & ONC & ONC & Both
\\ \cline{1-9} PH-B && N.O. & N.O. & N.O. & N.O. & N.O. & Both & N.O.
\\ \cline{1-9} II-A && ONC & N.O. & ONC & OC & ONC & ONC & ONC
\\ \cline{1-9} IIB-K && ONC & ONC & OC & ONC & Both & OC & ONC
\\ \cline{1-9} IIB-L && ONC & N.O. & OC & Both  & ONC & ONC & Both
\\ \cline{1-9}
\end{tabular}
\end{center}}
{\caption{\label{patterntable}\footnotesize{\bf The String Pattern
Table}\newline An ``$OC$'' for the $i^{th}$ row and $j^{th}$
column means that the signature is observable for many models of
the $i^{th}$ construction. The value of the $j^{th}$ signature for
the $i^{th}$ construction is (almost) always consistent with the
condition in the second row and $j^{th}$ column of the Table. An
``$ONC$'' also means that the signature is observable for many
models of $i^{th}$ construction. However, the value of the
signature (almost) always does \emph{not} consistent with the
condition as specified in the second row and $j^{th}$ column of
the Table. A ``$Both$'' means that some models of the $i^{th}$
construction have values of the $j^{th}$ signature which are
consistent the condition in the second row and the $j^{th}$ column
while other models of the $i^{th}$ construction have values of the
$j^{th}$ signature which are not consistent with the condition. An
``N.O.'' for the $i^{th}$ row and $j^{th}$ column implies that the
$j^{th}$ signature is \emph{not} observable for the $i^{th}$
construction, i.e. the values of the observable for all (most)
models of the construction are always below the observable limit
as defined by (\ref{observability}), for the given luminosity (5
$fb^{-1}$). So, the construction is not observable in the $j^{th}$
signature channel with the given amount of ``data''.}}
\end{table}

The logically simplest way to distinguish constructions on the
basis of their signature pattern would be to construct a
multi-dimensional plot which shows that all constructions occupy
different regions in the multi-dimensional space. Since this is
not practically feasible, we construct two dimensional projection
plots for various pairs of signatures. For simplicity in this
initial analysis, each observable signature has been divided into
two classes, based on the value the observable takes. The
observable value dividing the two classes is chosen so as to yield
good results. For a given two dimensional plot for two signatures,
we will have clusters of points representing various
constructions. Each point will represent a set of parameters for a
given construction, which we call a ``model". The cluster of
points representing a given construction may form a connected or
disconnected region. To distinguish any two given constructions,
we essentially look for conditions in this two dimensional plane
which are satisfied by all (most) models of one construction
(represented by one cluster of points) but not satisfied by all
(most) models of the other construction (represented by the other
cluster of points). In this way, it will be possible to
distinguish the two given constructions.

\begin{figure}[h!]
\center \epsfig{file=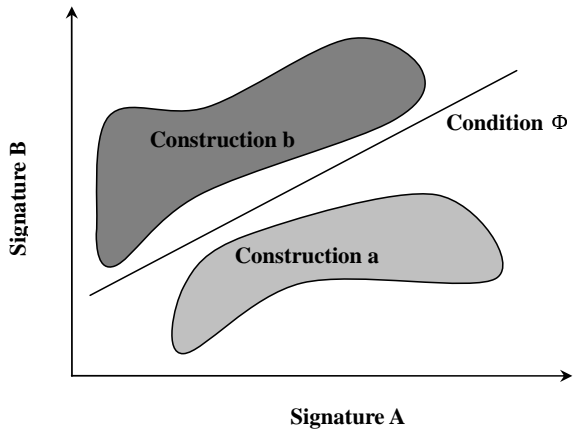,height=6cm,angle=0}
\caption{\footnotesize Cartoon to illustrate the method used to
distinguish constructions. } \label{cartoon}
\end{figure}

The cartoon in Figure \ref{cartoon} illustrates the above point in
a clear way. In a given two dimensional plot with axes given by
signatures $A$ and $B$, we will in general have two clusters of
points for two given constructions $a$ and $b$, as shown by the
light and dark regions respectively. If we define a condition
$\Phi$ on the signatures $A$ and $B$ such that it is given by the
line (or curve in general) shown in the cartoon, then it is
possible to distinguish constructions $a$ and $b$ by the above set
of signatures. To be clear, the above method of distinguishing
theoretical constructions has some possible technical limitations,
which will be addressed in section \ref{limitations}. Since the
purpose here is to explain the overall approach in a simple
manner, we have used the above method. One can make the approach
more sophisticated to tackle more complicated situations, as is
mentioned in section \ref{limitations}.

The counting signatures in Table \ref{patterntable} denote numbers
of events in excess of the Standard Model (SM). A description of
some of the SM backgrounds included is in the next section.

\vspace{0.2cm}

\hspace{3.5cm}\emph{Simple Criterion for Distinguishing Constructions}

\vspace{0.2cm}

Based on the above pattern table, our criterion for distinguishing any
given pair of constructions (a ``Yes'' in the pattern table) is
that their respective entries are very different in at least one
column, such as an $OC$ for one construction while an $ONC$ or
$N.O.$ for another construction. A $Both$ for one construction
while an $OC$ or $ONC$ or $N.O.$ for another does not distinguish the
constructions cleanly. If there is only a small region of overlap
between the two constructions for all signatures, then the two
constructions can be distinguished in the regions in which they
don't overlap. This would give a ``Probably Yes (PY)'' in the
pattern table, otherwise it would give a ``Probably No (PN)". Similarly,
an $ONC$ for one construction and an $N.O.$ for another also does not distinguish
the two constructions cleanly, and would give a ``PY'' or ``PN'' depending on their
overlap. Carrying out this procedure for all constructions and signatures
gives the result in Table \ref{resultstable}. It should be kept in mind though
that the result shown in Table \ref{resultstable} is only for a simple set of signatures.
Using more sophisticated signatures and analysis techniques could give better resulst.
Also, there are typically other useful signatures present than
what is listed in the Table. We have only shown the most useful
ones. In section \ref{distinguishibility}, we give a description
of the useful signatures and explain why these particular
signatures are useful in distinguishing the various constructions
in terms of the spectrum, the soft terms and in turn from the
underlying theoretical structure.

\section{Procedural Details}\label{procedure}

In this section  we enumerate the procedure to answer question (A)
in the Introduction, namely, how to go from a string construction
to the space of LHC signatures.

The first step concerns the spectrum of a given construction. Many
of the string-motivated constructions considered give a
semi-realistic spectrum which contains the MSSM, and perhaps also
some exotics. However for simplicity, in this initial analysis we
only consider the MSSM fields because mechanisms may exist which
project the exotic fields out or make them heavy. As already
explained, for the KKLT and Large Volume vacua, we just assume the
existence of an MSSM matter embedding. The weakly and strongly
coupled heterotic string constructions are naturally compatible
with gauge coupling unification at $M_{unif} \sim 2 \times
10^{16}$ GeV, but type II-A and type II-B constructions are not in
general. One can however try to impose that as an additional
constraint even for type II constructions since gauge unification
provides a very important clue to beyond-the-Standard Model
physics. Therefore, all constructions except the Large Volume
compactifications (IIB-L) used in our analysis either naturally
predict or consistently assume the existence of gauge coupling
unification at $M_{unif}$. The IIB-L construction does not have
the possibility of gauge coupling unification at $M_{unif}$
compatible with having a supersymmetry breaking scale of
$\mathcal{O}$(TeV) \cite{Balasubramanian:2005zx}. The string scale
for the IIB-L constructions is taken to be of
$\mathcal{O}(10^{10}-10^{11})$ GeV in order to have a
supersymmetry breaking scale of $\mathcal{O}$(TeV), making it
incompatible with standard gauge unification.

In order to connect to low energy four-dimensional physics, one
has to write the effective four dimensional action at the String
scale /11 dim Planck scale, which we will denote by $M_s$. $M_s$
will be equated to $M_{unif}$ for all constructions except for the
Large Volume compactifications (IIB-L). The four-dimensional
effective action for each of these constructions is determined by
a set of microscopic ``input" parameters of the underlying theory.
Soft supersymmetry breaking parameters for the MSSM fields are
calculated at $M_s$ as functions of these underlying input
parameters. A description of these input parameters can be found
in section \ref{examples} for the KKLT and Large Volume vacua, and
in the Appendix for all the string-motivated constructions. The
input parameters are taken to vary within appropriate ranges, as
determined by theoretical and phenomenological considerations. In
a more realistic construction, some of these parameters may
actually be fixed by the theory. Our approach therefore is broad
in the sense that we include a wide range of possibilities without
restricting too much to a particular one.

Each of the ``models'' for a particular construction is thus
defined by a list of input parameters which in turn translate to a
\emph{parameter space} of soft parameters at the unification
scale. In other words, the boundary conditions for the soft
parameters are determined by the underlying microscopic
constructions. As emphasized earlier, this is very different from
an \emph{ad hoc} choice of boundary conditions for the soft
parameters, as in many models like mSUGRA or minimal gauge mediation.

Once the soft parameters are \emph{determined} from the underlying
microscopic parameters, then the procedure to connect these soft
parameters to low-scale physics is standard, namely, the soft
parameters are evolved through the Renormalization Group (RG)
evolution programs (SuSPECT \cite{Djouadi:2002ze} and SOFTSUSY
\cite{Allanach:2001kg}) to the electroweak scale, and the spectrum
of particles produced is calculated. For concreteness and
simplicity, we assume that no intermediate scale physics exists
between the electroweak scale and the unification scale, though we
wish to study constructions with intermediate scale physics and
other subtleties in the future.

Of all the models thus generated, only some will be consistent
with low energy experimental and observational constraints. Some
of the most important low energy constraints are :

\begin{itemize}\footnotesize{
\item Electroweak symmetry breaking (EWSB). \item Experimental
bounds for superpartner masses. \item Experimental bound for the
Higgs mass. \item Constraints from flavor and CP physics.\item
Upper bound on the relic density from WMAP \cite{Bennett:2003bz}.}
\end{itemize}

In this analysis, we use $\tan \beta$ as an input parameter and
the RGE software package determines $\mu$ and $B\mu$ by requiring
consistent EWSB. This is because none of the constructions studied
are sufficiently well developed so as to predict these quantities
{\it a priori}. In the future, we hope to consider constructions
which predict $\mu$, $B\mu$ and the Yukawa couplings allowing us
to deduce $\tan \beta$ at the electroweak scale and explain EWSB.
The models we consider are consistent with constraints on particle
spectra, $b \rightarrow s\gamma$ \cite{Abe:2001hk}, $(g_{\mu}-2)$
\cite{Bennett:2002jb} and the upper bound on relic density. One
should not impose the lower bound on relic density since
non-thermal mechanisms can turn a small relic density into a
larger one. Also, the usually talked about lower bound on the LSP
mass ($\sim 50$ GeV) is only present for models with gaugino mass
unification. There is no general lower bound on the LSP mass,
especially if one also relaxes the constraint of a thermal relic
density. In our analysis, we have not imposed any lower bound on
the LSP mass. In our study at present, we have generated $\sim 50$
models for most constructions ($\sim 100$ models for the PH-A
construction)\footnote{However, not all 50 (or 100) models will be
above the observable limit in general.}. This might seem too small
at first. However from our analysis it seems that the results we
obtain are robust and do not change when more points are added.
For a purely statistical analysis this could be a weak point, but
because in each case we know the connection between the theory and
signatures and understand why the points populate the region they
do, we expect stable results. We simulated many more models for
two particular constructions and found that the qualitative
results do not change, confirming our expectations. This will be
explained in section \ref{limitations}.

Once one obtains the spectrum of superpartners at the low scale,
one calculates matrix elements for relevant physics processes at
parton level which are then evolved to ``long-distance" physics,
accounting for the conversion of quarks and gluons into jets of
hadrons, decays of tau leptons,etc. We carry out this procedure
using PYTHIA 6.324 \cite{Sjostrand:2003wg}. The resulting hadrons,
leptons and photons have to be then run through a detector
simulation program which simulates a real detector. This was done
by piping the PYTHIA output to a modified CDF fast detector
simulation program PGS \cite{pgs}. The modified version was developed by
John Conway, Stephen Mrenna and others and approximates an ATLAS or
CMS-like detector. The
output of the PGS program is in a format which is
also used in the LHC Olympics \cite{LHCO}. It consists of a list
of objects in each event labelled by their identity and their
four-vector. Lepton objects are also labelled by their charges and
b-jets are tagged. With this help of these, one can construct a
wide variety of signatures.

The precise definitions of jets and isolated leptons, criteria for
hadronically decaying taus, efficiencies for heavy flavor tagging
as well as trigger-level cuts imposed on objects are the same as
used for the ``blackbox" data files in the LHC Olympics
\cite{LHCO}. In addition, we impose event selection cuts as the
following:
\begin{itemize}
\item If the event has photons, electrons, muons or hadronic taus,
we only select the particles which satisfy the following -- Photon
$P_T\;> 10$ GeV; Electron, Muon $P_T\;
> 10$ GeV; (hadronic) Tau $P_T\; > 100$ GeV. \item For any event with jets, we
only select jets with $P_T\; > 100$ GeV. \item Only those events
are selected with $\notE_T\;$ in the event $> 100$ GeV.
\end{itemize}
These selection cuts are quite simple and standard. We have used a
simple and relatively ``broad-brush" set of cuts since for a
preliminary analysis, we want to analyze many constructions
simultaneously. As mentioned before, we have simulated 5 $fb^{-1}$
of data at the LHC for each model. A small luminosity was chosen
for two reasons -- first, in the interest of computing time and
second, in order to argue that our proposed technique is powerful
enough to distinguish between different constructions even with a
limited amount of data. Of course, to go further than
distinguishing classes of constructions broadly from relatively
simple signatures, such as getting more insights about particular
models within a given construction, one would in general require
more data and could sharpen the approach by imposing more
exclusive cuts and signatures.

In order to be realistic, one has to take the effects of the
standard model background into account. In our analysis, we have
simulated the $t\bar{t}$ background and the diboson ($WW,ZZ$)+
jets background. We have not included the uniboson ($W,Z$) + jets
background in the interest of time. However, from
\cite{Baer:1995va} we know that for a $\notE_T$ threshold of 100
GeV, as has been used in our analysis, the $t\bar{t}$ background
is either the largest background or comparable to the largest one
($W$ + jets typically). Therefore, we expect our analysis and
results to be robust against addition of the $W$ + jets
background. The criteria we employ for an observable signature is
:
\begin{equation}
\frac{N_{signal}}{\sqrt{N_{bkgd}}} > 4; \;\; \frac{N_{signal}}{N_{bkgd}} > 0.1; \;\;
N_{signal} > 5. \label{observability}
\end{equation}

These conditions are quite standard. A signature is observable
only when the most stringent of the three constraints is
satisfied.

Although the steps used in our analysis are quite standard, there are at least two respects
in which our analysis differs from that of previous ones in the
literature. First, our work shows for the first time that with a few reasonable assumptions,
one can study string theory constructions to the extent that reliable predictions for
experimental observables can be made, and more importantly, different string constructions
give rise to overlapping but distinguishable footprints in signature space. Moreover, it is
possible to understand \emph{why} particular combinations of signatures are helpful in
distinguishing different constructions, from the underlying theoretical structure of the
constructions. Therefore, even though we have done a simplified (but reasonably
realistic) analysis in terms of trigger and selection level cuts,
detection efficiencies of particles, detector simulation and
calculation of backgrounds, we expect that doing a more sophisticated analysis
will only change some of the details but not the qualitative results.
In particular, it will not affect the properties
that predictions for experimental observables can be made for
many classes of realistic string constructions, and that patterns of signatures
are sensitive to the structure of the underlying string
constructions, making it possible to distinguish among various
classes of string constructions.

\section{Distinguishibility of Constructions} \label{distinguishibility}

\subsection{General Remarks}\label{remarks}

For convenience, we present the signature pattern table again. As
was also mentioned in section \ref{results}, each signature has
been broadly divided into two main classes for simplicity. The
value of the observable dividing the two classes is chosen so as
to yield the best results. A description of the most useful
signatures is given below:

\begin{table}[h!]
{\begin{center}
\begin{tabular}{|l|c|p{1.7cm}|p{1.7cm}|p{1.7cm}|p{1.7cm}|p{1.7cm}|p{1.7cm}|p{1.7cm}|} \cline{1-9}
Signature && A & B & C & D & E & F & G
\\ \hline \hline Condition && $> 1200$ & $> 25$ & $> 1.6 $ & $> 0.54 $ & $> 0.05 $ & $ > 160 $GeV & $> 0.58 $
\\\hline \hline HM-A && OC& OC& OC& OC & OC & Both & OC
\\ \cline{1-9} HM-B && Both & Both & Both & Both & Both& Both &
Both
\\ \cline{1-9} HM-C && Both & Both & OC & Both & Both & Both & Both
\\ \cline{1-9} PH-A && ONC & N.O. & OC & Both & ONC & ONC & Both
\\ \cline{1-9} PH-B && N.O. & N.O. & N.O. & N.O. & N.O. & Both & N.O.
\\ \cline{1-9} II-A && ONC & N.O. & ONC & OC & ONC & ONC & ONC
\\ \cline{1-9} IIB-K && ONC & ONC & OC & ONC & Both & OC & ONC
\\ \cline{1-9} IIB-L && ONC & N.O. & OC & Both  & ONC & ONC & Both
\\ \cline{1-9}
\end{tabular}
\end{center}}
{\caption{\label{patterntable2}\footnotesize{\bf The String Pattern
Table} \newline An ``$OC$'' for the $i^{th}$ row and $j^{th}$
column means that the signature is observable for many models of
the $i^{th}$ construction. The value of the $j^{th}$ signature for
the $i^{th}$ construction is (almost) always consistent with the
condition in the second row and $j^{th}$ column of the Table. An
``$ONC$'' also means that the signature is observable for many
models of $i^{th}$ construction. However, the value of the
signature (almost) always does \emph{not} consistent with the
condition as specified in the second row and $j^{th}$ column of
the Table. A ``$Both$'' means that some models of the $i^{th}$
construction have values of the $j^{th}$ signature which are
consistent the condition in the second row and the $j^{th}$ column
while other models of the $i^{th}$ construction have values of the
$j^{th}$ signature which are not consistent with the condition. An
``N.O.'' for the $i^{th}$ row and $j^{th}$ column implies that the
$j^{th}$ signature is \emph{not} observable for the $i^{th}$
construction, i.e. the values of the observable for all (most)
models of the construction are always below the observable limit
as defined by (\ref{observability}), for the given luminosity (5
$fb^{-1}$). So, the construction is not observable in the $j^{th}$
signature channel with the given amount of ``data''.}}
\end{table}

\begin{itemize}
\footnotesize{ \item A -- Number of events with trileptons and
$\geq 2$ jets. The value of the observable dividing the signature
into two classes is 1200. \item B -- Number of events with clean
(not accompanied by jets) dileptons. The value of the observable
dividing the signature into two classes is 25. \item C -- $(Y/X)$
\footnote{The ratio $(Y/X)$ is computed only when both signatures
$X$ and $Y$ are above the observable limit.}; Y= Number of events
with 2 leptons, 0 b jets and $\geq 2$ jets, X= Number of events
with 0 leptons, 1 or 2 b jets and $\geq 6$ jets\footnote{This
signature is not very realistic in the first two years. Please
read the discussion in this subsection as to why this signature is still
used.}. The value of the observable dividing the signature into
two classes is 1.6. \item D -- $(Y/X)$; Y= Number of events with 2
leptons, 1 or 2 b jets and $\geq 2$ jets, X= Number of events with
2 leptons, 0 b jets and $\geq 2$ jets. The value of the observable
dividing the signature into two classes is 0.54. \item E -- The
charge asymmetry in events with one electron or muon and $\geq$ 2
jets ($A_{c}^{(1)}=\frac{N(l^+)-N(l^-)}{N(l^+)+N(l^-)}$). The
value of the observable dividing the signature into two classes is
0.065. \item F -- The peak of the missing energy distribution. The
value of the observable dividing the signature into two classes is
160 GeV. \item G -- $Y/X$; Y = Number of events with same sign
different flavor (SSDF) dileptons and $\geq 2$ jets, X = Number of
events with 1 tau and $\geq$ 2 jets. The value of the observable
dividing the signature into two classes is 0.5.}\end{itemize}

The signatures A-F turn out to be the most economic and useful in distinguishing all
constructions considered\footnote{It is important to understand
that these signatures were useful in distinguishing constructions
which have at least some models giving rise to observable
signatures with the given luminosity (5 $fb^{-1}$).With more
luminosity, many more models of these constructions would give
rise to observable signatures, so one would in general have a {\it
different} set of useful distinguishing signatures.}. We have also listed signature G
as an example to show that it is possible to construct other signatures which can distinguish
among some of the constructions. We understand that some of
these signatures are not very realistic. For example, the signature which
counts the number of events with 0 leptons, 1 or 2 b jets and
$\geq$ 6 jets is not very realistic initially because of
difficulties associated with calibrating the fake missing
$\notE_T$ from jet mismeasurement in events with six or more jets.
However, we have used this signature in our analysis at this stage because it
helps in explaining our results and the approach in an economic
way. Also, the fact that there are other useful
signatures\footnote{although they may be less economical in the sense that one would need
more signatures to distinguish the same set of constructions.} which
distinguish these constructions gives us additional confidence
about the robustness of our approach. These signatures were
hand-picked by experience and by trial-and-error. Once the set of
useful signatures was collected, the next task was to understand
\emph{why} the above set of signatures were useful in
distinguishing the constructions based on their spectrum, soft
parameters and their underlying theoretical setup. This is the
subject of sections \ref{spectrum}, \ref{fromsoft} and
\ref{fromtheory} respectively. We hope that carrying out the same
exercise for other constructions can help build intuition about
the kind of signatures which any given theory can produce. This
can eventually help in building a dictionary between structure of
underlying theoretical constructions and their collider
signatures.

The above list of signatures consists of counting signatures and
distribution signatures at the LHC. The counting signatures denote
number of events in excess of the Standard Model. Naively, one
would think that the number of signatures is very large if one
includes lepton charge and flavor information, and $b$ jet
tagging. However, it turns out that not all signatures are
independent. In fact, they can be highly correlated with each
other, drastically decreasing the effective dimensionality of
signature space. Thus, in order to effectively distinguish
signatures, one needs to use signatures sufficiently orthogonal to
each other. This has been emphasized recently in
\cite{Arkani-Hamed:2005px}. We will see that having an underlying
theoretical construction allows us to actually find those useful
signatures. Even though we have only listed a few useful
signatures in Table \ref{patterntable}, there are typically more
than one (sometimes many) signatures which distinguish any two
particular constructions. This is made possible by a knowledge of
the structure of the underlying theoretical constructions.

\subsection{Why is it possible to distinguish different Constructions?}\label{whypossible}

In view of the above comments, one would like to understand why it
is possible to distinguish different constructions in general and
why the signatures described in the previous section are useful in
distinguishing the various constructions in particular.

To understand the origin of distinguishibility of constructions,
one should first understand why each construction gives rise to a
specific pattern of soft supersymmetry breaking parameters and in
turn to a specific pattern of signatures. This is mainly due to
\emph{correlations in parameter space as well as in signature
space}. Let's explain this in detail. A construction is
characterized by its spectrum and couplings in general. These
depend on the underlying structure of the theoretical
construction, such as the form of the four-dimensional effective
action, the mechanism to generate the hierarchy, the details of
moduli stabilization and supersymmetry breaking, mediation of
supersymmetry breaking etc. At the end of the day, the theoretical
construction is defined by a small set\footnote{if they are indeed
``good" theoretical constructions.} of microscopic input
parameters in terms of which \emph{all} the soft supersymmetry
breaking parameters are computed. Since \emph{all} soft parameters
are calculated from the \emph{same} set of underlying input
parameters, this gives rise to correlations in the space of soft
parameters for any given construction. These correlations carry
through all the way to low energy experimental observables, as
will be explicitly seen later. The fact that there exist
correlations between different sets of parameters which have their
origin in the underlying theoretical structure allows us to gain
insights about the underlying theory, and is much more powerful
than completely phenomenological parameterizations such as mSUGRA,
minimal gauge mediation, etc.

Since any two \emph{different} theoretical constructions will
differ in their underlying structure in some way \emph{by
definition}, the correlations obtained in their parameter and
signature spaces will also be different in general. All these will
in general have \emph{different} effects on issues which influence
low energy phenomenology in an important manner, such as the scale
of supersymmetry breaking, unification of gauge couplings (or
not), flavor physics, origins of CP violation, etc. These factors
combined with experimental constraints allow different string
constructions to be distinguished from each other in general.

We now wish to understand why the particular signatures described
in section \ref{remarks} are useful in distinguishing the studied
constructions. In order to successfully do so, one has to
understand the relevant features of the various constructions and
their implications to hadron collider phenomenology, and devise
signatures which are sensitive to those features.

In the following subsections, we explain how to distinguish the
above constructions. In principle, one could directly try to
connect patterns of signatures to underlying string constructions.
However, in practice it is helpful to divide the whole process of
connecting patterns of signatures to theoretical constructions in
a few parts -- first, the results of the pattern table for each
construction are explained based on the spectrum of particles at
the low scale; second, important features of the spectrum of
superpartners at the low scale for the different constructions
(which give rise to their characteristic signature patterns) are
explained in terms of the soft supersymmetry breaking parameters
at the high scale; and third, the structure of the soft parameters
is explained in terms of the underlying theoretical structure of
the constructions.

For readers not interested in the details in the next subsection,
the main points to take away are that any given theoretical
construction only gives rise to a specific pattern of observable
signatures, and that one can understand and trust the regions of
signature space that are populated by a given theoretical
construction and that such regions are quite different for
different constructions, illustrating in detail that LHC
signatures can distinguish different theoretical constructions.

\subsection{Explanation of Signatures from the Spectrum}\label{spectrum}

In this subsection, we take the spectrum pattern for different
constructions as given and explain patterns of signatures based on
them. Then it is possible to treat all the constructions equally
as far as the explanation of the pattern of signatures from the
spectrum is concerned. The characteristic features of the spectrum
for the constructions considered are as follows:

\begin{itemize}{\footnotesize
\item HM-A  –- Universal soft terms. Bino LSP (``coannihilation
region"\footnote{Explained in
\ref{fromtheory}.\label{footnote6}}). Moderate gluinos (550-650
GeV), slightly lighter scalars. \item HM-B -- Universal soft
terms. Has two (disconnected) regions. Region I similar to HM-A.
Region II either ``focus point region"\footnote{Explained in
\ref{fromtheory}} or ``funnel region"\footnote{Explained in
\ref{fromtheory}.} Scalars much heavier ($> 800$ GeV) than
gauginos in region II. \item HM-C -- Non-universal soft terms.
Occupies a big region in signature space encompassing the two
regions mentioned for the HM-B construction. Heavy scalars. Can
have bino, wino or higgsino LSP. Spectrum and signature pattern
quite complicated. \item PH-A  --- Non-universal soft terms. Bino,
higgsino or mixed bino-higgsino LSP. Bino LSP has light gluino ($<
600$ GeV). Higgsino or mixed bino-higgsino LSP have gluinos
ranging from moderately heavy to heavy ($600-1200$ GeV). Heavy
scalar masses ($\geq 2$ TeV). \item PH-B -- Non-universal soft
terms. Wino and bino LSP. Light gluinos (200-550 GeV) always have
wino LSPs while heavier gluinos can have bino or wino LSPs.
Comparatively heavy LSP (can be upto 1 TeV), heavy scalar masses
($\geq 1$ TeV) except stau which is relatively light ($\geq 500$
GeV). \item II-A -- Non-universal soft terms. Can have bino, wino,
higgsino or mixed bino-higgsino LSP. Light or moderately heavy
gluino ($300-600$ GeV), scalar masses heavier than gluinos but not
very heavy ($< 1$ TeV). Stops can be as light as 500 GeV. Spectrum
and signature pattern quite complicated. \item IIB-K --
Non-universal soft terms. Heavy spectrum ($\geq$ 1 TeV) in
general, but possible to have light spectrum ($\leq$ 1 TeV). Bino
LSP\footnote{We have only analyzed $\alpha > 0$.}. Gluinos are
greater than about $450$ GeV while the lightest squarks
($\tilde{t}_1$) are greater than about $200$ GeV. For some models,
$\tilde{\tau}_1$ can be light as well. \item IIB-L --
Non-universal soft terms. Mixed bino-higgsino LSP. The gluinos
have a lower bound of about $350$ GeV, while the lightest squark
($\tilde{t}_1$) has a lower bound of about 700 GeV. }
\end{itemize} \label{spectra}

\begin{figure}[h!]
\begin{center}
 \epsfig{file=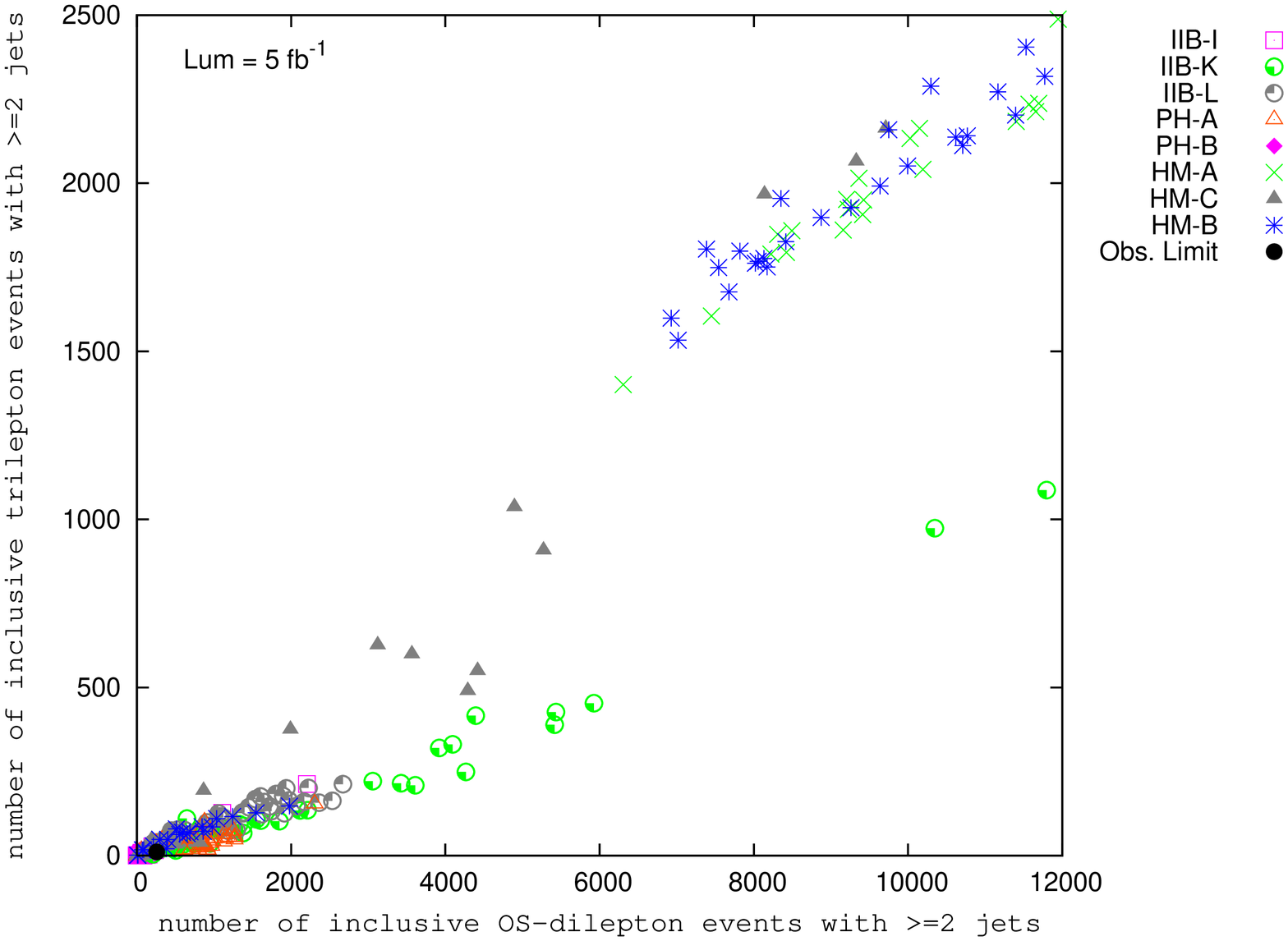,height=16cm, angle=-90}
\end{center}
\caption{\footnotesize  Plot of number of events with
opposite-sign dileptons and $\geq$ 2 jets and number of events
with three leptons and $\geq$ 2 jets. The black dot represents the
lower limit of observability of the two signatures, according to
conditions in equation (\ref{observability}). Note that the HM-A
and overlaping HM-B and HM-C construction can be distinguished
easily from the PH-A, PH-B, II-A, IIB-K and IIB-L constructions,
as they occupy very different regions. The plots are best seen in
color.} \label{OS}
\end{figure}

As mentioned in the list of characteristic features of the
spectrum above, the HM-B and HM-C constructions roughly occupy two
regions in signature space, as can be seen from Figure \ref{OS}.
One of these regions overlaps with the HM-A construction. This is
because the HM-B and HM-C constructions contain the HM-A
construction as a subset (see the Appendix for details). Since the
three constructions have the same theoretical structure in a
region of their high scale parameter space, the models of the
three constructions in that particular region cannot be
distinguished from each other from their signature pattern as
their signatures will always overlap. However, since the HM-B and
HM-C constructions have a bigger parameter space, they also occupy
a bigger region of signature space compared to the HM-A
construction. Thus, it is possible to distinguish the HM-A, HM-B
and HM-C constructions in regions in which they don't overlap,
i.e. in regions in which their underlying theoretical structure is
different. This is the origin of ``Probably Yes (PY)'' in Table
\ref{resultstable}.

\begin{figure}[h!]
\begin{center}
\epsfig{file=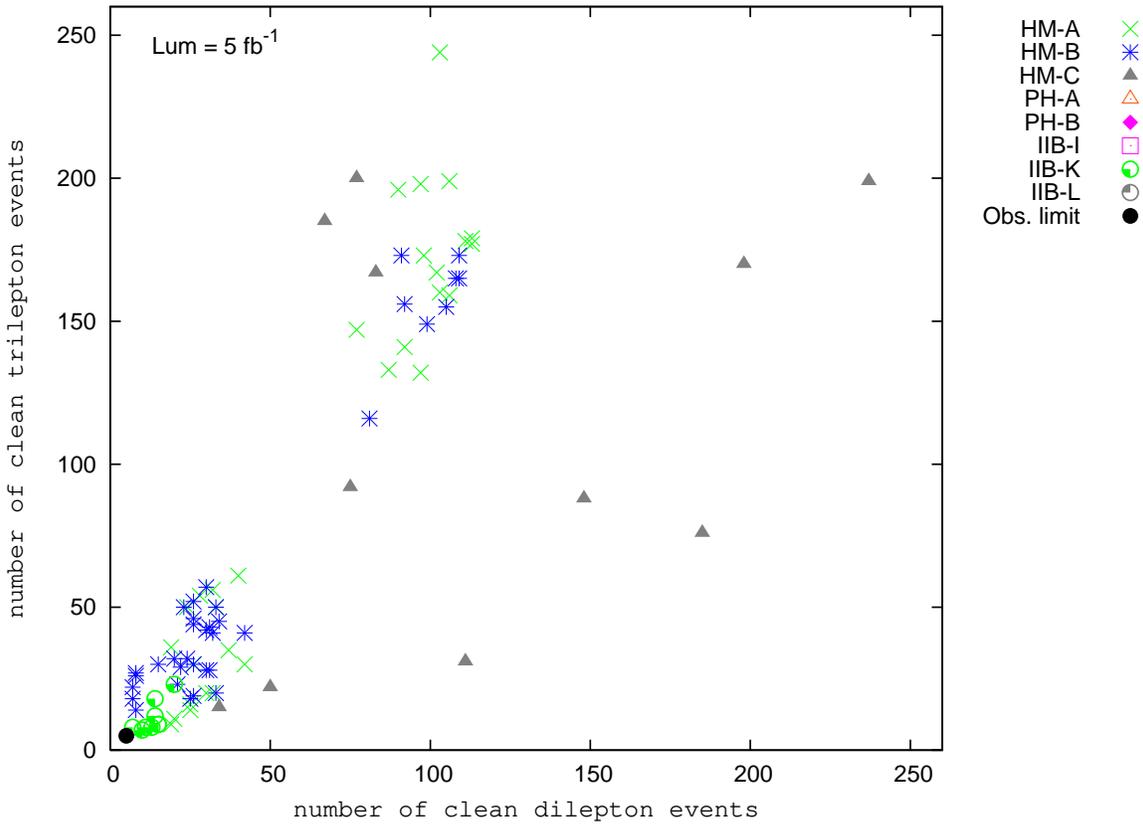,height=16cm, angle=-90}
\end{center}
\caption{\footnotesize  Plot of number of events with clean
dileptons and number of events with clean trileptons. ``clean"
means not accompanied by jets. The black dot represents the lower
limit of observability of the two signatures, according to
conditions in equation (\ref{observability}). The models below the
observable limit have not been shown. Note that the HM-A and
overlapping HM-B and HM-C constructions can be distinguished from
the PH-A, PH-B, II-A and IIB-L constructions, since the latter are
not observable with the given luminosity. The plots are best seen
in color.} \label{dilep}
\end{figure}

Figures \ref{OS} and \ref{dilep} shows two (very)\footnote{since
we do not take into account the charge and flavor information for
leptons, and the flavor information for jets (whether the jets are
of heavy flavor or not). } inclusive signature plots\footnote{The
figures are best seen in color.}. One can try to explain the
differences in these signatures among the HM-A construction and
the overlapping HM-B and HM-C constructions on the one hand and
the PH-A, PH-B, II-A, IIB-K and IIB-L constructions on the other,
from their spectra.

The HM-A construction and the overlapping HM-B, HM-C constructions
have a comparatively light spectrum at the low scale. These
give rise to a subset of the well known mSUGRA boundary
conditions, so after imposing constraints from low energy physics as
in section \ref{procedure}, one finds that the allowed spectrum
consists of light and moderately heavy gluinos, slightly lighter
squarks and light sleptons. Thus, $\tilde{g}\tilde{q}$ production
and $\tilde{q}\tilde{q}$ pair production are dominant with direct
$\tilde{N_2}\,\tilde{C_1}$ production also quite important.

Both squarks and gluinos ultimately decay to $\tilde{N_2}$ and
$\tilde{C_1}$ and since most sleptons (including selectrons,
smuons) are accessible, both $\tilde{N_2}$ and
$\tilde{C_1}$ decay to leptons and the LSP via sleptons. Since the
mass difference between $\tilde{N_2}$, $\tilde{C_1}$ and LSP
($\tilde{N_1}$) is big (because of universal boundary conditions,
$\Delta M_{\tilde{N_2}-\tilde{N_1}}, \Delta
M_{\tilde{C_1}-\tilde{N_1}} \sim M_{\tilde{N_1}}$), most of the
leptons produced pass the cuts. The $\tilde{N_2}$ and
$\tilde{C_1}$ by themselves are also comparatively light ($< 200$
GeV). On the other hand, the PH-A, PH-B, II-A and IIB-L
constructions are required to have heavy scalars and gluinos
varying in mass from light to heavy. So, the $\tilde{N_2}$ and
$\tilde{C_1}$ produced from gluinos decay to the LSP mostly
through a virtual $Z$ and $W$ respectively, which makes their
branching ratio to leptons much smaller. Because of
\emph{non-universal} soft terms, $\Delta
M_{\tilde{N_2}-\tilde{N_1}}$ and $\Delta
M_{\tilde{C_1}-\tilde{N_1}}$ can be bigger or smaller than in the
universal case. In the PH-A, PH-B, II-A and IIB-L constructions,
they are required to be comparatively smaller, leading to leptons
which are comparatively softer on average, many of which do not
pass the cuts.

For clean dilepton events, direct production of $\tilde{N_2}$ and
$\tilde{C_1}$ is required. The HM-A construction and overlapping
HM-B and HM-C constructions have comparatively lighter
$\tilde{N_2}$ and $\tilde{C_1}$, so
 $\tilde{N_2}$ and $\tilde{C_1}$ are directly produced. On the other hand, most models of the
PH-B and II-A constructions have heavier $\tilde{N_2}$ and
$\tilde{C_1}$ compared to the HM-A construction, making it harder
to produce them directly. The PH-A and IIB-L constructions have
some models with light $\tilde{N_2}$ and $\tilde{C_1}$, but the
other factors (decay via virtual W and Z, and smaller mass
separation between $\tilde{N_2}$, $\tilde{C_1}$ and LSP) turn out
to be more important, leading to no observable clean dilepton
events. Therefore, the result is that \emph{none} of the models of
the PH-A, PH-B, II-A and IIB-L constructions have observable clean
dilepton events. Thus, it is possible to distinguish the HM-A
construction and overlapping HM-B and HM-C constructions from the
PH-A, PH-B, II-A and IIB-L constructions by signatures A and B in
Table \ref{patterntable2} (shown in Figures \ref{OS} and
\ref{dilep}).

The case with the IIB-K construction is slightly different. These
constructions have many models with a heavy spectrum which implies
that those models do not have observable events with the given
luminosity of 5 $fb^{-1}$.  However, these constructions can also
have light gluinos and squarks with staus also being light in some
cases. So $\tilde{g}\tilde{q}$ production is typically dominant
for these models. The gluinos and squarks decay to $\tilde{N_2}$
and $\tilde{C_1}$ as for other constructions. Since for many IIB-K
models, the lightest stau is heavier than $\tilde{N_2}$ and
$\tilde{C_1}$ (even though it is relatively lighter than in the
PH-A, PH-B, II-A and IIB-L constructions), the $\tilde{N_2}$ and
$\tilde{C_1}$ decay to the LSP through a virtual $Z$ and $W$
respectively, making the branching fraction to leptons much
smaller than for the HM-A and overlapping HM-B and HM-C
constructions. In addition, the mass differences $\Delta
M_{\tilde{N_2}-\tilde{N_1}}$ and $\Delta
M_{\tilde{C_1}-\tilde{N_1}}$ are required to be smaller for the
IIB-K construction in general compared to that for the HM-A and
overlapping HM-B and HM-C constructions, making it harder for the
leptons to pass the cuts. Some IIB-K models have comparable mass
differences $\Delta M_{\tilde{N_2}-\tilde{N_1}}$ and $\Delta
M_{\tilde{C_1}-\tilde{N_1}}$ as the HM-A construction, but they
have much heavier gluinos compared to those for the HM-A
construction, making their overall cross-section much smaller.
Therefore, the IIB-K construction has fewer events for leptons in
general (in particular for trileptons) compared to that for the
HM-A and overlapping HM-B and HM-C constructions, as seen from
Figure \ref{OS}.

\begin{figure}[h!]
\begin{center}
 \epsfig{file=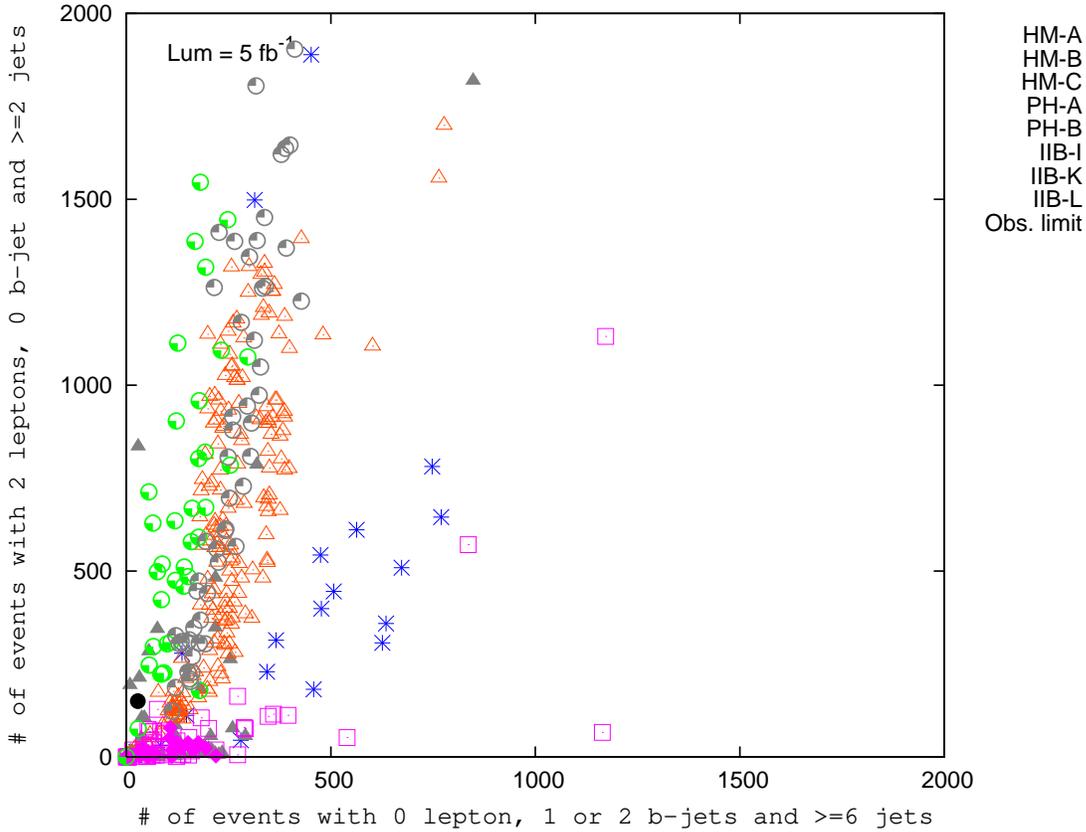,height=16cm, angle=-90}
\end{center}
\caption{\footnotesize Plot of number of events with 0 leptons, 1
or 2 b jets and $\geq$ 6 jets and number of events with 2 leptons,
0 b jets and $\geq$ 2 jets. The black dot represents the lower
limit of observability of the two signatures, according to
conditions in equation (\ref{observability}). The models below the
observable limit have also been shown to emphasize that the II-A
construction has very different number of events for these
signatures compared to other constructions even {\it without}
imposing the observability constraint. Note that the PH-B and II-A
constructions can be distinguished from the PH-A, IIB-K and IIB-L
constructions because they have very different slopes. The plots
are best seen in color.} \label{0l12b6j}
\end{figure}

Region II of the HM-B construction and the non-overlapping region
of the HM-C construction (with HM-A) cannot be cleanly
distinguished from the PH-A, PH-B, II-A, IIB-K and IIB-L
constructions from the above signatures. Region II of the HM-B
construction (the ``focus point'' or ``funnel region'' of mSUGRA)
however, can be distinguished from these constructions with the
help of other signatures\footnote{For example, the signatures
shown in Figure \ref{0l12b6j} can distinguish Region II of the
HM-B construction (the HM-B models distinct from the
PH-A,PH-B,IIB-K and IIB-L region all belong to Region II) with the
PH-A, IIB-K and IIB-L constructions. As another example, the ratio
of number of events with 0 b jets and $\geq$ 2 jets and number of
events with $\geq$ 3 b jets and $\geq$ 2 jets can distinguish
Region II of HM-B with PH-B, II-A and IIB-K constructions. These
can be explained on the basis of their spectra, but has not been
done here for simplicity. Also, the HM-B row in Table
\ref{patterntable2} has \emph{not} been divided into two parts (to
account for the two regions) to avoid clutter.}. The entries for
the HM-B construction in Table \ref{patterntable2} correspond to the overall
HM-B region, which explains the ``Both'' in all columns for the HM-B construction.
Since one can distinguish region II of HM-B with other constructions by using other
signatures as explained in the footnote, inspite of the ``Both'' entry in Table
\ref{patterntable2} one has a ``Yes'' for the HM-B row in the appropriate columns
in Table \ref{resultstable}. The non-overlapping region of the HM-C construction is a very big
region in signature space because of its big parameter space
(explained in the Appendix), making it relatively harder to
distinguish it from some of the other constructions. Since we have
not found signatures cleanly distinguishing the \emph{whole}
msoft-II region from some of the other constructions, we have put
a ``Probably Yes'' for the HM-C row in the appropriate columns in Table
\ref{resultstable}.

\begin{figure}[ht]
\begin{center}
 \epsfig{file=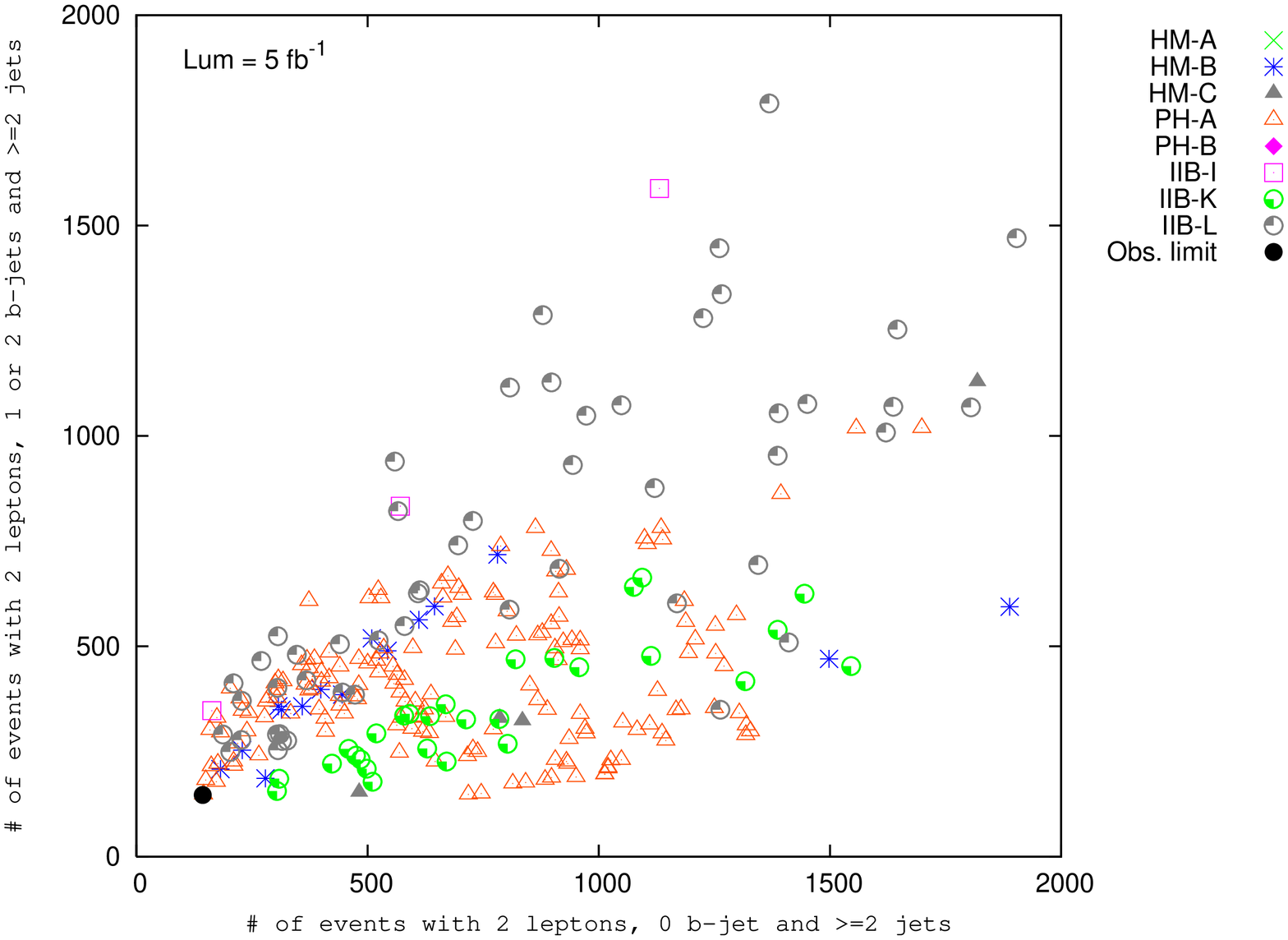,height=16cm, angle=-90}
\end{center}
\caption{\footnotesize  Plot of number of events with 2 leptons, 0
b jets and $\geq$ 2 jets and number of events with 2 leptons, 1 or
2 b jets and $\geq$ 2 jets. The black dot represents the lower
limit of observability of the two signatures, according to
conditions in equation (\ref{observability}). The models below the
observable limit have not been shown. Note that the PH-B and II-A
constructions can be distinguished from each other since the
former is not observable while the latter is observable. One can
also partially distinguish the PH-A and IIB-L constructions. The
plots are best seen in color.}\label{lepton}
\end{figure}

Now we explain the distinguishibility of the PH-A, PH-B, II-A,
IIB-K and IIB-L constructions. Figure \ref{0l12b6j} shows that the
PH-B and II-A constructions can be distinguished from the PH-A,
IIB-K and IIB-L constructions (signature $C$ in Table
\ref{patterntable2})\footnote{This signature may not be very
realistic. However, as explained in section \ref{remarks}, it has
been used here because it is very economical and illustrates the
approach in a simple manner.}. Figure \ref{lepton} shows that the
PH-B and II-A constructions can be distinguished from each other
and that the PH-A and IIB-L constructions can be {\it partially}
distinguished from each other (signature $D$ in Table
\ref{patterntable2}). To understand why it is possible to do so,
we look at these constructions in detail.

Let's start with the IIB-K construction. As explained earlier, the
IIB-K construction typically gives a heavy spectrum, although it
is possible to have a light spectrum with light gluinos and
squarks (stops) with the stau also being light in some cases. So,
$\tilde{g}\tilde{q}$ production is typically dominant. First and
second generation squarks are copiously produced. The first and
second generation squarks decay to non b-jets and the gluino also
decays more to non b jets than to b jets because the IIB-K
construction always has a bino LSP. Therefore, as seen from Figure
\ref{lepton}, the IIB-K construction has more events with 2
leptons, 0 b jets and $\geq$ 2 jets compared to those with 2
leptons, 1 or 2 b jets and $\geq$ 2 jets. Also, since the mass
difference between $\tilde{N_2},\tilde{C_1}$ and $\tilde{N_1}$ is
only big enough for leptons to pass the cuts but not for jets to
pass the cuts, the number of events with 0 leptons, 1 or 2 b jets
and $\geq$ 6 jets is smaller than those with 2 leptons, 0 b jets
and $\geq$ 2 jets, as seen from Figure \ref{0l12b6j}.

The PH-B construction has scalars which are quite
heavy ($> 1$ TeV). So, gluino pair production is dominant.  The
branching ratio of gluinos to $\tilde{C_1}$ + jets is typically
the largest for $\tan \beta \geq 20$, if it is kinematically
allowed \cite{Bartl:1994bu}, followed by $\tilde{N_i}$ + jets,
$i=1,2,3$, which are smaller. This is generally true for this
construction.

When the gluino is light ($\leq 550$ GeV), the PH-B construction
has wino LSP. Since $\mu$ is large ($> 1.3$ TeV), $m_{\tilde{C_1}}
\sim m_{\tilde{N_1}}$. Since $\tilde{C_1}$ and $\tilde{N_1}$ are
wino and $\tilde{N_2}$ is bino, the gluino decays mostly to non b
jets. Also, the leptons and jets coming from the decay of
$\tilde{C_1}$ to $\tilde{N_1}$ are very soft and do not pass the
cuts. When the gluino directly decays to $\tilde{N_1}$, there are
no leptons and only two jets. When the gluino is heavier ($> 550$
GeV), the PH-B construction can have wino as well as bino LSPs.
Because of a heavier gluino, the overall cross-section goes down.
For the wino LSP case, the same argument as above applies in
addition to the small cross-section implying even fewer lepton
events. For the bino LSP case, the gluino again decays mostly to
non b jets as $\tilde{C_1}$ and $\tilde{N_2}$ are wino and
$\tilde{N_1}$ is bino. Also, $m_{\tilde{C_1}} \sim
m_{\tilde{N_2}}$ but at the same time $\Delta
M_{\tilde{N_2}-\tilde{N_1}}$ and $\Delta
M_{\tilde{C_1}-\tilde{N_1}}$ are quite small ($\leq 20 $ GeV),
leading to soft leptons and jets many of which don't pass the
cuts. So the result is that PH-B models have very few events with
leptons and/or b jets. Therefore, as seen from Figures
\ref{0l12b6j} and \ref{lepton}, the PH-B construction does not
give rise to observable events for signatures $D$ and $E$ in Table
\ref{patterntable2}.

The PH-A construction is required to have bino, higgsino or mixed
bino-higgsino LSP with very heavy scalar masses ($\geq 2$ TeV).
Light gluinos ($\leq 600$ GeV) have bino LSP, while higgsino or
mixed bino-higgsino LSPs have heavier gluinos ($600-1200$ GeV).
$\tilde{N_2}$ and $\tilde{C_1}$ are also quite light ($< 250$
GeV).

For the bino LSP case, gluinos typically decay to $\tilde{N_1}$,
$\tilde{N_2}$ and $\tilde{C_1}$ and non b jets most of the time
compared to b jets, as $\tilde{N_1}$ is bino and $\tilde{N_2}$ and
$\tilde{C_1}$ are wino. The decays of $\tilde{N_2}$ and
$\tilde{C_1}$ can give rise to leptons passing the cuts. In
addition, direct production of $\tilde{N_2}$ and $\tilde{C_1}$ is
also important as they are light. They can also give rise to 2
leptons, 0 b jets and $\geq 2$ jets. Therefore, in this case, one
has more events with (two) leptons and non b jets compared to
those with (two) leptons and b jets. This can be seen clearly from
Figure \ref{lepton}. The decays of $\tilde{N_2}$ and $\tilde{C_1}$
also give jets (as they decay through virtual Z and W). Since the
gluino is light, the cross-section is quite big implying that
there are also a fair number of events with 0 leptons, 1 or 2 b
jets and $\geq 6$ jets. However, the number of events with 2
leptons, 0 b jets and $\geq 2$ jets is more than with  0
leptons, 1 or 2 b jets and $\geq 6$ jets due to the dominant
branching ratio of gluinos to non b jets. This can be seen from
Figure \ref{0l12b6j}.

For the higgsino LSP case, the gluino is comparatively heavier
($600-1200$ GeV), leading to a significant decrease in
cross-section. Now $\tilde{N_1}$, $\tilde{N_2}$ and $\tilde{C_1}$
are mostly higgsino, leading to a lot of production of b jets as
the relevant coupling is proportional to the mass of the
associated quark. The fact that $\tilde{N_1}$, $\tilde{N_2}$ and
$\tilde{C_1}$ are mostly higgsino also makes their masses quite
close to each other, implying that leptons and jets produced from
the decays of $\tilde{N_2}$ and $\tilde{C_1}$ to $\tilde{N_1}$ are
very soft and do not pass the cuts. Thus, these models have very
few events with leptons. Since the jets coming from the decays of
$\tilde{N_2}$ and $\tilde{C_1}$ to $\tilde{N_1}$ are also very
soft, the PH-A higgsino LSP models also have very few events with
0 leptons, 1 or 2 b jets and $\geq 6$ jets. Therefore, these do
not give rise to observable events for the signatures in Figures
\ref{0l12b6j} and \ref{lepton}.

For the mixed bino-higgsino LSP case, the gluino is again quite
heavy ($600-1200$ GeV), making the cross section much smaller
compared to the bino LSP case. Since $\tilde{N_1}$, $\tilde{N_2}$
and $\tilde{C_1}$ have a significant higgsino fraction, the
gluinos again decay more to b jets compared to non b jets. The
mass separation between $\{\tilde{N_2},\tilde{N_3},\tilde{C_1}\}$
and $\tilde{N_1}$ is such that the decays of $\tilde{N_2}$ and
$\tilde{C_1}$ to $\tilde{N_1}$ produce leptons which typically
pass the cuts and jets which only sometimes pass the cuts.
Therefore, these PH-A models have few events with 2 leptons,
0 b jets and $\geq 2$ jets. They give rise to events with 2
leptons, 1 or 2 b jets and $\geq 2$ jets but since the overall
cross-section is much smaller than for the bino LSP case, the
number of events for the above signature for these PH-A models is
just a little above the observable limit, as can be seen from
Figure \ref{lepton}. This is the origin of the ``Both'' entry for
signature $D$ in the row for the PH-A construction. Because of the
small overall cross-section as well as the fact that jets produced
from the decays of $\tilde{N_2}$ and $\tilde{C_1}$ only sometimes
pass the cuts, the number of events for 0 leptons, 1 or 2 b jets
and $\geq 6$ jets is also small for these PH-A models, as seen
from Figure \ref{0l12b6j}.

The IIB-L construction always has a mixed bino-higgsino LSP, for both light and heavy
gluino models. The light gluino IIB-L models have a large overall cross-section.
The gluinos decay both to non b jets and b jets owing to the mixed bino-higgsino
nature of the LSP. Also, the mass separation between $\tilde{N_2}$, $\tilde{C_1}$ and $\tilde{N_1}$
is not large which means that the leptons produced from the decays of  $\tilde{N_2}$ and $\tilde{C_1}$
pass the cuts but the jets produced seldom pass the cuts. So, the IIB-L construction has many events with 2 leptons,
0 b jets and $\geq$ 2 jets as well as with 2 leptons 1, or 2 b jets and $\geq$ 2 jets but not as many
with 0 leptons, 1 or 2 b jets and $\geq$ 6 jets, as seen from Figures \ref{0l12b6j} and \ref{lepton}.

From Figure \ref{lepton}, one sees that the IIB-L construction can
be distinguished \emph{partially} from the PH-A construction,
leading to a ``Probably Yes (PY)'' in the pattern table. One can
understand it as follows - as mentioned above, the IIB-L
construction always has a mixed bino-higgsino LSP while the PH-A
construction has a mixed bino higgsino LSP only when the gluino is
heavy (i.e. for a heavy spectrum). For light gluino models, as
mentioned before, the PH-A construction has a bino LSP. Therefore,
{\it for light gluino models}, the ratio of number of events with
2 leptons, 1 or 2 b jets and $\geq$ 2 jets and number of events
with 2 leptons 0 b jets and $\geq$ 2 jets is much more for the
IIB-L construction compared to the PH-A construction. These are
the models which differentiate the IIB-L and PH-A constructions in
Figure \ref{lepton}. It turns out that mixed bino-higgsino LSP
models with heavy gluinos in both constructions have very similar
spectra\footnote{This has been explicitly checked.}, leading to
very similar signatures in all studied channels. Therefore, the
IIB-L construction and the PH-A construction are not
distinguishable in this special region of spectrum and signature
space with the present set of signatures. Using more sophisticated
signatures may help distinguish these signatures more cleanly. As
already mentioned before, the PH-A construction also has models
with a pure higgsino LSP. Those models have very heavy gluinos
however, leading to no observable events in Figures \ref{0l12b6j}
and \ref{lepton}.

Moving on to the II-A construction, we note that it can have a bino, wino, higgsino or mixed
bino-higgsino LSP with light to moderately heavy gluino ($300-600
$ GeV) and moderately heavier scalars (stops can be specially
light). The spectrum and signature pattern are quite
complicated. Let's analyze all possible cases.

In this construction, the branching ratio of gluinos to
$\tilde{C_1}$ + jets is the largest as mentioned before, since
$\tan \beta \geq 20$. For the wino LSP case, since
$M_{\tilde{C_1}} \sim M_{\tilde{N_1}}$ the leptons and jets from
the decays of $\tilde{C_1}$ to $\tilde{N_1}$ are very soft and do
not pass the cuts. The decay of the gluino to $\tilde{C_1}$ is
accompanied by non b jets since $\tilde{N_1}$ and $\tilde{C_1}$
are wino and $\tilde{N_2}$ is bino. So, the II-A models with wino
LSP do not give rise to observable events with leptons and/or b
jets. This implies that the signatures in Figures \ref{0l12b6j}
and \ref{lepton} are not observable for these II-A models.

For the bino LSP case, $\tilde{N_2}$ and $\tilde{C_1}$ are quite
heavy ($> 350$ GeV), sometimes being even heavier than the gluino,
in which case only the decay of gluino to $\tilde{N_1}$ is allowed
leading to no leptons. Even when the decays of gluino to
$\tilde{C_1}$ and $\tilde{N_2}$ are allowed, they are mostly
accompanied by comparatively soft non b jets (due to kinematic
reasons). Since these II-A models are required to have
$\tilde{N_2}$ and $\tilde{C_1}$ much heavier than the PH-A bino
LSP models, the direct production of $\tilde{N_2}$ and
$\tilde{C_1}$ which could be a source of events with 2 leptons, 0
b jets and $\geq 2$ jets, is also relatively suppressed. Therefore
these models do not give rise to observable events with 2 leptons
0 b jets and $\geq 2$ jets as well as with 2 leptons, 1 or 2 b
jets and $\geq 2$ jets. However, there are some bino LSP II-A
models which also have light squarks (stops mostly) and light
gluinos in addition to having heavy $\tilde{N_2}$ and
$\tilde{C_1}$ as above. For these bino LSP models,
$\tilde{q}\tilde{q}$ pair production is quite important. These
squarks mostly decay to a gluino and quarks, followed by the decay
of the gluino to mostly the LSP and jets (both b and non b jets).
Thus, these bino LSP II-A models have many events with 0 leptons,
1 or 2 b jets and $\geq 6$ jets but no observable events with 2
leptons, 1 or 2 b jets and $\geq 2$ jets.

For the higgsino LSP case, since $\tan \beta \geq 20$, the gluino
mostly decays to $\tilde{C_1}$ + b jets as the associated coupling
is proportional to the mass of the relevant quark. Also, in this
case $\tilde{N_1}$, $\tilde{N_2}$ and $\tilde{C_1}$ are all
higgsino like and very close to each other. So, leptons from the
decays of $\tilde{N_2}$ and $\tilde{C_1}$ to $\tilde{N_1}$ are
very soft and do not pass the cuts. Therefore II-A models with
higgsino LSP do not give rise to observable events with leptons
and/or non b jets. For some of these higgsino LSP II-A models,
there are still a fair number of events with 0 leptons 1 or 2 b
jets and $\geq 6$ jets. This is because even though
$\tilde{q}\tilde{q}$ pair production is less important compared to
those in bino LSP II-A models, the branching ratio of gluinos to b
jets is much bigger (due to a higgsino LSP).

For the mixed bino-higgsino LSP case, the decay of gluino to
$\tilde{C_1}$ + b jets is dominant since $\tilde{C_1}$ is mostly
higgsino. The next important decays are to $\tilde{N_1}$,
$\tilde{N_2}$ and $\tilde{N_3}$ + b jets followed by a small
fraction to non b jets. The mass separation between
$\{\tilde{N_2},\tilde{N_3},\tilde{C_1}\}$ and $\tilde{N_1}$ is
such that leptons produced can easily pass the cuts, while the
jets produced sometimes pass the cuts. Some of these mixed
bino-higgsino LSP II-A models also have comparatively light
squarks, implying that $\tilde{q}\tilde{q}$ and
$\tilde{q}\tilde{g}$ production are also important. The squark
decays to a quark and a gluino, followed by the usual decays of
the gluino. For these models, $\tilde{N_2}$ and $\tilde{C_1}$ are
also light, implying that in such cases direct production of
$\tilde{N_2}$ and $\tilde{C_1}$ is also possible. The decays of
$\tilde{N_2}$ and $\tilde{C_1}$ can give rise to events with 2
leptons, 0 b jets and $\geq 2$ jets.

Therefore, the conclusion is that for II-A models with mixed
bino-higgsino LSP and light squarks, there are observable events
with 2 leptons, 1 or 2 b jets and $\geq 2$ jets; with 0 leptons, 1
or 2 b jets and $\geq 6$ jets as well as with 2 leptons 0 b jets
and $\geq 2$ jets. The number of events with 2 leptons, 1 or 2 b
jets and $\geq 2$ jets is greater than those with 2 leptons, 0 b
jets and $\geq 2$ jets because of the dominant branching fraction
of gluinos to b jets. Therefore, signature $D$ (ratio of the above
two type of events - Figure \ref{lepton}) can distinguish the II-A
and PH-B constructions as the II-A construction has observable
events while the PH-B construction does not give rise to
observable events. The number of events for 0 leptons 1 or 2 b
jets and $\geq 6$ jets will be larger than those with 2 leptons, 0
b jets and $\geq 2$ jets, again due to the dominant branching
ratio of the gluino to b jets. So, signature $C$ (ratio of the
above two type of events - Figure \ref{0l12b6j}) can distinguish
the PH-A,IIB-K and IIB-L constructions from the II-A and PH-B
constructions. The above results can be seen from Figures
\ref{0l12b6j} and \ref{lepton}, where the qualitative difference
between the constructions is clear.
\begin{figure}[h!]
\begin{center}
 \epsfig{file=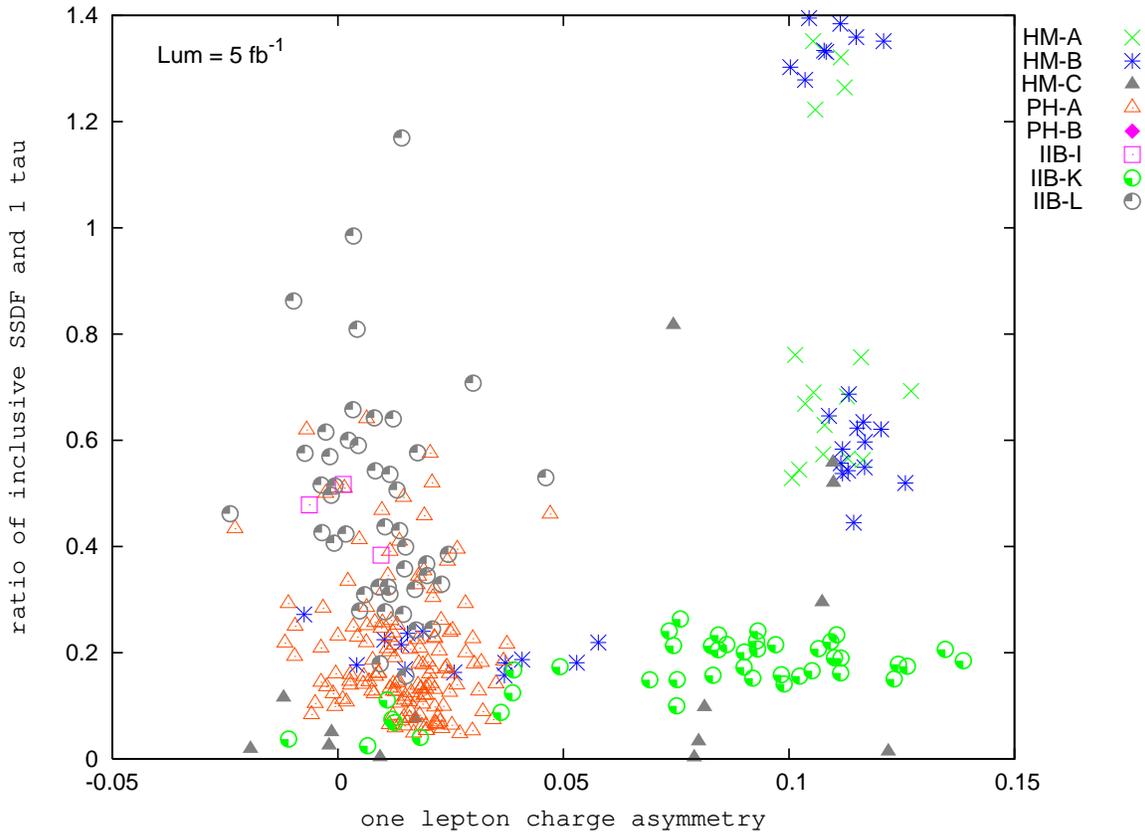,height=16cm, angle=-90}
\end{center}
\caption{\footnotesize  Plot of the charge asymmetry in events
with a single electron or muon $\&\;\geq$ 2 jets $\&$ the ratio of
number of events with same sign different flavor (SSDF) dileptons
and $\geq$ 2 jets and number of events with 1 tau and $\geq$ 2
jets. The models which are below the observable limit as defined
by (\ref{observability}) are not shown. Note that the IIB-K
construction can be distinguished from the PH-A and IIB-L
constructions, as the former occupies a mostly horizontal region
while the latter occupy a mostly vertical region. The overlapping
IIB-K and PH-A models can be distinguished from Figure
\ref{metpk}. The plots are best seen in color.}\label{ac1-SSDF}
\end{figure}
The II-A models shown above the observable limit have mixed
bino-higgsino LSP with lighter squarks than in other II-A cases.

\begin{figure}[h!]
\begin{center}
 \epsfig{file=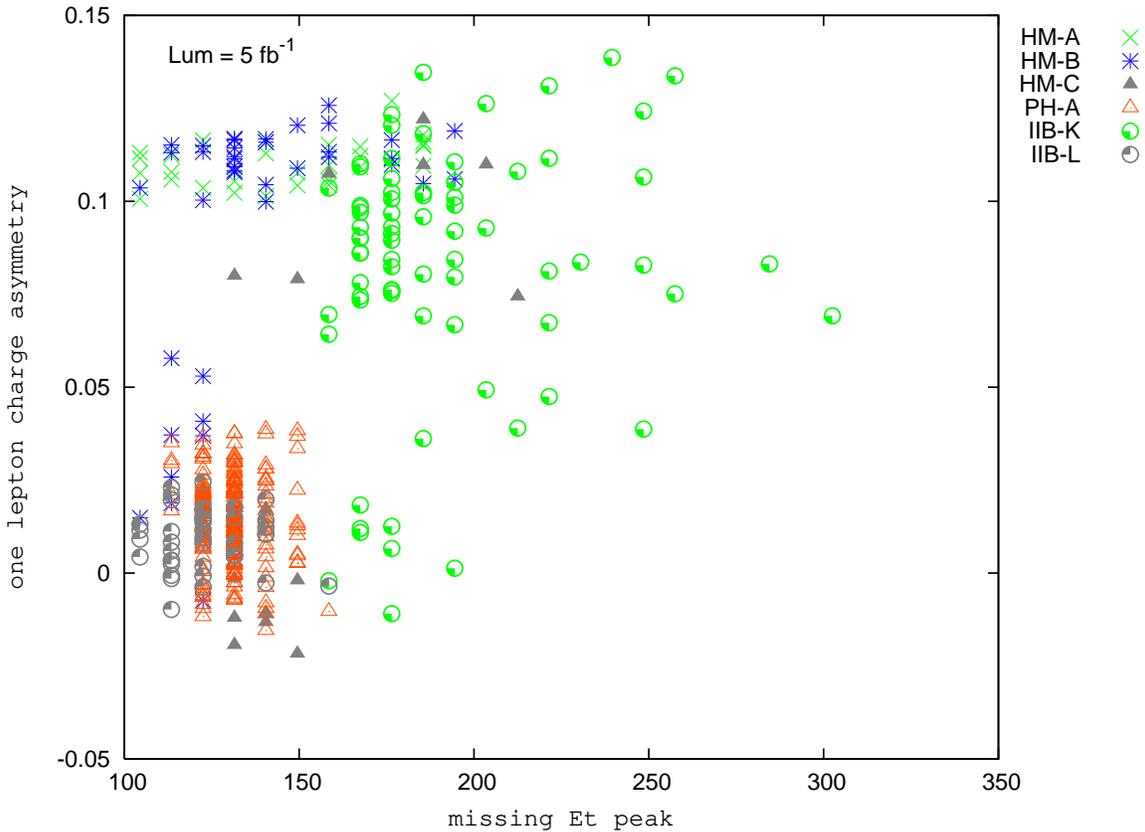,height=16cm, angle=-90}
\end{center}
\caption{\footnotesize  Plot of the peak of the $\notE_T$
distribution and the charge asymmetry in events with a single
electron or muon $\&\;\geq$ 2 jets. The models which are below the
observable limit as defined by (\ref{observability}) are not
shown. Note that this plot distinguishes the overlapping PH-A and
IIB-K models in Figure \ref{ac1-SSDF}. The plots are best seen in
color.}\label{metpk}
\end{figure}

We are now left with explaining the distinguishibility of the PH-A
and IIB-L constructions on the one hand and the IIB-K construction
on the other. Figures \ref{ac1-SSDF} and \ref{metpk} show that the
IIB-K construction can be distinguished from the PH-A and IIB-L
constructions. The reason is as follows - As explained earlier,
the IIB-K construction can have a light spectrum with light
gluinos, light stop and sometimes a light stau. So,
$\tilde{g}\tilde{q}$ production is typically dominant. Since
up-type squarks are produced preferentially at the LHC (as it is a
$pp$ collider), they decay preferentially to a positive chargino
$\tilde{C}_1^+$, which in turn decays preferentially to a
positively charged lepton $l^+$ (in its leptonic decays).
Therefore, the asymmetry in number of events with a single
electron or muon and $\geq$ 2 jets
($A_c^{(1)}$)\footnote{$A_c^{(1)}
\equiv\frac{N(l^+)-N(l^-)}{N(l^+)+N(l^-)}$, where for example
$N(l^+)$ is the number of events with a single positively charged
electron or muon and $\geq$ 2 jets.} is much greater than in the
case of PH-A and IIB-L constructions where $\tilde{g}\tilde{g}$
pair production is dominant. There are a few IIB-K models which
have a small $A_c^{(1)}$ and which overlap with some PH-A models
(seen in Figure \ref{ac1-SSDF}) even though $\tilde{g}\tilde{q}$
production is dominant. This is due to some special features of
their spectrum, such as the lightest stop and/or the lightest stau
being very light. These features either suppress the production of
$\tilde{C}_1^+$ or suppress the decay of $\tilde{C}_1^+$ to
electrons or muons. However, as seen from Figure \ref{metpk},
these overlapping IIB-K models can be distinguished from the PH-A
and IIB-L constructions by the peak of the $\notE_T$ distribution.
This is related to the mass of the LSP. The IIB-K constructions
(which are observable with 5 $fb^{-1}$) have a comparatively
heavier LSP than the IIB-L constructions in general, making the
peak of the $\notE_T$ distribution larger than those for the IIB-L
constructions\footnote{This is because the IIB-K models with a
light spectrum have the lightest stop correlated with the mass of
the LSP ($m_{\tilde{t}_1} \gtrsim m_{\tilde{N}_1}$)
\cite{Choi:2004sx}; $\tilde{t}_1$ cannot be too light, else it
would be directly seen at the Tevatron.}. The PH-A models which
overlap with the small $A_c^{(1)}$ IIB-K models have bino LSPs
which are lighter than that of the IIB-K models. So, the PH-A
models have a smaller $\notE_T$ peak than the overlapping IIB-K
models in Figure \ref{ac1-SSDF}, as can be seen from Figure
\ref{metpk}.
We have thus explained the distinguishibility of all constructions
based on the spectrum at the low scale. Typically, there are more
than one (sometimes many) signatures which can distinguish any two
given constructions. This redundancy gives us confidence that our
analysis is robust and that the conclusions will not be affected
with more sophisticated analysis. For simplicity, we have only
explained one signature distinguishing a pair of constructions but
all such signatures can be understood similarly.

\subsection{Explanation of Spectrum from the Soft
Parameters}\label{fromsoft}

We now turn to understanding the origin of the spectrum of
particles at the low scale (which are responsible for the
signature pattern) for the constructions in terms of pattern of
soft parameters. For illustrative purposes, we carry out this
exercise for two constructions -- the HM-A construction and the
PH-B construction. If we take the soft parameters at the string (or unification) scale as given,
then it does not matter that these are really only toy constructions. The kind of analysis carried out
for these constructions
below can also be carried out for more well motivated constructions
such as the KKLT and Large Volume ones, as well. However, to go to
the final step, i.e. from the soft parameters to the structure of
the underlying theoretical construction, it makes sense to stick to the
more well motivated constructions -- the KKLT and Large Volume
constructions, as we will in the following subsection.

Starting with the HM-A construction, one would like to understand
its characteristic spectrum, viz. $m_{\tilde{g}} \sim
m_{\tilde{q}}
> m_{\tilde{l}}$. Why does the gluino mass lie in the range $550-650$ GeV?
We note that the HM-A construction is a heterotic M theory
construction compactified on a Calabi-Yau with only one K\"{a}hler
modulus. This implies that the soft terms obtained at the
unification scale are universal \cite{Choi:1997cm}. Thus, the soft
terms obtained at the unification scale are a special case of the
well studied mSUGRA boundary condition. Now, phenomenological
studies of the mSUGRA boundary condition have shown that in order
to get a small relic density (satisfying the WMAP upper bound
\footnote{Typically, a lower bound on the relic density is also
imposed. However, we have only used the upper bound in our
analysis, as explained in section \ref{procedure}. The area
covered by the three regions can change depending on whether a
lower bound is also imposed.}), there are three allowed regions in
the $m-M$ plane \cite{Ellis:2003cw} \footnote{One usually assumes
$A_0$ = 0 in these plots.}. Here $m$ stands for the universal
scalar mass parameter while $M$ stands for the universal gaugino
mass parameter. These three regions are the following:
\begin{itemize}
\item The stau coannihilation region -- In this region, the stau is almost degenerate with the LSP
which is a bino. One gets an acceptable relic density because of coannihilation of the stau and the
LSP to a tau. This requires $m<M$ with $m$ roughly between 100 and 150 GeV and $M$
roughly between 150 and 300 GeV (assuming $A_0$ = 0).
\item The focus point region -- This region requires a large scalar mass parameter ($m > M$)
at the unification scale, and gives rise to a higgsino LSP with acceptable relic density.
\item The funnel region -- In this region, the LSP is annihilated by a $s$-channel pole, with $m_{LSP}
 \approx m_A/2$. This also requires $m>M$.
\end{itemize}

\noindent In the case of the HM-A construction, the soft mass
parameters always have the hierarchy $M>m$
\cite{Choi:1997cm}, which implies that only the stau
coannihilation region is possible for the HM-A construction. Also,
the allowed ranges for the $m$ and $M$ parameters roughly
explains the mass scale of the gluino and squarks at the low scale
from standard RG evolution. Therefore, one has to now understand
the origin of the allowed values of the $m$ and $M$
parameters from the nature of the expressions for soft terms and
the ``theory'' input parameters.

The expressions for the soft terms depend on three input
parameters -- the goldstino angle $\theta$, the gravitino mass
$m_{3/2}$ and the parameter $\alpha(t+\bar{t})$ with $t$ as the
K\"{a}hler modulus, in addition to $\tan{\beta}$. For futher
details, the reader is referred to \cite{Choi:1997cm}. The
expressions for the soft parameters are given by : \ba
\label{soft-msoftI} M &=&
\frac{\sqrt{3}Cm_{3/2}}{(s+\bar{s})+\alpha(t+\bar{t})}\{(s+\bar{s})\sin(\theta)e^{-i\gamma_s}+
\frac{\alpha(t+\bar{t})}{\sqrt{3}}\cos(\theta)e^{-i\gamma_t}\} \\
m^2&=& V_0+m_{3/2}^2-\frac{3m_{3/2}^2C^2}{3(s+\bar{s})+\alpha(t+\bar{t})}\,\{\alpha(t+\bar{t})\,
(2-\frac{\alpha(t+\bar{t})}{3(s+\bar{s})+\alpha(t+\bar{t})})\sin^2(\theta)
+ \nonumber\\
& & (s+\bar{s})\,(2-\frac{3(s+\bar{s})}{3(s+\bar{s})+\alpha(t+\bar{t})})\cos^2(\theta)-
\frac{2\sqrt{3}\alpha(t+\bar{t})(s+\bar{s})}{3(s+\bar{s})+\alpha(t+\bar{t})}
\sin(\theta)\cos(\theta)\cos(\gamma_s-\gamma_t)\}
\nonumber\\
A&=& \sqrt{3}Cm_{3/2}\{-1+\frac{3\alpha(t+\bar{t})}{3(s+\bar{s})+\alpha(t+\bar{t})}\sin(\theta)e^{-i\gamma_s}
+\sqrt{3}(-1+\frac{3(s+\bar{s})}{3(s+\bar{s})+\alpha(t+\bar{t})})\cos(\theta)e^{-i\gamma_t}\}\nonumber
\ea

\begin{figure}
  \begin{center}
    \begin{tabular}{cc}
      \resizebox{80mm}{!}{\includegraphics{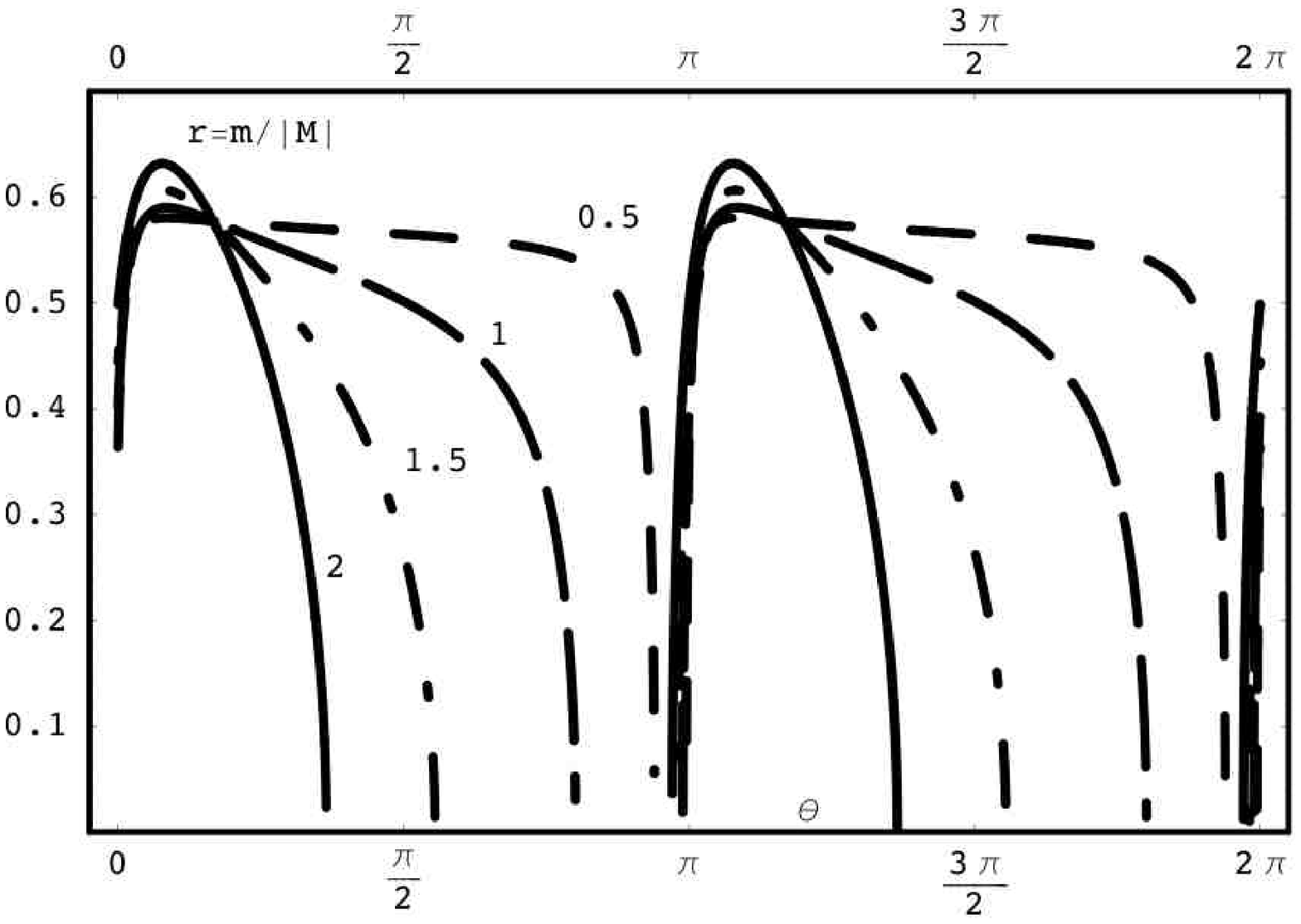}} &
      \resizebox{80mm}{!}{\includegraphics{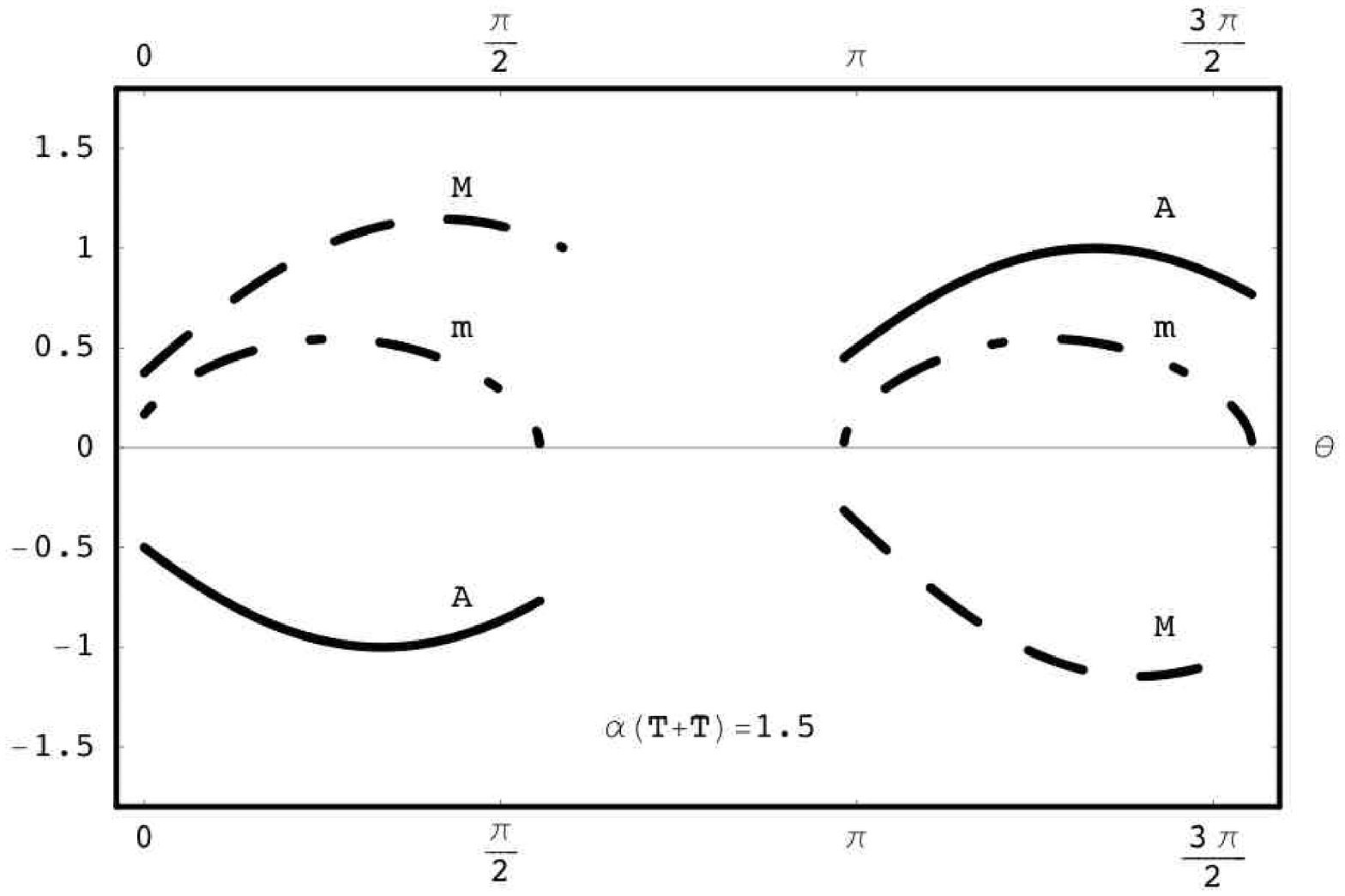}} \\
    \end{tabular}
    \caption{\footnotesize \textbf{Left} : The ratio $r\equiv m/|M|$ as a
function of $\theta$ for four different values of
$\alpha(T+\bar{T})$ represented by various curves. \textbf{Right}:
The universal soft parameters as a function of $\theta$ for
$\alpha(T+\bar{T}) = 1.5$. The solid curve stands for trilinears,
dotted dashed curve for scalars and dashed curve for gauginos. }
    \label{msoftI}
  \end{center}
\end{figure}

\noindent where we are using the following parameterization, which define $F$
terms for the moduli \cite{Brignole:1997dp}: \ba
F^s&=&\sqrt{3}m_{3/2}C(s+\bar{s})\sin(\theta)e^{-i\gamma_s}\\
F^t&=&m_{3/2}C(t+\bar{t})\cos(\theta)e^{-i\gamma_t}\nonumber\\
C^2&=&1+\frac{V_0}{3m_{3/2}^2}\nonumber \ea The ratio of the
scalar to the gaugino mass parameter $r \equiv m/|M|$ is shown in
the first plot in Figure \ref{msoftI} as a function of the
goldstino angle $\theta$. For the stau coannihilation region, the
ratio $r$ has to be roughly $0.5-0.6$. We see that for this value
of $r$, one set of allowed values of $\theta$ lie near
$0,\pi,2\pi$ for all allowed values of $\alpha(T+\bar{T})$. This
means that the supersymmetry breaking is moduli dominated ($F_s
\approx 0$). In addition, there also exist other values of
$\theta$ which are closer to ($\frac{1}{2}\pi,\frac{3}{2}\pi$)
rather than to ($0,\pi,2\pi$). However, these values are ruled out
by constraints on low energy phenomenology, as the trilinear
parameter $A_0$ for these values is pretty large, as seen from the
second plot in Figure \ref{msoftI}. This is because a very large
value of the trilinear parameter makes the scalar mass squared run
negative at the low scale and also causes problems for EWSB. Once
the correct ratio $r$ of the gaugino mass parameter to the scalar
mass parameter is obtained, one can get their correct absolute
scales by tuning $m_{3/2}$ as all the soft parameters are
proportional to them. One thus gets a gluino in the $550-650$ GeV
range. The allowed values of $m_{3/2}$ lie in the TeV range.

Moving on to the PH-B construction, one would again like to
understand the origin of the characteristic features of its
spectrum, viz. heavy squarks ($\geq 1$ TeV), moderately heavy
sleptons except the stau which is considerably lighter, and
gluinos which can vary from being light ($250-450$ GeV) to heavy
($\geq 1000$ GeV). Light gluinos ($< 450 $GeV) in this
construction \emph{always} give rise to a wino LSP while the
heavier ones give rise to bino or wino LSPs, as we explain below.

The PH-B construction is a weakly coupled heterotic string
construction with a tree level K\"{a}hler potential and two
gaugino condensates. The soft terms for this construction depend
on the ``theory'' input parameters -- the gravitino mass
$m_{3/2}$, the Green-Schwarz coefficient $\delta_{GS}$ and the
$vev$ of the K\"{a}hler modulus $t$, in addition to $\tan{\beta}$.
In these kind of constructions, it was further noted that a
minimum with $F_s=0$  and $F_t \neq 0$ is preferred with $t$ being
stabilized at values slightly greater than 1. For details, refer
to \cite{Casas:1990qi}. The result is that all soft terms are zero
at tree level. The expressions for the soft parameters are
approximately given by \cite{Kane:2002qp}: \ba
\label{soft-racetrack} M_a &\approx&
\frac{g_a^2}{2}\,[(2\frac{\delta_{GS}}{16{\pi}^2}+b_a)\,G_2(t,\bar{t})+2b_am_{3/2}];
\;\;G_2(t,\bar{t})\equiv (2\zeta(t)+\frac{1}{t+\bar{t}}) \\
m_i^2 &\approx& \gamma_i\,m_{3/2}^2 \nonumber \\
A_{ijk} &\approx& m_{3/2}\,(\gamma_i+\gamma_j+\gamma_k) \nonumber
\ea
\noindent where $\zeta(t)$ is the Riemann zeta function. The
dominant contribution to the soft scalar mass parameters is
proportional to the gravitino mass with the proportionality
constant being the anomalous dimension of the respective fields
($\gamma_i$). Since the anomalous dimension of the quarks is
bigger than that of the leptons, the squarks turn out to be
heavier than the sleptons. Also, the anomalous dimension of the
stau ($\tilde{\tau}$) is the smallest (smaller by a factor of
about 3 compared to that for $\tilde{Q}_3$), leading to the stau
as the lightest slepton. To get the absolute scale correct, one
has to realize that soft terms in this case arise from one loop
contributions. Thus, they are suppressed and therefore need a much
heavier $m_{3/2}$ $(\sim 20 $ TeV) in order to evade the chargino,
neutralino and higgs mass lower bounds. This is the reason for the
heavy squarks, moderately heavy sleptons and a light stau at the
low scale.
\begin{figure}
  \begin{center}
    \begin{tabular}{cc}
      \resizebox{70mm}{!}{\includegraphics{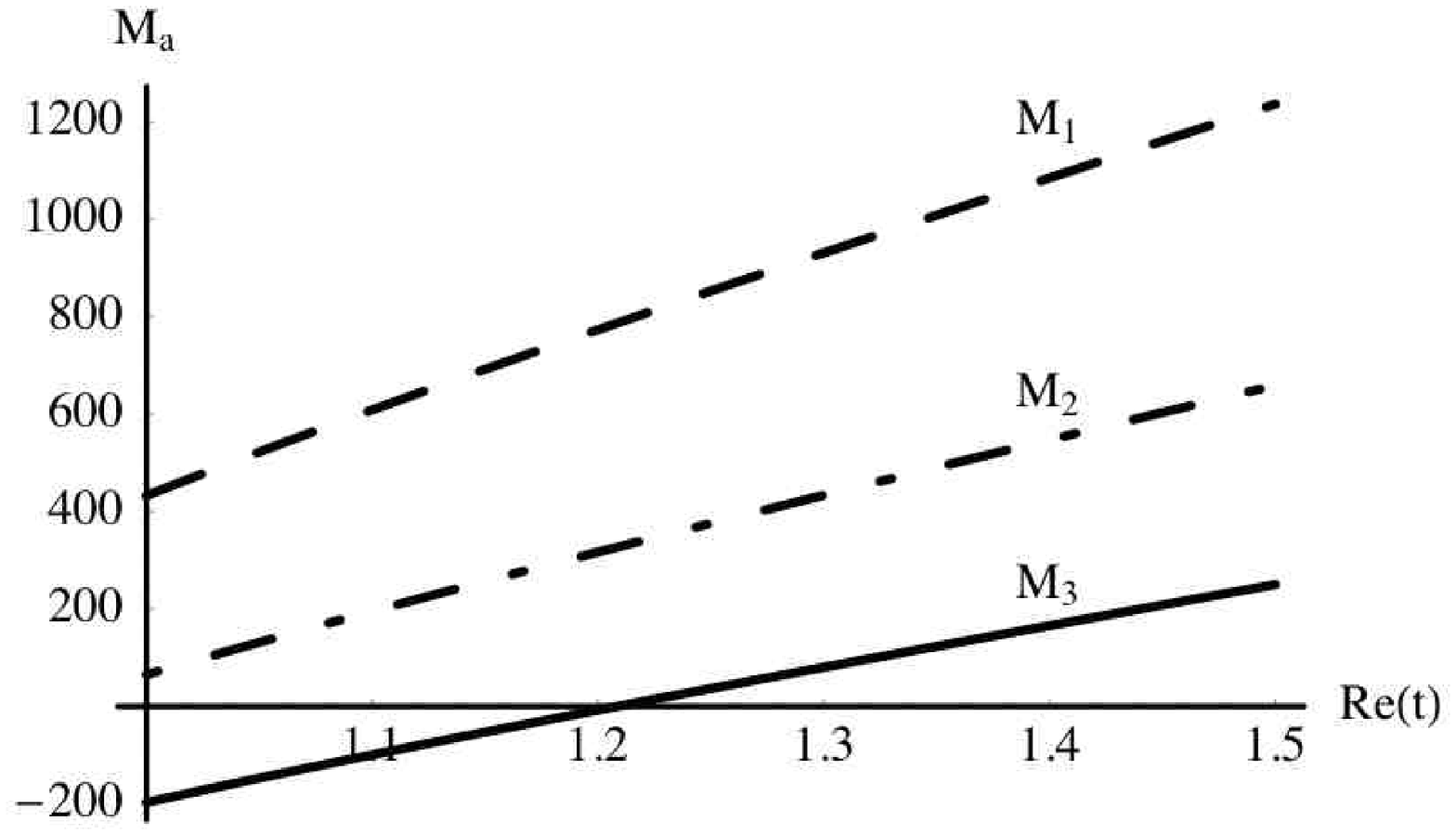}} &
      \resizebox{70mm}{!}{\includegraphics{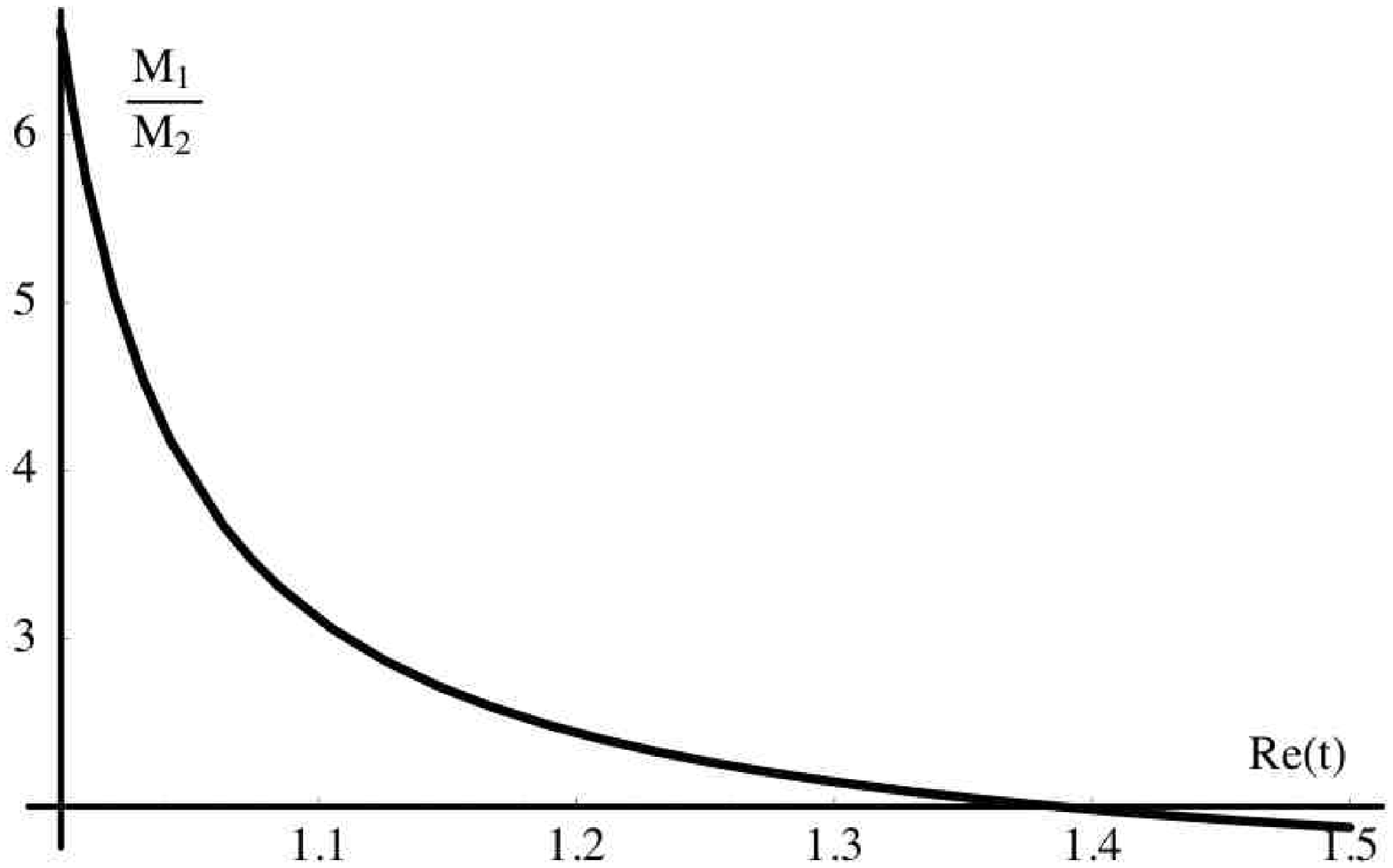}} \\
    \end{tabular}
    \caption{\footnotesize \textbf{Left} : Plots for gaugino mass parameters as a function of Re($t$):
The solid curve stands for $M_3$, dotted-dashed for $M_2$ and
dashed for $M_1$. \textbf{Right} : The ratio $M_1/M_2$ as a
function of $Re(t)$. Re($t$) varies from 1 to 1.5. The plots are
shown for a given value of $\delta_{GS}$ and $m_{3/2}$.
$\delta_{GS}$ is -15 and $m_{3/2}$ is 20 TeV.}
    \label{racetrack}
  \end{center}
\end{figure}

For the gaugino sector, it is instructive to look at the plots of
the variation of gaugino mass parameters as a function of Re($t$)
and the ratio of bino and wino mass parameter ($\frac{M_1}{M_2}$)
as functions of Re($t$) for a given value of $\delta_{GS}$ and
$m_{3/2}$. Choosing different values of $\delta_{GS}$ does not
change the qualitative feature of the plots. Since all gaugino
mass parameters are proportional to $m_{3/2}$, changing $m_{3/2}$
changes the overall scale of all gaugino mass parameters. We first
explain why light gluinos give rise to a wino LSP while heavier
ones to a bino LSP for a fixed $m_{3/2}$. We then consider the
effects of changing $m_{3/2}$. The first plot in Figure
\ref{racetrack} shows that the gluino mass parameter is the
smallest of the three (taking the sign into account) at the
unification scale. This arises from the fact that the combination
($G_2(t,\bar{t})+2m_{3/2}$) in (\ref{soft-racetrack}) is negative
and the one loop beta function coefficient $b_3$ is the largest
for $M_3$ \cite{Kane:2002qp}. From the second plot, we see that
ratio $\frac{M_1}{M_2}$ is greater than 2 for Re($t$) smaller than
a certain value, which is around 1.4 for the value of
$\delta_{GS}$ chosen in the figure. Since we roughly have :

\begin{eqnarray}
M_{1_{low}} \approx 0.45\,M_{1_{unif}}; \;\; M_{2_{low}} \approx 0.9\,M_{2_{unif}},
\end{eqnarray}

\noindent it implies that if the ratio
$\frac{M_{1_{unif}}}{M_{2_{unif}}}$ is greater than 2, we have
$M_{1_{low}} > M_{2_{low}}$, leading to a wino LSP. Therefore for
Re($t$) smaller than a certain value (1.4 in the figure), one
obtains a wino LSP while for greater values of Re($t$) one obtains
a bino LSP. From the first plot, one now sees that for values of
Re($t$) smaller than the critical value, the gluino mass is quite
small at the unification scale. Thus for a given $m_{3/2}$, PH-B
models with small gluino masses have wino LSPs while those with
heavy gluinos have bino LSPs. If we now change the gravitino mass,
we change the overall scale of the gaugino mass parameters. Since
the scalars are also proportional to $m_{3/2}$, it is not possible
to make $m_{3/2}$ very small as the higgs mass bound will be
violated. But one can have a large gravitino mass giving rise to a
large gluino mass, with both wino and bino LSPs. Bino LSP models
however have heavier gluino masses than those with wino LSP as
$M_3$ is bigger for the bino LSP models, as explained above.  One
also finds that for Re($t$) around a particular value (1.2 in the
above figure), the gluino mass almost vanishes leading to a gluino
LSP at the low scale, which is not considered in our analysis.
Therefore that region in Re($t$) is not allowed. Another region
which is not allowed by low energy constraints is the region near
Re($t$)=1, where $M_2$ becomes very small, leading to
incompatibilities with the lightest chargino and LSP bounds.

\subsection{Explanation of Soft Parameters from the Underlying Theoretical Construction}\label{fromtheory}

One finally needs to explain the structure of soft terms (which
explains the spectrum pattern and hence the signature pattern)
from the structure of the underlying theoretical construction.
This would complete the sequence of steps to go from LHC
signatures to string theory. As explained before, we carry out
this exercise for the KKLT and Large Volume constructions, since
they are well defined from a microscopic point of view, and have a
reasonably well understood mechanism of supersymmetry breaking and
moduli stabilization.

Both of the constructions have complex structure moduli and
dilaton stabilized by turning on generic fluxes. The K\"{a}hler
moduli are stabilized by including non-perturbative corrections to
the superpotential. In Large Volume (IIB-L) vacua, a certain kind
of $\alpha'$ correction is also taken into account in the scalar
potential unlike that in the KKLT (IIB-K) vacua. In type IIB-K
constructions, the flux superpotential has to be fine-turned so as
to give a small ($\sim$ TeV) gravitino mass. For the IIB-L
construction however, no fine-tuning of the flux superpotential is
required. This gives rise to a relatively low (intermediate scale)
string scale if one wants a small gravitino mass.

The common feature of these two constructions is that the
K\"{a}hler moduli are stabilized mostly by non-peturbative
corrections. This leads to a particular feature in the gaugino
sector. It was shown in \cite{Conlon:2006us} that the gaugino
masses are suppressed relative to the gravitino mass $(\sim
m_{3/2}/\ln(m_{pl}/m_{3/2}))$ in all type IIB vacua with matter
residing on stacks of D7-branes and with all K\"{a}hler moduli
stabilized mostly by non-perturbative corrections to the
superpotential.

For the IIB-K constructions studied mostly in the literature,
there is only one overall K\"{a}hler modulus, the F-term of which
is suppressed. Since both the scalar masses and trilinear terms
are proportional to this F-term, they are both suppressed relative
to the gravitino in the IIB-K construction \cite{Choi:2004sx}.
Anomaly mediated contributions to soft terms have to be added as
they are comparable with those at tree level, leading to mixed
modulus-anomaly soft supersymmetry breaking terms. The above
feature survives for cases with more K\"{a}hler moduli, if all of
them are stabilized mostly by non-perturbative effects
\cite{Denef:2005mm}. On the other hand, the IIB-L constructions
require the presence of a large volume limit -- this means that
the Calabi-Yau manifold must have at least two K\"{a}hler moduli-
one of which is small ($T_s$) and the other is big
($T_b$)\cite{Balasubramanian:2005zx}. The presence of the
perturbative $\alpha'$ correction in the K\"{a}hler potential
gives a contribution to the scalar potential of the same order as
the non-perturbative corrections for the ``big'' K\"{a}hler
modulus, in contrast to that in the IIB-K construction. The F-term
of the small K\"{a}hler modulus ($F_s$) is suppressed by
$\ln(m_{pl}/m_{3/2})$, while that of the big K\"{a}hler modulus
($F_b$) is not suppressed. Since only D7-branes wrapping the small
4-cycle (represented by the small modulus) give a reasonable gauge
coupling, the visible sector gaugino masses are proportional to
$F_s$ and are suppressed relative to $m_{3/2}$. However both $F_s$
and $F_b$ enter into the expression for scalar masses and
trilinear terms. Since $F_b$ is not suppressed, therefore for the
IIB-L construction, only the gaugino sector has mixed
modulus-anomaly terms, with the scalars and trilinears generically
of the same order as the gravitino mass. This characteristic
feature is also true for Calabi-Yaus with more K\"{a}hler moduli
provided they admit a large volume limit, though the explicit soft
terms are hard to obtain. Another difference between the IIB-K and
IIB-L constructions is that the soft terms for the IIB-K
construction are first computed at the unification scale ($\sim
10^{16}$ GeV) while those for the IIB-L construction are computed
at the intermediate string scale ($\sim 10^{11}$ GeV).

The above analysis thus explains the origin of the pattern of soft
parameters in terms of the structure of the underlying theoretical
construction for the two constructions which leads to a
distinguishable signature pattern at the LHC. The important thing to take
home from this analysis is that different constructions lead to
different effective actions and therefore to different expressions
for the soft terms in terms of the underlying microscopic
input parameters. In addition, the relations \emph{among} the different
soft parameters ($M_a,m^2_i\,\&\,A_{ijk}$) also change for
different constructions. Therefore, a proper understanding of
these relations and their implications for relevant features of
the phenomenology is the key to relating high scale theory and
data. In this sense, we think that the approach advocated here is
likely to work even if one has much more realistic constructions
from different parts of the M-theory amoeba which stabilize
all the moduli, generate a stable hierarchy and also give a
realistic spectrum and couplings.

\section{Distinguishing Theories Qualitatively}\label{qualitative}

In the previous section, we analyzed the eight constructions in
great detail -- in particular, we computed the LHC signatures of
these constructions and understood the origin of these signatures
from features of the theoretical constructions. It is worthwhile
to ask whether one can abstract important lessons from this
exercise so that one could use them to analyze other classes of
constructions, and to draw qualitative reliable conclusions from
data.

For example, one could try to first extract relevant phenomenological features of the effective
beyond-the-Standard Model theory from data and then focus on classes of M theory vacua which give rise
to those particular features. This alternative may also be more helpful to people who are interested
primarily in understanding general features of beyond-the-Standard
Model (BSM) physics from LHC data rather than connecting it to an
underlying high scale theory like string theory. From our studies, we
find that a combination of features of any construction crucially determine
the broad pattern of LHC signatures. For concreteness,
we write our results in the framework of low-scale supersymmetry as BSM physics, similar results will
hold true for other approaches as well. The important features we find\footnote{There could be more such features.}
are:

\begin{itemize}
\item The universality (or not) of gaugino masses at the
unification (or compactification) scale. \item If gauginos are
non-universal - the origin of the non-universality, i.e. whether
the non-universality is present at tree-level itself as opposed to
arising mostly due to one-loop anomaly mediated contributions.
\item If gauginos are non-universal - the hierarchy between $M_1$,
$M_2$,  $M_3$ and $\mu$. \item The relative hierarchy between the
scalars and gauginos at the string scale, i.e. whether the scalars
are of the same order as the gauginos as opposed to being heavier
or lighter than the gauginos. \item Nature and content of the LSP.
\item Hierarchy among scalars at the string scale, particularly
third family {\it vs} the first and second families.
\end{itemize}

Some comments are in order. These features are not always
independent of each other. For example, the hierarchy between
$M_1$, $M_2$ and $M_3$ determines the nature of the LSP (combined
with a knowledge of $\mu$). Also, if tree-level gaugino masses are
small so that non-universality arises only due to the anomaly
mediated contributions, then the gauginos are typically suppressed
relative to the scalars and the hierarchy between $M_1$, $M_2$ and
$M_3$ is fixed. Another important fact which should be kept in
mind is that a combination of all the features above gives rise to
the observed \emph{pattern} of LHC signatures, not just a
particular one. Therefore, once one obtains data, the task boils
down to figuring out the correct combination of ``relevant
features" which reproduces the data (at least roughly). Let's
explain this with two examples - the HM-A construction and the
IIB-L construction. Since we are only concerned with relevant
features of the effective BSM theory, all constructions considered
can be treated equally.

The HM-A and overlapping HM-B and HM-C constructions have
universal gaugino masses at the unification scale and have a bino
LSP. They also have scalars of the same order as the gauginos at
the unification scale, so RGE effects make the scalars (the third
generation in particular) lighter than the gauginos at the low
scale. This combination of ``relevant features" determines the
broad pattern of LHC signatures for these constructions. Since the
gauginos are universal at the unification scale, the ratio of
$M_1$, $M_2$ and $M_3$ at the low scale is $1 : 2 : 7$, which
controls the lower bound on the gluino mass and the LSP mass due
to experimental constraints on the chargino mass. Also, since
scalars are slightly lighter than gluinos at the low scale, both
$\tilde{g}\tilde{q}$ and $\tilde{q}\tilde{q}$ pair production are
comparable. The fact that the LSP is bino-like is also due to
universal gaugino masses as well as the fact that the scalars are
comparable to gauginos at the unification scale. A bino LSP can
then reduce its relic density by stau coannihilation, which can
only happen if the stau is light and almost degenerate with the
LSP. All these factors give rise to a very specific set of
signature pattern, as analyzed in section \ref{spectrum}.

In contrast, the IIB-L construction has a different set of
``relevant features" which determine its broad pattern of LHC
signatures and also allow it to be distinguishable from the other
constructions. The gaugino masses are non-universal at the string
scale\footnote{One does not have standard gauge unification at
$2\times10^{16}$ GeV in these models with $W_0=O(1)$
\cite{Balasubramanian:2005zx}.} and the scalars are heavier than
the gauginos at the low scale\footnote{This implies that the
scalars are also heavier than the gauginos at the string scale.}.
One also a mixed bino-higgsino LSP in this case. Since the scalars
are heavier than the gauginos, $\tilde{g}\tilde{g}$ pair
production is the dominant production mechanism. Also, the only
way to decrease the relic density of a bino-like LSP is to have a
significant higgsino fraction as the stau (or stop) coannihilation
channel is not open. Again, these features result in a very
specific set of signature pattern, as analyzed in section
\ref{spectrum}.

We therefore see that the features mentioned above are crucial in
determining the pattern of signatures at the LHC. Having said
that, it is important to remember that these features only
determine the broad pattern of LHC signatures and one needs more
inputs to explain the entire signature pattern in detail.

\section{Possible Limitations}\label{limitations}

One may raise questions about a few aspects of our analysis. The
first concerns the sampling of the parameter space of each
construction. One may worry that by only considering $\sim$50
models ($\sim$100 for the PH-A construction)\footnote{Not all
50(or 100) models simulated will be above the observable limit in
general.}, the parameter space of each construction is sampled
very sparsely and adding more models could qualitatively change
the overall signature pattern of the constructions. We think
however, that it is reasonable to expect that that is not true.

\begin{figure}[h!]
  \begin{center}
      \epsfig{file=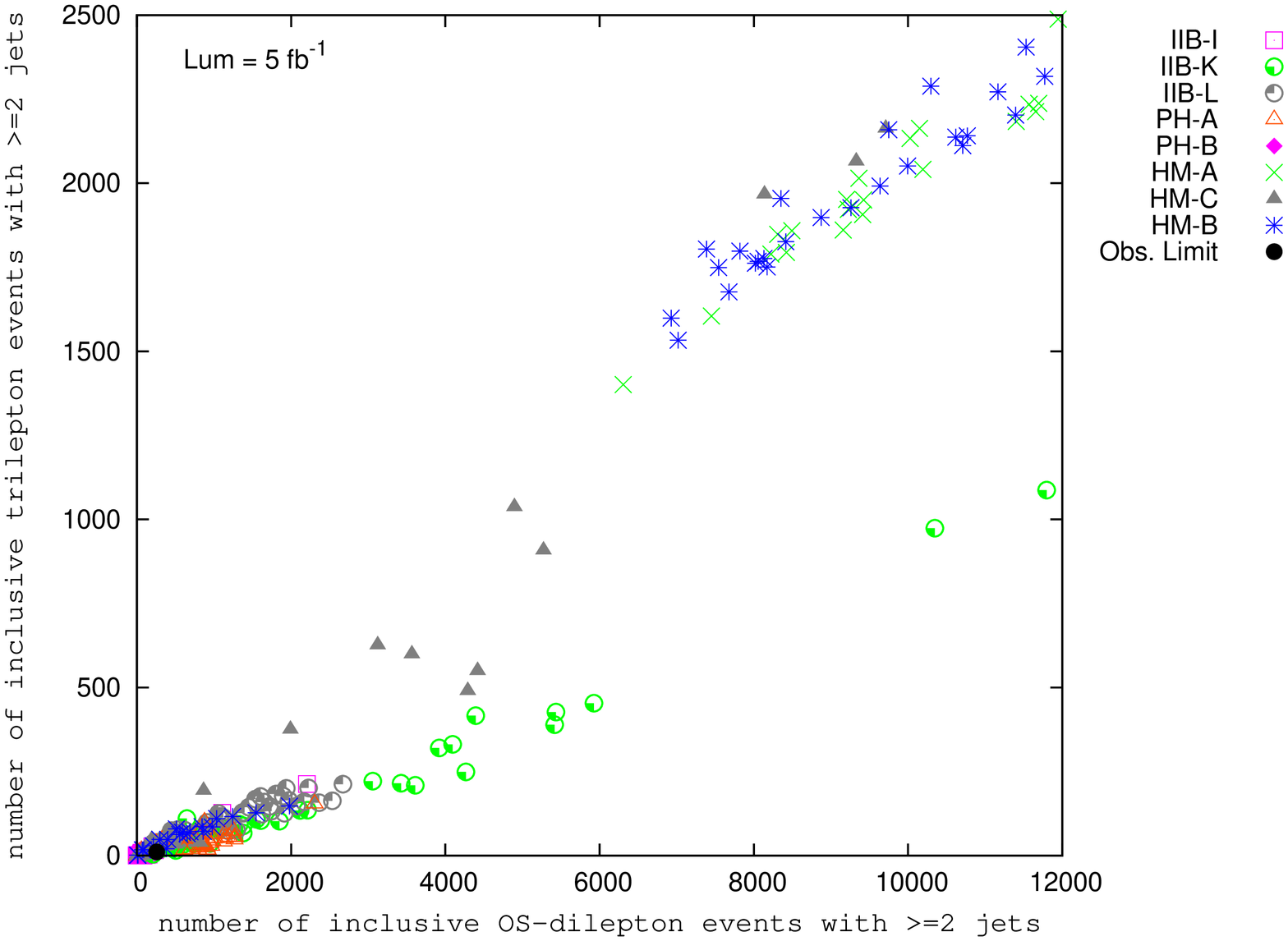,height=16cm, angle=-90}
    \caption{\footnotesize Plot of number of events with 1 lepton and $\geq$ 2 jets and number of
events with opposite sign dileptons and $\geq$ 2 jets, each
sampled with $\sim 50$ models, except PH-A ($\sim$100 models).}
    \label{OS-3l}
  \end{center}
\end{figure}

\begin{figure}[h!]
  \begin{center}
 \epsfig{file=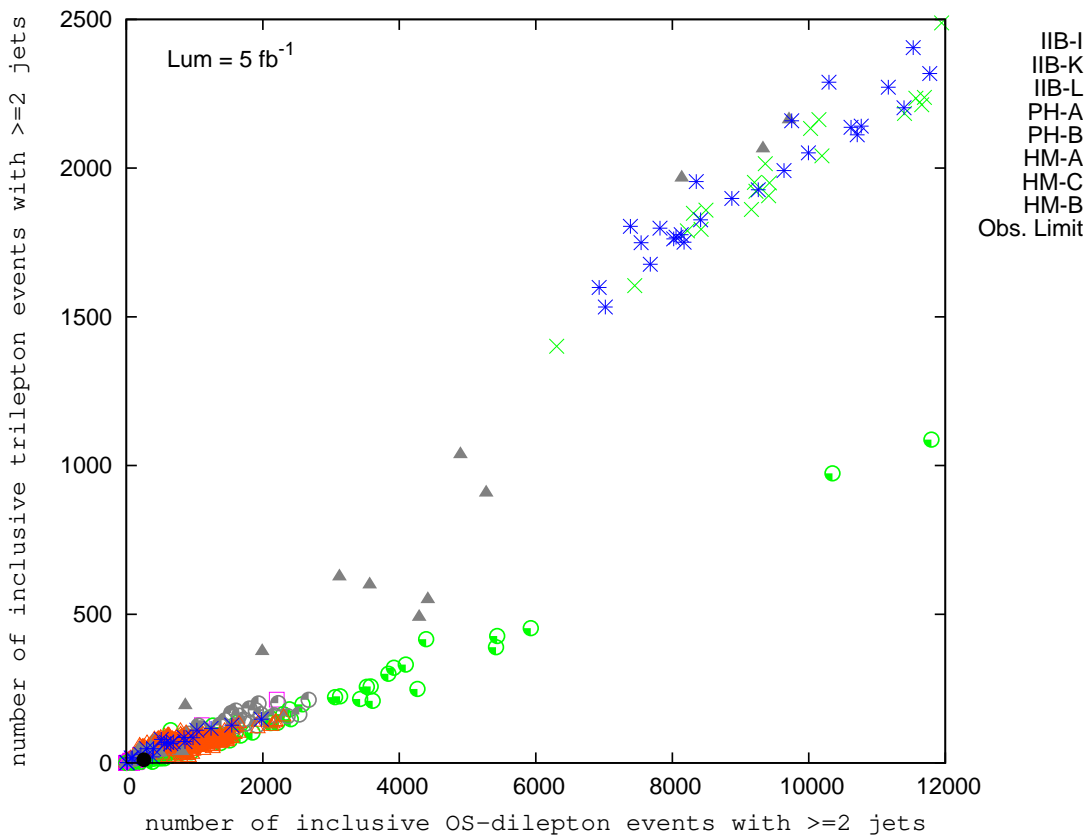,height=16cm, angle=-90}
 \caption{\footnotesize The same
    plot as in Figure \ref{OS-3l}, in which the IIB-K construction is
sampled with $\sim 100$ models and the PH-A construction with
$\sim$400 models.}\label{OS-3l-new}
  \end{center}
\end{figure}

\begin{figure}[h!]
  \begin{center}
      \epsfig{file=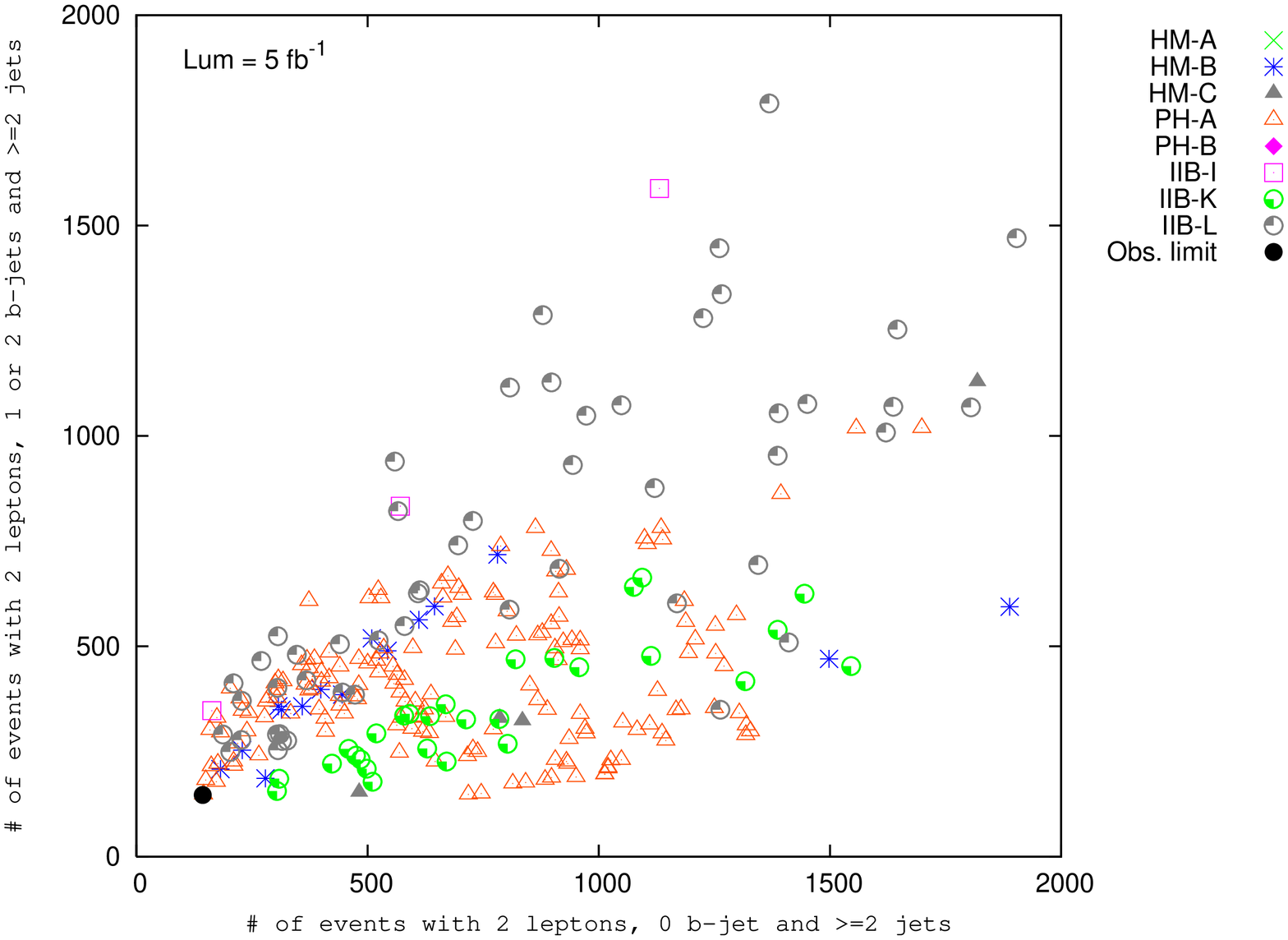,height=16cm, angle=-90}
    \caption{\footnotesize Plot with x axis showing number of events with 2 leptons, 0 b jets and $\geq$2 jets,
      and y axis showing the number of events with 2 leptons, 1 or 2 b jets and $\geq 2$ jets
     each sampled with $\sim 50$ models, except PH-A ($\sim$100 models).}
    \label{ratio1}
  \end{center}
\end{figure}

This is because, as explained in the previous sections, we have
outlined the origin of the pattern of signatures of the various
constructions on the basis of their spectrum and in turn on the
basis of their underlying theoretical setup. Since the dependence
of the soft terms on the microscopic input parameters as well as
relations between the soft terms are known and have been
understood, we expect our results for the pattern table to be
robust even when the parameter space is sampled more densely. This
will be strictly true only if one understands the theoretical
construction well enough so that one has a ``representative
sample" of the entire parameter space of that construction. We
expect this to be true for our constructions. In order to confirm
our expectation, we simulated $\sim$400 models for the PH-A
construction and $\sim$100 for the IIB-K construction and we found
that the results obtained with $\sim$100 (and $\sim$50) models did
not change when other models were added in our analysis. This can
be seen from Figures \ref{OS-3l} and \ref{OS-3l-new} as well as
Figures \ref{ratio1} and \ref{ratio2}. The other signatures also
do not change the final result. In the future, we plan to do a
much more comprehensive analysis with a dense sampling of the
parameter space for all the other constructions.

Another possible objection could be that the procedure of dividing
a signature into two classes arbitrarily and distinguishing them
on the basis of falling into one class or the other is too naive
and may lead to misleading results arising from intrinsic
statistical uncertainties and background uncertainties in the
value of each signature and impreciseness of the boundary. While
this is a valid concern in general, we think that this does not
affect the main results of our analysis at this level. This can
again be attributed to the fact that the pattern of signatures is
understood on the basis of their underlying theoretical structure.
We are also encouraged as there are typically more than one
(sometimes many) signatures distinguishing any two particular
constructions.

Another possible limitation which one could point out is that
distinguishing theoretical constructions on the basis of two
dimensional signature plots is not very powerful. Since we are
only looking at various two-dimensional projections of a
multi-dimensional signature space, it is possible that two
different theoretical constructions occupy different regions in
the multi-dimensional signature space even though they overlap in
all the two-dimensional projections. One would then not be able to
cleanly distinguish two constructions by this approach even though
they are intrinsically distinguishible. However, one can get
around this limitation by tagging individual models of each
theoretical construction. It would then be possible to figure out
if two different constructions are distinguishible even if they
overlap in all two dimensional signature plots. As already noted
in section \ref{results}, our purpose was to outline the approach
in a simple manner. It is clear that the approach has to be made
more sophisticated for more complicated situations.

For a mathematically precise way of distinguishing pairs of
constructions, one could use the following procedure. Imagine
dividing the parameter space of the two constructions into a
coarse grid with coordinates given by their parameter vectors
$\vec{x}$. For example, for the PH-A construction, $\vec{x} =
\{m_{3/2},a_{np},\tan(\beta)\}$ \cite{Kane:2002qp}. We can then construct
a ${\chi}^2$-like variable defined as follows:
\begin{figure}[ht]
\begin{center}
 \epsfig{file=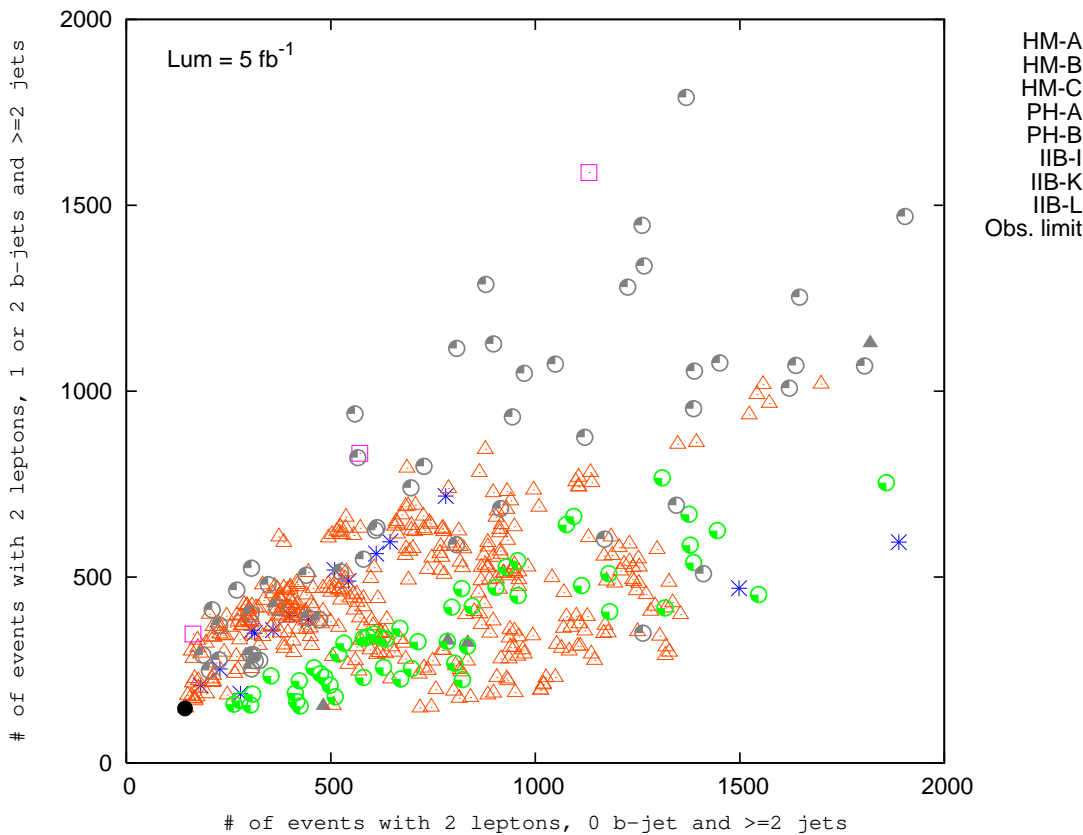,height=16cm, angle=-90}
 \caption{\footnotesize  The same
    plot as in Figure \ref{ratio1}, in which the IIB-K construction is
sampled with $\sim 100$ models and the PH-A construction with
$\sim$400 models.}\label{ratio2}
  \end{center}
\end{figure}

\begin{eqnarray} \label{chisq}
(\Delta S)_{AB} =
\mathrm{min}_{\{\vec{x},\vec{y}\}}\sum_{i=1}^{n_{sig}}\left(\frac{s_{A}^i(\vec{x})-s_{B}^i(\vec{y})}{\sigma_{AB}^i}\right)^2
\end{eqnarray}
\noindent where $A$ and $B$ stand for the two constructions, $s_i$
stands for the $i$-th signature and $\sigma_{AB}^i$ stands for the
error bar assigned between the $A$ and $B$ constructions for the
$i$-th signature. $\sigma_{AB}^i$ can be determined from
statistical errors of the $i$-th signature for constructions $A$
and $B$ as well as the standard model background error for the
$i$-th signature.

Since we minimize with respect to the parameter vectors $\vec{x}$
and $\vec{y}$ of the two constructions, equation (\ref{chisq}) can
be geometrically visualized as the ``minimum distance squared" in
the full multi-dimensional signature space between the two
constructions $A$ and $B$ with an appropriate ``metric"
\footnote{The inverse square of the error $\sigma^i_{AB}$ acts
like the metric in signature space.}, and serves as a measure of
the difference between the two constructions $A$ and $B$. Carrying
out the minimization procedure in practice is quite a non-trivial
task, especially when the parameter vectors have many components,
as the time required to complete the minimization procedure in a
satisfactory way is too large. However techniques have been
introduced to get around this problem, for example see
\cite{Allanach:2004my}. In the future, we plan to do a
comprehensive analysis with more statistics and a precise method
to distinguish models, as explained above.

Finally, one might object to the approach of figuring out
experimental predictions for various classes of underlying
theoretical constructions and distinguishing different theoretical
constructions from each other before actual data, instead of the
more ``standard" approach of comparing each theoretical
construction with actual data. Many reasons can be given in this
regard. From a conceptual point of view, this approach fits will
with the philosophy of addressing the Inverse Problem in a general
framework. It is also crucial to addressing the question of
predictivity of an underlying theory like string theory in general
and the particular string vacuum we live in, in particular. From a
more practical point of view, studying the various subtle aspects
of connecting an underlying theoretical construction with collider
signatures is quite a non-trivial and subtle exercise and requires
considerable time and investment. It is therefore very helpful to
build knowledge and intuition in this regard and be prepared for
actual data. Carrying out this exercise could help discover
important properties of the low-energy implications of various
classes of theoretical constructions, in turn pointing to new
classes of collider observables as well as helping design new
analysis techniques.

\section{Summary and Future Directions}\label{conclude}
In this work, we have tried to address the goal of learning about
the underlying theory from LHC data - the deeper Inverse Problem.
We have proposed an approach by which it can be shown that the two
prerequisites to addressing the deeper Inverse Problem, namely, a)
To reliably go from a microscopic theory to the space of
experimental observables, and b) To distinguish among the various
classes of microscopic theoretical constructions on the basis of
their experimental signatures, can be satisfied for many
semi-realistic string constructions which can be described within
the supergravity approximation. In our opinion, the paper is
seminal in the sense that it proposes a new way of thinking about
fundamental theory, model-building and collider phenomenology such
that there is a better synergy between each of these subfields.
Perhaps the most important result, which has never before been
presented, is that different classes of string constructions give
\emph{finite} footprints in signature space and that different
string constructions give practically overlapping but different
and distinguishable footprints.

The reason it is possible to distinguish theoretical constructions
is that patterns of experimental observables (for eg. signatures
at the LHC) are \emph{sensitive} to the structure of the
underlying theoretical constructions, because of correlations (see
section \ref{whypossible}). More precisely, this means that a
given theoretical construction occupies a {\it finite} region of
signature space which is in general different from another
theoretical construction. Moreover, the origin of this difference
can be understood from the underlying structure of the
construction. Therefore, even though we have carried out a
simplified analysis in terms of the imposition of cuts, detection
efficiencies of particles, detector simulation and calculation of
backgrounds, we still have confidence in the robustness of our
results. We have analyzed two classes of string vacua and six
other string-motivated constructions in detail. The point of this
exercise was to illustrate our approach, the same procedure should
be carried out for more classes of realistic vacua so that the
procedure becomes more-and-more useful. If the approach fails, it
will not be because of these and similar issues discussed above,
but rather because many regions of the entire M theory amoeba
cannot yet be analyzed by the approach we use. However, we think
that rather than giving up ahead of time, the best attitude is to
continue to expand both theoretical understanding as well as our
approach, and confront them with data.

There are two directions in which the approach advocated in this
paper can be generalized and sharpened further. The first concerns
theoretical issues -- efforts should be made to go beyond toy
models focussing on few aspects of theory and phenomenology to
more holistic ones that address (if not solve) all of the issues
an underlying string theoretic construction might be expected to
explain. For instance, a better understanding of the theory can
fix some (or all) of the microscopic input parameters of a given
construction, increasing the predictivity of the construction. The
approach advocated by M. Douglas and W. Taylor, {\it viz.} to look
for correlations in the space of observables by analyzing
different classes of vacua is very similar to our approach in
principle \cite{douglas-taylor}.

The second concerns the analysis and interpretation of data and
its connection to the underlying theory. Creative thinking is
needed in identifying collider observables which more directly
probe the key features of the Beyond-the-Standard Model (BSM)
lagrangian and its connection to the underlying theory. We were
able to identify some useful observables by examining specific
constructions. In addition, one should find ways in which
observables from all fields -- collider physics, flavor physics,
cosmology, etc. could be used in conjunction to distinguish among,
and favor or exclude, many classes of string constructions in a
quick and robust manner. Our proposed technique is very useful in
this regard as it is very easy to add non-collider observables --
such as from flavor physics, cosmology, etc. to the collider
observables such that they are all treated in a uniform manner.

It is important to understand that the proposed approach should be
applied at various stages, with different tools and techniques
useful for each stage. The first stage would consist of
distinguishing many classes of constructions with limited amount
of data by using simple signatures and simple analysis techniques.
This has the advantage that one can rule out various classes of
constructions with relative ease. However, in order to go further,
it is important to develop more specialized analysis techniques
and use more exclusive signatures. This is best done in subsequent
stages, when one zooms in to a more limited set of constructions
and also obtains more data. Since one has better statistics, one
can use optimized and more exclusive signatures as well as use
more sophisticated analysis techniques to get more detailed
information about the constructions. Many of these sophisticated
analysis techniques already exist in the literature
\cite{Kneur:1998gy}, although they have been applied to very
special scenarios like minimal supergravity, minimal gauge
mediation, etc. One would now need to apply similar techniques
(suitably modified) to the set of constructions consistent with
limited data. Also, in the past year, a lot of progress has been
made towards uncovering the low energy spectrum and parameters
from (simulated) LHC signatures in the form of ``blackboxes''
constructed by some groups. This has been the program of the
\emph{LHC Olympics} Workshops in the past year \cite{LHCO-talks}.

Combining these sophisticated techniques with (some) knowledge of
the connection between theoretical constructions and data obtained
in the first stage, we hope that one can further distinguish the
remaining constructions, learn more about underlying theoretical
issues, like supersymmetry breaking and mediation, moduli
stabilization, inflation, etc.

On a more philosophical note, we understand that in many cases the
structure of string theory is not understood well enough to permit
a connection to low energy physics in general and collider data in
particular. In such a situation, we think that the most useful
approach one can take is to compute predictions for low energy
experimental observables for as many classes of realistic string
vacua as possible and try to learn how information from
experimental data may favor some regions of the M-theory amoeba
over others. Doing so will lead to learning more about string
theory, and could be crucial to learning how or if string theory
can be related to the real world.

\section{Acknowledgments}
\noindent The authors appreciate helpful conversations with and
suggestions from Marcus Berg, Joseph Lykken, David Morrissey,
Brent Nelson, Fernando Quevedo, Michael Schulz, Washington Taylor,
Jesse Thaler, Lian-Tao Wang, Ting Wang and James Wells. GLK and PK thank
the Kavli Insitute for Theoretical Physics (KITP), UCSB for its hospitality
where part of the research was conducted. The
research of GLK and PK supported in part by the US
Department of Energy and in part by the National Science Foundation under Grant No. PHY99-07949.
The research of JS is supported by the US Department of Energy.

\appendix
\section{Description of ``String-motivated" Constructions}
Here we give a description of the string motivated constructions
used in our study. As stated earlier, we work in the framework of
string theory giving rise to semi-realistic constructions with
low-energy supersymmetry. Taking the extraordinary (apparent)
unification of gauge couplings in the MSSM to be an important clue
to fundamental physics, we only consider constructions which can
lead to unification of gauge couplings\footnote{either naturally
as in heterotic constructions or by imposing this as a constraint
as in type II constructions.}. In addition for simplicity, in this
paper we assume that the low energy spectrum is that of the MSSM
with no intermediate scale physics between the TeV scale and the
unification scale.

The constructions used in our study come from different corners of
the string/M theory amoeba. As explained in section
\ref{examples}, these constructions are not complete from a
theoretical point of view, especially because aspects of moduli
stabilization and supersymmetry breaking have not been taken into
account in a comprehensive manner. Supersymmetry breaking, for
instance, is only parameterized. The constructions are described
below:

\vspace{0.5cm}

\hspace{3.50cm} {\textit{\textbf{Weakly coupled Heterotic string
constructions}}}

\vspace{0.5cm}

Weakly coupled heterotic string constructions are the oldest
branch of string phenomenology, with many papers in the literature
\cite{Dixon:1986qv}. A good review of the subject can be found in
\cite{Font:1989aj}. In most examples of such constructions, one
compactifies on a six dimensional toroidal orbifold with
$\mathcal{N}=1$ SUSY and calculates the exact spectrum and some
couplings. Many examples are known, with a spectrum close to that
the MSSM, albeit with some exotics. There are also constructions
with compactifications on Calabi-Yaus, although one has less
control on the spectrum and couplings. The effective lagrangian
thus obtained is given by $\mathcal{N }=1, d=4$ supergravity,
which is encoded by three functions - the Kahler potential ($K$),
the superpotential ($W$) and the gauge kinetic function ($f$).
These functions depend in general on moduli fields which basically
parametrize the size and shape of the compactified dimensions.
Among the moduli fields is the dilaton which is unique in the
sense that it appears at low energies in a uniform way. Its vacuum
expectation value gives the value of the string coupling constant
($g_{str}$). For the weakly coupled heterotic string, it also
represents the tree level value of the gauge kinetic function.

\vspace{0.5cm}

\hspace{1cm}{\textit{Constructions with non-perturbative
contributions to the K\"{a}hler Potential}}

\vspace{0.5cm}

In this class of constructions, one considers compactification on
a toroidal orbifold with $\mathcal{N}=1$ SUSY and a spectrum
similar to that of the MSSM. The SUSY breaking mechanism is
thought to be provided by the non-perturbative phenomenon of
gaugino condensation which generates a non-perturbative
superpotential for the dilaton. The susy breaking mechanism in the
closed string (moduli) sector is mediated to the visible sector by
gravity. However, if one takes the tree level K\"{a}hler
potential, it is known that one gaugino condensate cannot
stabilize the dilaton with vanishing vacuum energy without
resorting to strong coupling. In this class of models, one
considers a non-perturbative contribution to the K\"{a}hler
potential \'{a} l{a} \cite{Shenker:1990uf} to get around this
problem. Explicit models have been constructed which incorporate
this K\"{a}hler stabilization mechanism with a realistic model of
modular invariant gaugino condensation in the hidden sector
\cite{Binetruy:1996xj}. Instead of constructing explicit detailed
models of supersymmetry breaking, one can generate benchmark
models by parameterizing the supersymmetry breaking by the F-term
for the dilaton and treating the gravitino mass and the
nonperturbative contribution to the K\"{a}hler potential as
tunable parameters (within appropriate ranges) \cite{Kane:2002qp}.

\vspace{0.5cm}

\hspace{1cm}{\textit{Constructions with tree level K\"{a}hler
Potential and multiple gaugino condensates}

\vspace{0.5cm}

As in the previous case, one again considers compactification on
toroidal orbifolds leading to a semi-realistic spectrum. However,
here one considers a tree level K\"{a}hler potential with multiple
gaugino condensates (typically two) to stabilize the dilaton and
break supersymmetry. Supersymmetry breaking is mediated by
gravity. In addition, one also has to take into account a Green
Schwarz counterterm, which is inherited from the underlying string
theory and can be thought of as a loop correction. Explicit models
stabilizing the dilaton and the untwisted moduli at reasonable
values and giving a gravitino mass of the order of 1-10 TeV have
been constructed \cite{Casas:1990qi}. This typically leads to a
minimum with $\langle F^S \rangle = 0$ and $\langle F^{T_i}
\rangle \neq 0$. For generating benchmark models, one can take the
gravitino mass, the Green-Schwarz coefficient and the vev of the
untwisted moduli of the orbifold as tunable parameters (within
appropriate ranges). For further details, please refer to
\cite{Kane:2002qp}.

\vspace{0.5cm}

\hspace{3.5cm} {\textit{\textbf{Strongly coupled Heterotic string constructions}}}

\vspace{0.5cm}

In this class of constructions, as the name suggests one looks at
the strongly coupled limit of $E_8 \times E_8$ heterotic string
theory. Horava and Witten showed that this limit of heterotic
string theory can be described by eleven-dimensional supergravity
on a manifold with boundaries where the two $E_8$ gauge multiplets
are restricted to the two ten dimensional boundaries \cite{HW1}.
So, it is also known as Heterotic M theory, as eleven dimensional
supergravity is the low energy description of M theory. To go to
four dimensions, one compactifies the eleven dimensional M theory
on a Calabi-Yau manifold times an interval. The effective action
for this class of constructions, which is again four dimensional
$\mathcal{N}=1$ supergravity, has been studied up to the first
order in the eleven dimensional parameter $\epsilon$
\cite{Witten}. Heterotic M theory has some nice phenomenological
properties, for eg. it is possible to lower the eleven dimensional
Planck scale ($M_{11}$) to the phenomenologically favored scale of
unification of the known gauge couplings ($M_{unif}$) in a natural
manner \cite{BD}, thereby achieving a truly unified theory.
However, concrete model building such as getting the precise
spectrum and couplings and stabilizing all the moduli with the
desired properties, is difficult.

\vspace{0.5cm}

\hspace{3.5cm} {\textit{Heterotic M theory Constructions with one modulus}}

\vspace{0.5cm}

The simplest situation occurs when one considers compactification
on a Calabi-Yau with Hodge-Betti number $h_{1,1}=1$ and
$h_{2,1}=0$, as the relevant expressions are simple. In this class
of constructions, the gauge coupling of the $E_8$ on the hidden
boundary typically becomes large at around the unification scale
and so it is reasonable to expect gaugino condensation in the
hidden $E_8$ sector. This phenomenon breaks supersymmetry which is
transmitted to the visible boundary by gravity. For constructing
benchmark models, one can parametrize the supersymmetry breaking
by F terms for the dilaton ($S$) and the radius modulus of the
eleventh dimension ($T$). This gives rise to universal soft terms
\cite{Choi:1997cm}.

\vspace{0.5cm}

\hspace{3.5cm} {\textit{Heterotic M theory Constructions with many moduli}}

\vspace{0.5cm}

Here one considers compactification on a more general Calabi-Yau,
with $h_{1,1} > 1$, i.e. more than one K\"{a}hler modulus. While
from a conceptual point of view this should not
be treated differently from the previous case, it turns out that
from the phenomenological point of view it gives rise to
different consequences. The basic reason is that in this case, one
loses universality of soft terms. In addition, most of the
explicit semi-realistic compactifications with a spectrum similar
to that of the MSSM have been on Calabi-Yaus with $h_{1,1} > 1$.
Therefore, it makes sense to consider this more general case
separately. Soft terms have been computed in the supergravity
limit for Calabi-Yau compactifications in ten dimensions with more
than one K\"{a}hler modulus \cite{BIMS}. Using the above and the
general result for the effective action up to the first order in
the expansion parameter $\epsilon$, the (numerical) computation of
soft parameters can be extended for some Calabi-Yau manifolds with
two K\"{a}hler moduli \cite{jing}.

\vspace{0.5cm}

\hspace{3.5cm} {\textit{Heterotic M theory Constructions with five-branes}}

\vspace{0.5cm}

In Heterotic M theory, one has non-perturbative objects called M-5
branes which are sources for the dual of the four-form field
strength present in eleven dimensional supergravity. It can be
shown that under certain circumstances, these M5 branes survive
the orbifold projection of the Horava-Witten construction,
permitting much more freedom to play with matter fields, gauge
groups and yukawa textures. The functions encoding the effective
action of $\mathcal{N}=1$ supergravity taking the effect of
five-brane moduli into account have been computed
\cite{Lukas:1998hk}. In this setup, it can also be shown that
supersymmetry can still be broken by the same global mechanism (as
in the case without five-branes) and is transmitted to the visible
sector by gravity. For benchmark models, one can assume
supersymmetry braking to be parametrized by F terms for the
dilaton, K\"{a}hler moduli and the five-brane moduli. An explicit
calculation of soft-terms has been done in \cite{Cerdeno:1999ur}.

\vspace{0.5cm}

\hspace{3.5cm} {\textit{\textbf{Type II String theory Constructions}}}

\vspace{0.5cm}

This class of constructions is relatively more recent than the
previously considered heterotic constructions. It has been shown
that there exist new classes of perturbative $\mathcal{N}=1$,
$D=4$ vacua which have their origin in type II string theory \cite{Ibanez:2001jhep}.
\cite{Blumenhagen:2005mu} is a good review. The development of D-brane physics is
crucial for these constructions. Here one considers type II (A or B) string theory
compactified on a six dimensional manifold $X$. For explicit
calculations of the couplings, one typically takes $X$ to be a
toroidal orientifold, although Calabi-Yau manifolds can also be
used in general. In addition, one has various stacks of Dp-branes
wrapping different cycles. The spectrum for this setup naturally
consists of non-abelian gauge theory coupled to chiral matter. The
dependence of the quantities - $K,W,f$ encoding the
$\mathcal{N}=1$ effective action on the moduli has been derived
for many cases \cite{Cvetic:2003ch}.

\vspace{0.5cm}

{\textit{Type II String theory Constructions on toroidal orientifolds with Intersecting D-branes}}

\vspace{0.5cm}

In type IIA language, in this class of constructions one
compactifies on toroidal orientifolds with stacks of intersecting
IIA D-branes wrapping intersecting cycles in the compact space.
Various examples have been constructed with a spectrum close to
that of the MSSM, although with exotics \cite{Ibanez:2001jhep}. In
type IIB language, which is related by T-duality to the previous
one, one has background magnetic fluxes on IIB D-brane world
volumes. Supersymmetry in type II models can be broken either by
strong gauge dynamics in hidden sector D-brane stacks
\cite{Cvetic:2003yd} or by supergravity fluxes \cite{gkp01}. The
supersymmetry breaking is transmitted to the visible D-brane
stacks by gravity, if one assumes that the there is no
intersection between the visible and hidden stacks. For benchmark
models, one can parametrize the supersymmetry breaking by F terms
for the moduli. For concreteness, we will use the soft terms
calculated for a particular intersecting D-brane setup in terms of
the moduli and their F terms \cite{Kane:2004hm}. Even though gauge
coupling unification is not unified in D-brane models in general,
the parameter space can be constrained by imposing such a
requirement. This was done in \cite{Kane:2004hm}.


\vspace{0.5cm}


\begin{thebibliography}{99}


\bibitem{Battaglia:2005zf}
  M.~Battaglia, A.~Datta, A.~De Roeck, K.~Kong and K.~T.~Matchev,
  JHEP {\bf 0507}, 033 (2005)
  [arXiv:hep-ph/0502041].
\\
  J.~M.~Smillie and B.~R.~Webber,
  JHEP {\bf 0510}, 069 (2005)
  [arXiv:hep-ph/0507170].
\\
  A.~Datta, K.~Kong and K.~T.~Matchev,
  Phys.\ Rev.\ D {\bf 72}, 096006 (2005)
  [Erratum-ibid.\ D {\bf 72}, 119901 (2005)]
  [arXiv:hep-ph/0509246].
\\
  A.~Datta, G.~L.~Kane and M.~Toharia,
  arXiv:hep-ph/0510204.
\\
  H.~C.~Cheng, I.~Low and L.~T.~Wang,
  arXiv:hep-ph/0510225.
\\
  P.~Meade and M.~Reece,
  arXiv:hep-ph/0601124.

\bibitem{Binetruy:2003cy}
  P.~Binetruy, G.~L.~Kane, B.~D.~Nelson, L.~T.~Wang and T.~T.~Wang,
  Phys.\ Rev.\ D {\bf 70}, 095006 (2004)
  [arXiv:hep-ph/0312248].

\bibitem{Weinberg:1975gm}
  S.~Weinberg,
  Phys.\ Rev.\ D {\bf 13}, 974 (1976).
\\
  T.~W.~Appelquist, D.~Karabali and L.~C.~R.~Wijewardhana,
  Phys.\ Rev.\ Lett.\  {\bf 57}, 957 (1986).
\\
  B.~Holdom and J.~Terning,
  Phys.\ Lett.\ B {\bf 247}, 88 (1990).
\\
  C.~T.~Hill,
  Phys.\ Lett.\ B {\bf 345}, 483 (1995)
  [arXiv:hep-ph/9411426].
  M.~A.~Luty and T.~Okui,
  arXiv:hep-ph/0409274.
\\
  T.~Appelquist, N.~Christensen, M.~Piai and R.~Shrock,
  Phys.\ Rev.\ D {\bf 70}, 093010 (2004)
  [arXiv:hep-ph/0409035].
\\
  M.~Perelstein,
  JHEP {\bf 0410}, 010 (2004)
  [arXiv:hep-ph/0408072].

\bibitem{Arkani-Hamed:1998rs}
  N.~Arkani-Hamed, S.~Dimopoulos and G.~R.~Dvali,
  Phys.\ Lett.\ B {\bf 429}, 263 (1998)
  [arXiv:hep-ph/9803315].
 \\
  N.~Arkani-Hamed, S.~Dimopoulos and G.~R.~Dvali,
  Phys.\ Rev.\ D {\bf 59}, 086004 (1999)
  [arXiv:hep-ph/9807344].
\\
  N.~Arkani-Hamed, S.~Dimopoulos, G.~R.~Dvali and J.~March-Russell,
  Phys.\ Rev.\ D {\bf 65}, 024032 (2002)
  [arXiv:hep-ph/9811448].
\\
  N.~Arkani-Hamed, S.~Dimopoulos, G.~R.~Dvali and N.~Kaloper,
  Phys.\ Rev.\ Lett.\  {\bf 84}, 586 (2000)
  [arXiv:hep-th/9907209].
\\
Phys.\ Lett.\ B {\bf 436}, 257 (1998)  [arXiv:hep-ph/9804398].
\\
I.~Antoniadis,  
Phys.\ Lett.\ B {\bf 246}, 377 (1990). 


\bibitem{Randall:1998uk}
  L.~Randall and R.~Sundrum,
  Nucl.\ Phys.\ B {\bf 557}, 79 (1999)
  [arXiv:hep-th/9810155].
  \\
  L.~Randall and R.~Sundrum,
  Phys.\ Rev.\ Lett.\  {\bf 83}, 3370 (1999)
  [arXiv:hep-ph/9905221].
 \\
  L.~Randall and R.~Sundrum,
  Phys.\ Rev.\ Lett.\  {\bf 83}, 4690 (1999)
  [arXiv:hep-th/9906064].


\bibitem{Nomura:2003du}
  Y.~Nomura,
  JHEP {\bf 0311}, 050 (2003)
  [arXiv:hep-ph/0309189].
\\
  C.~Csaki, C.~Grojean, J.~Hubisz, Y.~Shirman and J.~Terning,
  Phys.\ Rev.\ D {\bf 70}, 015012 (2004)
  [arXiv:hep-ph/0310355].
\\
  C.~Csaki, C.~Grojean, H.~Murayama, L.~Pilo and J.~Terning,
  Phys.\ Rev.\ D {\bf 69}, 055006 (2004)
  [arXiv:hep-ph/0305237].
\\
  C.~Csaki, C.~Grojean, L.~Pilo and J.~Terning,
  Phys.\ Rev.\ Lett.\  {\bf 92}, 101802 (2004)
  [arXiv:hep-ph/0308038].
  \\
  R.~S.~Chivukula, E.~H.~Simmons, H.~J.~He, M.~Kurachi and M.~Tanabashi,
  Phys.\ Lett.\ B {\bf 603}, 210 (2004)
  [arXiv:hep-ph/0408262].

\bibitem{Kaplan:1983sm}
  D.~B.~Kaplan, H.~Georgi and S.~Dimopoulos,
  Phys.\ Lett.\ B {\bf 136}, 187 (1984).
  \\
  H.~Georgi and D.~B.~Kaplan,
  Phys.\ Lett.\ B {\bf 145}, 216 (1984).
  \\
  R.~S.~Chivukula, C.~Hoelbling and N.~J.~Evans,
  Phys.\ Rev.\ Lett.\  {\bf 85}, 511 (2000)
  [arXiv:hep-ph/0002022].
  \\
  K.~Agashe, R.~Contino and A.~Pomarol,
  Nucl.\ Phys.\ B {\bf 719}, 165 (2005)
  [arXiv:hep-ph/0412089].

\bibitem{Arkani-Hamed:2002qx}
  N.~Arkani-Hamed, A.~G.~Cohen, E.~Katz, A.~E.~Nelson, T.~Gregoire and J.~G.~Wacker,
  JHEP {\bf 0208}, 021 (2002)
  [arXiv:hep-ph/0206020].
  \\
  N.~Arkani-Hamed, A.~G.~Cohen, E.~Katz and A.~E.~Nelson,
  JHEP {\bf 0207}, 034 (2002)
  [arXiv:hep-ph/0206021].
 \\
  S.~Chang and J.~G.~Wacker,
  Phys.\ Rev.\ D {\bf 69}, 035002 (2004)
  [arXiv:hep-ph/0303001].
  \\
  M.~Schmaltz,
  JHEP {\bf 0408}, 056 (2004)
  [arXiv:hep-ph/0407143].
  \\
  M.~Schmaltz and D.~Tucker-Smith,
  arXiv:hep-ph/0502182.


\bibitem{Arkani-Hamed:2004fb}
  N.~Arkani-Hamed and S.~Dimopoulos,
  JHEP {\bf 0506}, 073 (2005)
  [arXiv:hep-th/0405159].
\\
  G.~F.~Giudice and A.~Romanino,
  Nucl.\ Phys.\ B {\bf 699}, 65 (2004)
  [Erratum-ibid.\ B {\bf 706}, 65 (2005)]
  [arXiv:hep-ph/0406088].
 \\
  N.~Arkani-Hamed, S.~Dimopoulos, G.~F.~Giudice and A.~Romanino,
  Nucl.\ Phys.\ B {\bf 709}, 3 (2005)
  [arXiv:hep-ph/0409232].

\bibitem{kklt03}
S.~Kachru, R.~Kallosh, A.~Linde, and S.~P. Trivedi, ``De Sitter
vacua in string
  theory.'' {\em Phys. Rev.} {\bf D68} (2003) 046005,

\bibitem{Balasubramanian:2005zx}
  V.~Balasubramanian, P.~Berglund, J.~P.~Conlon and F.~Quevedo,
   ``Systematics of moduli stabilisation in Calabi-Yau flux
  JHEP {\bf 0503}, 007 (2005)
  [arXiv:hep-th/0502058].
  J.~P.~Conlon, F.~Quevedo and K.~Suruliz,
   ``Large-volume flux compactifications: Moduli spectrum and D3/D7 soft
  JHEP {\bf 0508}, 007 (2005)
  [arXiv:hep-th/0505076].
  B.~C.~Allanach, F.~Quevedo and K.~Suruliz,
  arXiv:hep-ph/0512081.

\bibitem{Burgess:2003ic}
  C.~P.~Burgess, R.~Kallosh and F.~Quevedo,
  JHEP {\bf 0310}, 056 (2003)
  [arXiv:hep-th/0309187].

\bibitem{Choi:2004sx}
  K.~Choi, A.~Falkowski, H.~P.~Nilles, M.~Olechowski and S.~Pokorski,
  JHEP {\bf 0411}, 076 (2004)
  [arXiv:hep-th/0411066].
\\
  K.~Choi, A.~Falkowski, H.~P.~Nilles and M.~Olechowski,
  Nucl.\ Phys.\ B {\bf 718}, 113 (2005)
  [arXiv:hep-th/0503216].
\\
  A.~Falkowski, O.~Lebedev and Y.~Mambrini,
  JHEP {\bf 0511}, 034 (2005)
  [arXiv:hep-ph/0507110].
\\
  K.~Choi, K.~S.~Jeong, T.~Kobayashi and K.~i.~Okumura,
  Phys.\ Lett.\ B {\bf 633}, 355 (2006)
  [arXiv:hep-ph/0508029].
\\
  H.~Baer, E.~K.~Park, X.~Tata and T.~T.~Wang,
   ``Collider and dark matter searches in models with mixed modulus-anomaly
  arXiv:hep-ph/0604253.

\bibitem{GaNeWu99}
  M.K.~Gaillard, B.D.~Nelson and Y.-Y.~Wu,
  Phys. Lett. B {\bf 459} {\rm (1999) 549}.
\\
  J.~Bagger, T.~Moroi and E.~Poppitz,
  {\it JHEP} {\bf 0004} (2000) 009.

\bibitem{Quevedo06}
F.~Quevedo, Talk at KITP:
http://online.itp.ucsb.edu/online/strings06/quevedo/\\Private
Communication.

\bibitem{Conlon:2006us}
  J.~P.~Conlon and F.~Quevedo,
  JHEP {\bf 0606}, 029 (2006)
  [arXiv:hep-th/0605141].

\bibitem{Acharya:2006ia}
  B.~Acharya, K.~Bobkov, G.~Kane, P.~Kumar and D.~Vaman,
  Phys.\ Rev.\ Lett.\  {\bf 97}, 191601 (2006)
  [arXiv:hep-th/0606262].
\\
  B.~S.~Acharya, K.~Bobkov, G.~L.~Kane, P.~Kumar and J.~Shao,
  arXiv:hep-th/0701034.


\bibitem{Djouadi:2002ze}
  A.~Djouadi, J.~L.~Kneur and G.~Moultaka,
  [arXiv:hep-ph/0211331]. 

\bibitem{Allanach:2001kg}
  B.~C.~Allanach,
  Comput.\ Phys.\ Commun.\  {\bf 143}, 305 (2002)
  [arXiv:hep-ph/0104145].

\bibitem{Bennett:2003bz}
  C.~L.~Bennett {\it et al.},
  Astrophys.\ J.\ Suppl.\  {\bf 148}, 1 (2003)
  [arXiv:astro-ph/0302207].  
  \\
  D.~N.~Spergel {\it et al.}  [WMAP Collaboration],
  Astrophys.\ J.\ Suppl.\  {\bf 148}, 175 (2003)
  [arXiv:astro-ph/0302209]. 
\bibitem{Abe:2001hk}
  K.~Abe {\it et al.}  [Belle Collaboration],
  Phys.\ Lett.\ B {\bf 511}, 151 (2001)
  [arXiv:hep-ex/0103042].
  \\
  D.~Cronin-Hennessy {\it et al.}  [CLEO Collaboration],
  Phys.\ Rev.\ Lett.\  {\bf 87}, 251808 (2001)
  [arXiv:hep-ex/0108033].
  \\
  R.~Barate {\it et al.}  [ALEPH Collaboration],
  Phys.\ Lett.\ B {\bf 429}, 169 (1998). 

\bibitem{Bennett:2002jb}
  G.~W.~Bennett {\it et al.}  [Muon g-2 Collaboration],
  Phys.\ Rev.\ Lett.\  {\bf 89}, 101804 (2002)
  [Erratum-ibid.\  {\bf 89}, 129903 (2002)]
  [arXiv:hep-ex/0208001].
 \\
  G.~W.~Bennett {\it et al.}  [Muon g-2 Collaboration],
  Phys.\ Rev.\ Lett.\  {\bf 92}, 161802 (2004)
  [arXiv:hep-ex/0401008]. 

\bibitem{Sjostrand:2003wg}
  T.~Sjostrand, L.~Lonnblad, S.~Mrenna and P.~Skands,
LU TP 06-13, FERMILAB-PUB-06-052-CD-T [hep-ph/0603175].

\bibitem{pgs}
http://www.physics.ucdavis.edu/~conway/research/software/pgs/pgs4-general.htm.

\bibitem{LHCO}
http://physics.princeton.edu/~verlinde/research/lhco/ \\
http://ph-dep-th.web.cern.ch/ph-dep-th/lhcOlympics/lhcolympicsII.html. \\
http://www.phys.washington.edu/users/strasslr/LHCO.BBpage.html.

\bibitem{Baer:1995va}
  H.~Baer, C.~h.~Chen, F.~Paige and X.~Tata,
  Phys.\ Rev.\ D {\bf 53}, 6241 (1996)
  [arXiv:hep-ph/9512383].
\\
  H.~Baer, C.~h.~Chen, F.~Paige and X.~Tata,
  Phys.\ Rev.\ D {\bf 52}, 2746 (1995)
  [arXiv:hep-ph/9503271].

\bibitem{Arkani-Hamed:2005px}
  N.~Arkani-Hamed, G.~L.~Kane, J.~Thaler and L.~T.~Wang,
  JHEP {\bf 0608}, 070 (2006)
  [arXiv:hep-ph/0512190].

\bibitem{Bartl:1994bu}
  A.~Bartl, W.~Majerotto and W.~Porod,
  Z.\ Phys.\ C {\bf 64}, 499 (1994)
  [Erratum-ibid.\ C {\bf 68}, 518 (1995)].

\bibitem{Choi:1997cm}
  K.~Choi, H.~B.~Kim and C.~Munoz,
  Phys.\ Rev.\ D {\bf 57}, 7521 (1998)
  [arXiv:hep-th/9711158].

\bibitem{Ellis:2003cw}
  J.~R.~Ellis, K.~A.~Olive, Y.~Santoso and V.~C.~Spanos,
  Phys.\ Lett.\ B {\bf 565}, 176 (2003)
  [arXiv:hep-ph/0303043].
\\
  J.~L.~Feng,
  eConf {\bf C0307282}, L11 (2003)
  [arXiv:hep-ph/0405215].

\bibitem{Brignole:1997dp}
  A.~Brignole, L.~E.~Ibanez and C.~Munoz,
  arXiv:hep-ph/9707209.

\bibitem{Casas:1990qi}
  J.~A.~Casas, Z.~Lalak, C.~Munoz and G.~G.~Ross,
  Nucl.\ Phys.\ B {\bf 347}, 243 (1990).
\\
  B.~de Carlos, J.~A.~Casas and C.~Munoz,
  Nucl.\ Phys.\ B {\bf 399}, 623 (1993)
  [arXiv:hep-th/9204012].
\\
  B.~de Carlos, J.~A.~Casas and C.~Munoz,
  Phys.\ Lett.\ B {\bf 299}, 234 (1993)
  [arXiv:hep-ph/9211266].
\\
  A.~de la Macorra and G.~G.~Ross,
  Nucl.\ Phys.\ B {\bf 404}, 321 (1993)
  [arXiv:hep-ph/9210219].

\bibitem{Denef:2005mm}
 F.~Denef, M.~R.~Douglas, B.~Florea, A.~Grassi and S.~Kachru,
arXiv:hep-th/0503124.

\bibitem{Kane:2002qp}
  G.~L.~Kane, J.~D.~Lykken, S.~Mrenna, B.~D.~Nelson, L.~T.~Wang and T.~T.~Wang,
  Phys.\ Rev.\ D {\bf 67}, 045008 (2003)
  [arXiv:hep-ph/0209061].


\bibitem{Allanach:2004my}
  B.~C.~Allanach, D.~Grellscheid and F.~Quevedo,
  JHEP {\bf 0407}, 069 (2004)
  [arXiv:hep-ph/0406277].

\bibitem{douglas-taylor}
  M.~R.~Douglas and W.~Taylor,
  arXiv:hep-th/0606109.
\\
Talk at KITP : http://online.itp.ucsb.edu/online/strings06/taylor/ \\
Private Communication.

\bibitem{Kneur:1998gy}
  J.~L.~Kneur and G.~Moultaka,
  Phys.\ Rev.\ D {\bf 59}, 015005 (1999)
  [arXiv:hep-ph/9807336].
\\
  A.~Djouadi {\it et al.}  [MSSM Working Group],
  arXiv:hep-ph/9901246.
\\
  J.~L.~Kneur and G.~Moultaka,
  Phys.\ Rev.\ D {\bf 61}, 095003 (2000)
  [arXiv:hep-ph/9907360].
\\
  C.~G.~Lester, M.~A.~Parker and M.~J.~White,
  JHEP {\bf 0601}, 080 (2006)
  [arXiv:hep-ph/0508143].
\\
  J.~A.~Aguilar-Saavedra {\it et al.},
  arXiv:hep-ph/0511344.
\\
  B.~C.~Allanach {\it et al.},
  arXiv:hep-ph/0602198.

\bibitem{LHCO-talks}
http://physics.princeton.edu/~verlinde/research/lhco/\\
http://ph-dep-th.web.cern.ch/ph-dep-th/lhcOlympics/2ndWin06/program.html


\bibitem{Dixon:1986qv}
  L.~J.~Dixon, D.~Friedan, E.~J.~Martinec and S.~H.~Shenker,
  Nucl.\ Phys.\ B {\bf 282}, 13 (1987);\\
  S.~Hamidi and C.~Vafa,
  Nucl.\ Phys.\ B {\bf 279}, 465 (1987);\\
  I.~Antoniadis, J.~R.~Ellis, J.~S.~Hagelin and D.~V.~Nanopoulos,
  Phys.\ Lett.\ B {\bf 231}, 65 (1989);\\
  G.~Cleaver, M.~Cvetic, J.~R.~Espinosa, L.~L.~Everett, P.~Langacker and J.~Wang,
  Phys.\ Rev.\ D {\bf 59}, 115003 (1999)
  [arXiv:hep-ph/9811355];\\
  G.~B.~Cleaver, A.~E.~Faraggi, D.~V.~Nanopoulos and J.~W.~Walker,
  Mod.\ Phys.\ Lett.\ A {\bf 15}, 1191 (2000)
  [arXiv:hep-ph/0002060];\\
  L.~J.~Dixon, V.~Kaplunovsky and J.~Louis,
  Nucl.\ Phys.\ B {\bf 329}, 27 (1990);\\
  L.~J.~Dixon, V.~Kaplunovsky and J.~Louis,
  Nucl.\ Phys.\ B {\bf 355}, 649 (1991);\\
  L.~E.~Ibanez and D.~Lust,
  Nucl.\ Phys.\ B {\bf 382}, 305 (1992)
  [arXiv:hep-th/9202046];\\
H.~P.~Nilles and S.~Stieberger,
Nucl.\ Phys.\ B {\bf 499}, 3 (1997) [arXiv:hep-th/9702110].
\\
P.~Binetruy, M.~K.~Gaillard and B.~D.~Nelson,
Nucl.\ Phys.\ B {\bf 604}, 32 (2001) [arXiv:hep-ph/0011081].

\bibitem{Font:1989aj}
  A.~Font, L.~E.~Ibanez, F.~Quevedo and A.~Sierra,
  Nucl.\ Phys.\ B {\bf 331}, 421 (1990).

\bibitem{Shenker:1990uf}
  S.~H.~Shenker,
RU-90-47
{\it Presented at the Cargese Workshop on Random Surfaces, Quantum
Gravity and Strings, Cargese, France, May 28 - Jun 1, 1990}
\\
  J.~Polchinski,
  Phys.\ Rev.\ D {\bf 50}, 6041 (1994)
  [arXiv:hep-th/9407031].
\\
  E.~Silverstein,
  Phys.\ Lett.\ B {\bf 396}, 91 (1997)
  [arXiv:hep-th/9611195].

\bibitem{Binetruy:1996xj}
  P.~Binetruy, M.~K.~Gaillard and Y.~Y.~Wu,
  Nucl.\ Phys.\ B {\bf 481}, 109 (1996)
  [arXiv:hep-th/9605170].
\\
  P.~Binetruy, M.~K.~Gaillard and Y.~Y.~Wu,
  Nucl.\ Phys.\ B {\bf 493}, 27 (1997)
  [arXiv:hep-th/9611149].
\\
  P.~Binetruy, M.~K.~Gaillard and Y.~Y.~Wu,
  Phys.\ Lett.\ B {\bf 412}, 288 (1997)
  [arXiv:hep-th/9702105].

\bibitem{HW1} P. Horava and E. Witten, Nucl. Phys. B {\bf 460} (1996)
506.
\\
P. Horava and E. Witten, Nucl. Phys. B {\bf 475}(1996) 94.
\\
P. Horava, Phys. Rev. D {\bf 54}(1996) 7561.

\bibitem{Witten} E. Witten, Nucl. Phys. B
               {\bf 471}(1996) 135.
\\
  A.~Lukas, B.~A.~Ovrut and D.~Waldram,
  Nucl.\ Phys.\ B {\bf 532}, 43 (1998)
  [arXiv:hep-th/9710208].
\\
A. Lukas, B.~A. Ovrut, D. Waldram, JHEP
              {\bf 9904} (1999) 009.

\bibitem{BD}
T. Banks and M. Dine, Nucl. Phys. B {\bf 479} (1996) 137.
\\
  T.~Banks and M.~Dine,
  arXiv:hep-th/9609046.

\bibitem{BIMS}
A. Brignole, L.E. Ib\'a\~nez, C. Mu\~noz and C. Scheich,
               Z. Phys.  C74 (1997) 157.
\\
H.~B. Kim and C. Mu\~noz, Z. Phys. C {\bf 75} (1997) 367.

\bibitem{jing}
G.L. Kane, P. Kumar and J. Shao, \textit{To appear}.

\bibitem{Lukas:1998hk}
  A.~Lukas, B.~A.~Ovrut and D.~Waldram,
  Phys.\ Rev.\ D {\bf 59}, 106005 (1999)
  [arXiv:hep-th/9808101].
\\
  A.~Lukas, B.~A.~Ovrut and D.~Waldram,
  Fortsch.\ Phys.\ {\bf 48}, 167 (2000)
  [arXiv:hep-th/9903144].

\bibitem{Cerdeno:1999ur}
  D.~G.~Cerdeno and C.~Munoz,
  Phys.\ Rev.\ D {\bf 61}, 016001 (2000)
  [arXiv:hep-ph/9904444].
\\
  D.~G.~Cerdeno and C.~Munoz,
  Phys.\ Rev.\ D {\bf 66}, 115007 (2002)
  [arXiv:hep-ph/0206299].

\bibitem{Ibanez:2001jhep}
L.~E.~Ibanez, F.~Marchesano and R.~Rabadan, JHEP {\bf 0111}, 002
(2001) [arXiv:hep-th/0105155];
\\
C.~Angelantonj and A.~Sagnotti,
Phys.\ Rept.\ {\bf 371} (2002) 1 [Erratum-ibid.\  {\bf 376} (2003)
339] [hep-th/0204089].
\\
R.~Blumenhagen, L.~G\"orlich and B.~K\"ors,
Nucl.\ Phys.\ B {\bf 569} (2000) 209 [hep-th/9908130];
\\
R.~Blumenhagen, L.~G\"orlich and B.~K\"ors,
JHEP {\bf 0001} (2000) 040 [hep-th/9912204].
\\
G.~Pradisi,
Nucl.\ Phys.\ B {\bf 575} (2000) 134 [hep-th/9912218].
\\
R.~Blumenhagen, L.~G\"orlich and B.~K\"ors,
hep-th/0002146.
\\
C.~Angelantonj, I.~Antoniadis, E.~Dudas and A.~Sagnotti,
Phys.\ Lett.\ B {\bf 489} (2000) 223 [hep-th/0007090].
\\
S.~F\"orste, G.~Honecker and R.~Schreyer,
Nucl.\ Phys.\ B {\bf 593} (2001) 127 [hep-th/0008250].
\\
C.~Angelantonj and A.~Sagnotti,
hep-th/0010279.
\\
S.~F\"orste, G.~Honecker and R.~Schreyer,
JHEP {\bf 0106} (2001) 004 [hep-th/0105208].
\\
Nucl.\ Phys.\ B {\bf 616}, 3 (2001) [arXiv:hep-th/0107138];
\\
G.~Shiu and S.~H.~H.~Tye,
Phys.\ Rev.\ D {\bf 58}, 106007 (1998) [arXiv:hep-th/9805157];
\\
M.~Cvetic, P.~Langacker and G.~Shiu,
Phys.\ Rev.\ D {\bf 66}, 066004 (2002) [arXiv:hep-ph/0205252];
\\
D.~Lust and S.~Stieberger,
[arXiv:hep-th/0302221];
\\
I.~Antoniadis, E.~Kiritsis and T.~N.~Tomaras,
Phys.\ Lett.\ B {\bf 486}, 186 (2000) [arXiv:hep-ph/0004214];
\\
D.~Cremades, L.~E.~Ibanez and F.~Marchesano,
JHEP {\bf 0207}, 009 (2002) [arXiv:hep-th/0201205];
\\
R.~Blumenhagen, D.~Lust and S.~Stieberger,
JHEP {\bf 0307}, 036 (2003) [arXiv:hep-th/0305146];
\\
D.~Cremades, L.~E.~Ibanez and F.~Marchesano,
JHEP {\bf 0307}, 038 (2003) [arXiv:hep-th/0302105];
\\
F.~Marchesano,
JHEP {\bf 0405}, 079 (2004) [arXiv:hep-th/0404229];
\\
M.~Cvetic and I.~Papadimitriou,
Phys.\ Rev.\ D {\bf 68}, 046001 (2003) [Erratum-ibid.\ D {\bf 70},
029903 (2004)] [arXiv:hep-th/0303083];
\\
S.~A.~Abel and A.~W.~Owen,
Nucl.\ Phys.\ B {\bf 682}, 183 (2004) [arXiv:hep-th/0310257];
Nucl.\ Phys.\ B {\bf 664}, 3 (2003)[arXiv:hep-th/0303124];
\\
B.~Kors and P.~Nath,
Nucl.\ Phys.\ B {\bf 681}, 77 (2004) [arXiv:hep-th/0309167];
\\
D.~Lust, P.~Mayr, R.~Richter and S.~Stieberger,
Nucl.\ Phys.\ B {\bf 696}, 205 (2004) [arXiv:hep-th/0404134].
\\
I.~Antoniadis, E.~Kiritsis, J.~Rizos and T.~N.~Tomaras,
Nucl.\ Phys.\ B {\bf 660}, 81 (2003) [arXiv:hep-th/0210263].
\\
P.~Anastasopoulos, T.~P.~T.~Dijkstra, E.~Kiritsis and A.~N.~Schellekens,
arXiv:hep-th/0605226.

\bibitem{Blumenhagen:2005mu}
  R.~Blumenhagen, M.~Cvetic, P.~Langacker and G.~Shiu,
  [arXiv:hep-th/0502005].


\bibitem{Cvetic:2003ch}
M.~Bertolini, M.~Billo, A.~Lerda, J.~F.~Morales and R.~Russo,
Nucl.\ Phys.\ B {\bf 743}, 1 (2006) [arXiv:hep-th/0512067].
\\
M.~Cvetic and I.~Papadimitriou,
Phys.\ Rev.\ D {\bf 68}, 046001 (2003) [Erratum-ibid.\ D {\bf 70},
029903 (2004)] [arXiv:hep-th/0303083];
\\
D.~Cremades, L.~E.~Ibanez and F.~Marchesano,
JHEP {\bf 0307}, 038 (2003) [arXiv:hep-th/0302105].
\\
D.~Lust, P.~Mayr, R.~Richter and S.~Stieberger,
Nucl.\ Phys.\ B {\bf 696}, 205 (2004) [arXiv:hep-th/0404134];

\bibitem{Cvetic:2003yd}
  M.~Cvetic, P.~Langacker and J.~Wang,
  Phys.\ Rev.\ D {\bf 68}, 046002 (2003)
  [arXiv:hep-th/0303208].


\bibitem{gkp01}
S.~B. Giddings, S.~Kachru, and J.~Polchinski, ``Hierarchies from
fluxes in
  string compactifications,'' {\em Phys. Rev.} {\bf D66} (2002) 106006;
\\
P.~G. C{\' a}mara, L.~E. Ib{\' a}{\~ n}ez, and A.~M. Uranga,
``Flux-induced
  SUSY-breaking soft terms;'' {\em Nucl. Phys.} {\bf B689} (2004) 195--242;
\\
J.~Louis, ``Soft supersymmetry
  breaking in Calabi-Yau orientifolds with D-branes and fluxes,'' {\em Nucl.
  Phys.} {\bf B690} (2004) 21--61;
\\
P.~G. C{\' a}mara, L.~E. Ib{\'a}{\~ n}ez, and A.~M. Uranga,
``Flux-induced
  SUSY-breaking soft terms on D7-D3 brane systems;''
\\
M.~Cveti{\v c}, T.~Li, and T.~Liu, ``Standard-like Models as Type
IIB Flux
  Vacua;''

\bibitem{Kane:2004hm}
  G.~L.~Kane, P.~Kumar, J.~D.~Lykken and T.~T.~Wang,
  Phys.\ Rev.\ D {\bf 71}, 115017 (2005)
  [arXiv:hep-ph/0411125].

\end{thebibliography}
\end{document}